\documentclass[12pt, letterpaper, oneside]{book} 
\usepackage[top=3cm,bottom=3cm,right=2cm,left=2cm]{geometry}
\usepackage[utf8]{inputenc}
\usepackage{graphicx}
\usepackage{url}
\usepackage{hyperref}
\usepackage{amsmath}
\usepackage{amsfonts}
\usepackage{amssymb}
\usepackage{amsthm}
\usepackage{breqn}
\usepackage{float}
\usepackage{pdfpages}

\begin{document}


\thispagestyle{plain}
\includegraphics[width=.74in, height=1in,keepaspectratio=true]{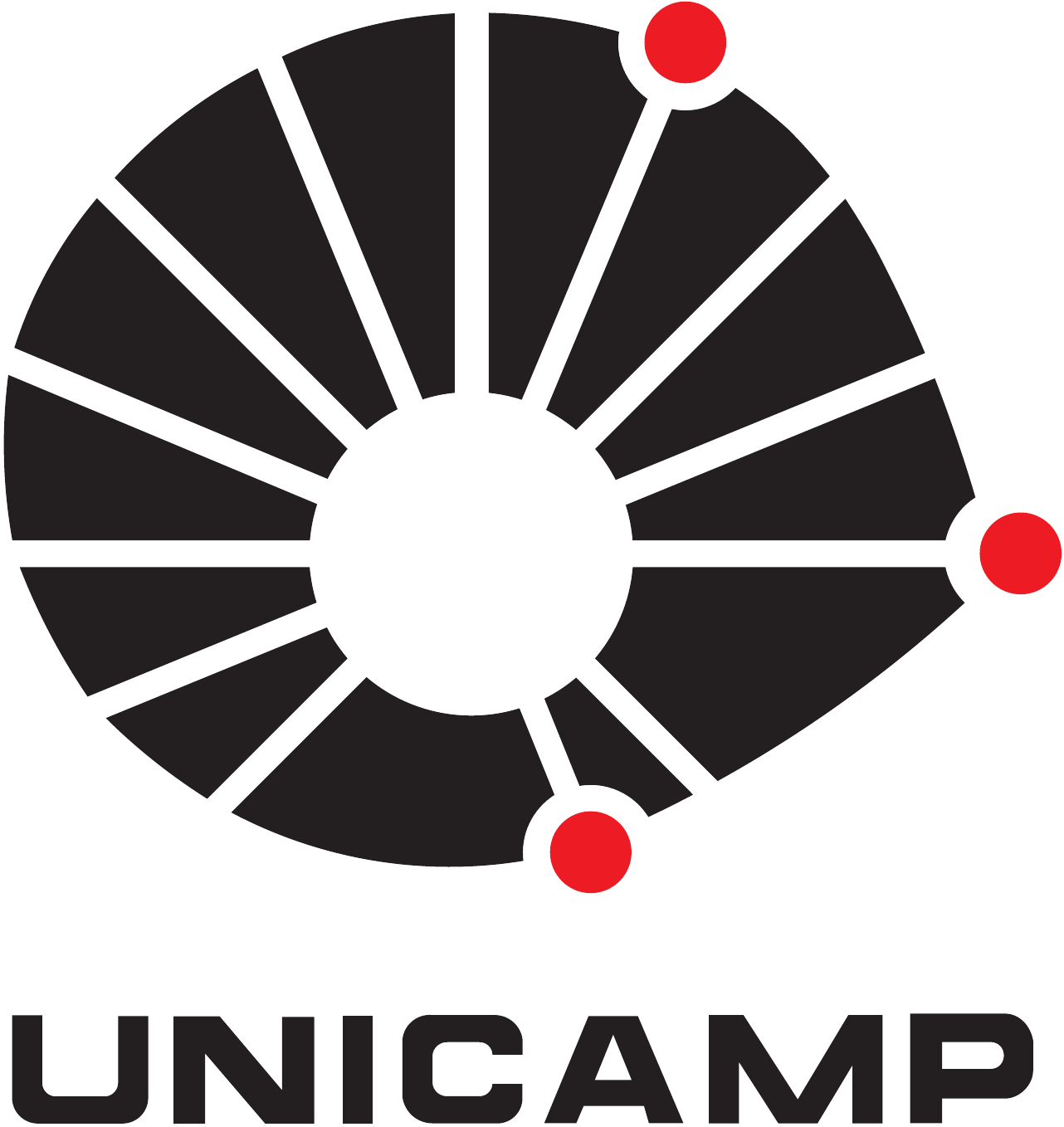}
\begin{center}
  {\large\textbf{\textsc{Universidade Estadual de Campinas}}
  \vspace{.5cm}

 Instituto de Física ``Gleb Wataghin''}
\end{center}
\vspace*{1cm}
\begin{center}

  {\Large\textsc{Carolina Arruda Moreira}}
\end{center}
\vspace{4cm}

\begin{center}
  {\Large\textbf{\textsc{Synchronization induced by external forces in modular networks}}}
\end{center}
\vfill

\begin{center}
  {\Large\textbf{\textsc{Sincronização induzida por forças externas em redes modulares}}}
\end{center}
\vfill

\vspace{4cm}

\begin{center}
  \textbf{CAMPINAS \\ 2020}
\end{center}

\vspace{4cm}

\thispagestyle{plain}

\vfill
\begin{center}
  {\large\textbf{\textsc{Carolina Arruda Moreira}}}
\end{center}
\vfill

\begin{center}
  {\Large\textbf{\textsc{Synchronization induced by external forces in modular networks}}}
\end{center}
\vfill

\begin{center}
  {\Large\textbf{\textsc{Sincronização induzida por forças externas em redes modulares}}}
\end{center}
\vfill

\begin{flushright}
  \begin{minipage}[c]{.5\textwidth}
    Thesis presented to the Institute of Physics ``Gleb Wataghin'' in the University of Campinas in partial fulfillment of the requirements for the degree of Doctor in Sciences, with emphasis on Physics.
  \end{minipage}
\end{flushright}
\vspace{1.5cm}

\begin{flushright}
  \begin{minipage}[c]{.5\textwidth}
    Tese apresentada ao Instituto de Física ``Gleb Wataghin'' da Universidade
    Estadual de Campinas como parte dos requisitos exigidos para a obtenção do título de Doutora em Ciências, na área de Física.
  \end{minipage}
\end{flushright}
\vspace{1.5cm}

\noindent
\textbf{Supervisor/orientador: \\ Marcus A. M. de Aguiar
}
\vspace{.25cm}

\vspace{.5cm}

\noindent
\begin{minipage}[c]{.3\textwidth}
  {\footnotesize\textsc{Este exemplar corresponde à versão final da tese
  defendida pela aluna Carolina Arruda Moreira e orientada pelo Prof. Dr. Marcus A. M. de Aguiar.}}
\end{minipage}
\vspace{.5cm}

\begin{center}
  {\small\textbf{\textsc{ Campinas \\ 2020}}}
\end{center}

\newpage
\begin{center}
  \large{\textbf{Resumo}}
\end{center}  
  Neste trabalho estudamos a sincronização de osciladores de Kuramoto sujeitos a forças externas em redes modulares complexas. A motivação está na dinâmica neuronal que ocorre durante o processamento de informação no córtex cerebral que parece estar relacionada ao disparo síncrono de grupos de neurônios. A organização dos neurônios é modular, com agrupamentos associados a diferentes funções e estruturas cerebrais, e precisa responder constantemente a estímulos externos. Anormalidades no processo de sincronização, como a ativação de múltiplos módulos têm sido associadas à doenças como epilepsia e Alzheimer. Nesse contexto, estudamos o comportamento de osciladores de Kuramoto forçados, onde apenas uma fração deles é submetida a uma força externa periódica. Quando todos os osciladores recebem o estímulo externo o sistema sempre sincroniza com a força externa se a sua intensidade for suficientemente grande. Mostramos que as condições para a sincronização global dependem da fração de nós forçada e da topologia da rede e das intensidades do acoplamento interno e da força externa. Desenvolvemos cálculos numéricos e analíticos para a força crítica que leva a rede à sincronização global em função da fração de osciladores forçados. Como uma aplicação estudamos a resposta da rede de junções elétricas do \textit{C. elegans} ao estímulo externo usando o modelo de Kuramoto parcialmente forçado, aplicando a força a grupos específicos de neurônios. Os estímulos foram aplicados a três módulos topológicos, dois gânglios, especificados por sua localização anatômica, e aos grupos funcionais compostos por todos os neurônios sensoriais e motores. Encontramos que os módulos topológicos não contêm grupos puramente anatômicos ou classes funcionais e que estimular diferentes classes neuronais leva a respostas muito diferentes, medidas em termos de sincronização e correlações de velocidade de fase. Em todos os casos a estrutura modular impede a sincronização global, protegendo o sistema de falhas. As respostas aos estímulos aplicados aos módulos topológicos e funcionais mostram padrões pronunciados de correlação ou anti-correlação com outros módulos que não foram observados quando o estímulo foi aplicado a um gânglio com neurônios funcionais mistos. Todos os códigos e dados utilizados nesta tese estão disponível em \cite{github}.
  
\newpage
\begin{center}
  \large{\textbf{Abstract}}
\end{center}

In this work we study the synchronization of Kuramoto oscillators driven by external forces in complex modular networks. The motivation is the neuronal dynamics that takes place during information processing in the neural cortex, which seems to be related to the synchronous firing of groups of neurons. The neuron organization is modular, with clusters associated to different functions and brain structures, and need to constantly respond to external stimuli. Abnormalities in the process of synchronization, such as the activation of multiple modules, have been associated with epilepsy and Alzheimer's disease. In this context, we study the behavior of forced Kuramoto oscillators where only a fraction of them is subjected to a periodic external force. When all oscillators receive the external drive the system always synchronize with the periodic force if its intensity is sufficiently large. We show that the conditions for global synchronization depend on the fraction of nodes being forced and on network topology, strength of internal couplings and intensity of external forcing. We develop numerical and analytical calculations for the critical force for global synchronization as a function of the fraction of forced oscillators. As an application we study the response of the electric junction \textit{C. elegans} network to external stimuli using the partially forced Kuramoto model and applying the force to specific groups of neurons. Stimuli were applied to three topological modules, two ganglia, specified by their anatomical localization, and to the functional groups composed of all sensory and motoneurons. We found that topological modules do not contain purely anamotical groups or functional classes, and that stimulating different classes of neurons lead to very different responses, measured in terms of synchronization and phase velocity correlations. In all cases the modular structure hindered full synchronization, protecting the system from seizures. The responses to stimuli applied to topological and functional modules showed pronounced patterns of correlation or anti-correlation with other modules that were not observed when the stimulus was applied to a ganglion with mixed functional neurons. All codes and data used in this thesis are available in \cite{github}.

\tableofcontents

\chapter{Introduction}

Nature is full of oscillatory systems. Many of them exhibit regular behavior, as atoms vibrating around their equilibrium positions and planets orbiting around a center of gravity, while others show chaotic dynamics, as temperature and atmospheric pressure variations, electrical currents in specific circuits and fluctuations in stock exchanges. In biological sciences, oscillatory systems are also abundant and often need to work in synchrony to regulate physical activities, such as pacemaker cells in the heart \cite{Michaels704} and fireflies flashing collectively to help females find suitable mates \cite{Moiseff181,Buck1976}. There are evidences that synchronization also plays a key role in information processing in areas on the cerebral cortex \cite{steriade1997,kelso2014, Pikovsky2003,gray1994,Mackay1997}. Even the brain rest state activity is characterized by local rhythmic synchrony that induces spatiotemporally organized spontaneous activity at the level of the entire brain \cite{Deco2011}. Artificial systems, such as electrochemical oscillators \cite{Kiss2008} and coupled metronomes \cite{Pantaleone2002}, have also been studied. Another very common collective behavior is the incoherent claps of an audience starting to become a single pulse, where everyone applauds in the same time. All these examples are universal and emerge naturally, because the elements of the system produce rhythms by interacting with each other \cite{Pikovsky2003}.

One of the first observations of the synchronization phenomenon was reported by the Dutch scientist Christiaan Huygens in the middle of the 17th century when he noticed that a pair of pendulum clocks had their oscillations exactly out-of-phase when they were suspended in the same support. Three centuries later, radio engineers observed that two electrical coupled devices with initial different frequencies vibrate together after some time. In 1967, the biologist A. T. Winfree was the first to propose a mathematical model to describe synchronization \cite{Winfree}, but his equations were too difficut to solve. It was in 1974 that the Japanese physicist Y. Kuramoto proposed a useful simplification of the math \cite{Kuramoto1975}.

Kuramoto's model has become a paradigm in the study of synchronization and has been explored in connection with biological systems, neural networks and the social sciences \cite{Rodrigues2016, Acebron2005}. It describes a set of coupled harmonic oscillators with independent natural frequencies. Kuramoto demonstrated that for small values of the coupling the oscillators continued to move as if they were independent, but as the coupling increased beyond a critical value, a finite fraction of oscillators started to move together as if they were a single unit. The transition between the non-synchronized and the synchronized states characterizes a second order phase transition in the thermodynamic limit, where the system has infinite elements. This phenomenon can be seen in analogy to a ferromagnetic phase transition, where the magnetization increases continuously from zero as the temperature is lowered below a critical value, known as the Curie temperature.

Until recently all systems that had spontaneous synchronization exhibited a second-order phase transition. However, under specific conditions the Kuramoto system has an abrupt change on order parameter, which is a first order phase transition. This behavior is termed explosive synchronization and it has been studied in several works \cite{Kim2017, Kim2016, Wang2016, Wang2017, Leyva2012, Deco2013}, where the dynamic is dependent on the system's topology. This phenomenon is observed in real-world systems, occurring from electronic devices in the field of engineering, to neuroscience, as reported in \cite{Kim2017, Kim2016} the conscious-unconscious transition when the brain is awaking from anesthesia.

Synchronization in many biological systems, however, is not spontaneous, but frequently depends on external stimuli. Information processing in the brain, for example, might be triggered by visual, auditory or olfactory inputs \cite{Pikovsky2003}. Different patterns of synchronized neuronal firing are observed in the mammalian visual cortex when subjected to stimuli \cite{gray1994}. In the sensomotor cortex synchronized oscillations appear with amplitude and spatial patterns that depend on the task being performed \cite{gray1994, Mackay1997}. Synchronization of brain regions that are not directly related to the task in question can be associated to disorders like epilepsy, autism, schizophrenia and Alzheimer \cite{Uhlhaas2006,Schmidt2015}. In the heart, cardiac synchronization is induced by specialized cells in the sinoatrial node or by an artificial pacemaker that controls the rhythmic contractions of the whole heart \cite{Reece2012}. The periodic electrical impulses generated by pacemakers can be seen as an external periodic force that synchronizes the heart cells. Another example of driven system is the daily light-dark cycle on the organisms \cite{Liu1997}. In mammalians, cells specialized on the sleep control exhibit intrinsic oscillatory behavior whose connectivity is still unknown \cite{Liu2011}. The change in the light-dark cycle leads to a response in the circadian cycle mediated by these cells, which synchronize via external stimulus. Although the biological dynamics are quite complex, it is possible to map, under some circumstances, simplified models as the Kuramoto system, using known models of complex networks.

The phenomenon of induced synchronization has been studied by many authors since the late 80's \cite{Sakaguchi1988,Ott2008,Antonsen2008,Childs2008,Hindes2015}, where is natural to extend the Kuramoto model by including the influence of an external periodic force acting on the system. In these works the force is applied to all oscillators in a structure equivalent to a fully connected network. The motivation for this thesis, therefore, is to understand the response of synthetic and real complex networks to a localized stimuli using the forced Kuramoto model. In particular, we are interested in the conditions for global synchronization when the force acts only on a fraction of the oscillators and in applications of this theory to neural networks.

Understanding the network of neuronal connections in the brain is key to unravel the way it works and processes information. The complexity of these networks has been emphasized by many authors \cite{Sporns2013}, and characterized with different measures, such as degree distribution, transitivity and betweenness centrality \cite{Rubinov2010}. An important feature of neural networks is their high degree of heterogeneity, in the sense that the number of connections per neuron varies considerably and typically displays some sort of power law distribution. Moreover, neurons tend to form communities, where the density of connections is higher within than among communities. Because connections are constrained by anatomical features, neurons are also organized into physically arranged clusters, such as lobes or ganglia, where neurons with different functional roles coexist \cite{sporns-betzel2016,bacik2016,antono2015}. 

Communities are often related to specialized areas of the brain and their number and structure are an indication of how many different tasks it can perform  \cite{kim2014}. The integration of communities, on the other hand, measures how well the outcomes of these different processes can combined to build a global view of the inputs \cite{sporns-betzel2016}. When triggered by external stimuli, such as visual or olfactory inputs, the information processing occurs by the synchronized firing of neurons responsible to process those specific tasks \cite{Schmidt2015, gray1994}. Synchronization of larger sets of neurons, or even global synchronization, indicates cerebral disorders \cite{Uhlhaas2006} such as epilepsy \cite{global2012} and Alzheimer's disease \cite{global2016}, causing a general breakdown in the neuronal network. Lack of synchronization, on the other hand, suggests difficulty to respond to the stimulus or to function properly, as reported in unsuccessful overnight memory consolidation in old people \cite{helfrich2018}, deficiency in the auditory-motor connections \cite{sowinskia2013} or brain disorders in autistic individuals \cite {autism2013, autism2011}. In this context, the knowledge of the organization of different types of neurons in the network and their segregation into modules or communities is fundamental to understand how stimuli affect the target module and under what conditions it propagates to other regions leading to global or poor responses. In this sense, it is possible to use the forced Kuramoto model and apply the external stimuli only on a specific group of the neural network, which can be functional or anatomical. In this work we analyse this issue using synthetic networks and applying the generalized results to the \textit{C. elegans} neural network.

\textbf{Outline of the Thesis  } The Kuramoto model is considered the simplest mathematical model of synchronization phenomena. In Chapter 2 we review the analytical derivation made by Kuramoto and show an extension on complex networks followed by a brief discussion of explosive synchronization. In Chapter 3 we analyse the Kuramoto model subjected to an external periodic force acting in all oscillators based on the work of Childs and Strogatz \cite{Childs2008} using the techniques of Chapter 2.

The study of the forced Kuramoto model on complex networks is explored in Chapter 4, where we consider the force acting only on a fraction of oscillators. In this context, we show the conditions for global synchronization as a function of the fraction of nodes being forced and how it depends on network structure. We present analytical and numerical calculations on synthetic networks, exploring the fully connected, random and scale-free topologies. In Chapter 5 we use a real complex network and study the response of the \textit{C. elegans}' neural electrical junction network to external localized stimuli using the partially forced Kuramoto model developed in Chapter 4. We also analyse the network's topology and use a modularization procedure in order to understand how the system is organized. We show that the modular structure hinders the global synchronization, revealing the complexity of the brain's wiring and function. Finally, in Chapter 6 we summarize our main results and discuss further extensions of this work.

\chapter{The Kuramoto Model}
\label{chapter2}

In this chapter we review the synchronization model proposed by Y. Kuramoto in 1975 \cite{Kuramoto1975}, which is considered the simplest model of synchronization phenomena. Kuramoto considered a system composed of identical oscillators interacting with each other via a coupling parameter. He showed that for small values of the coupling the oscillators continued to move as if they were independent. However, as the coupling increased beyond a critical value, a finite fraction of oscillators started to move together, a behavior termed spontaneous synchronization. This fraction increases smoothly with the coupling, characterizing a second order phase transition in the limit of infinite oscillators. For large enough coupling the whole system oscillates with the same frequency, as if it were a single element. In the first section of this chapter, we will introduce the mathematical model and reproduce the analytical calculations made by Kuramoto.

In the subsequent sections we will also show that the original model can be extended to complex networks with a slightly change of mathematical parameters and will briefly discuss the phenomena of explosive synchronization on networks using the Kuramoto model.

\section{The Kuramoto Model}

\begin{figure}[H]
\center
\includegraphics[scale=0.16]{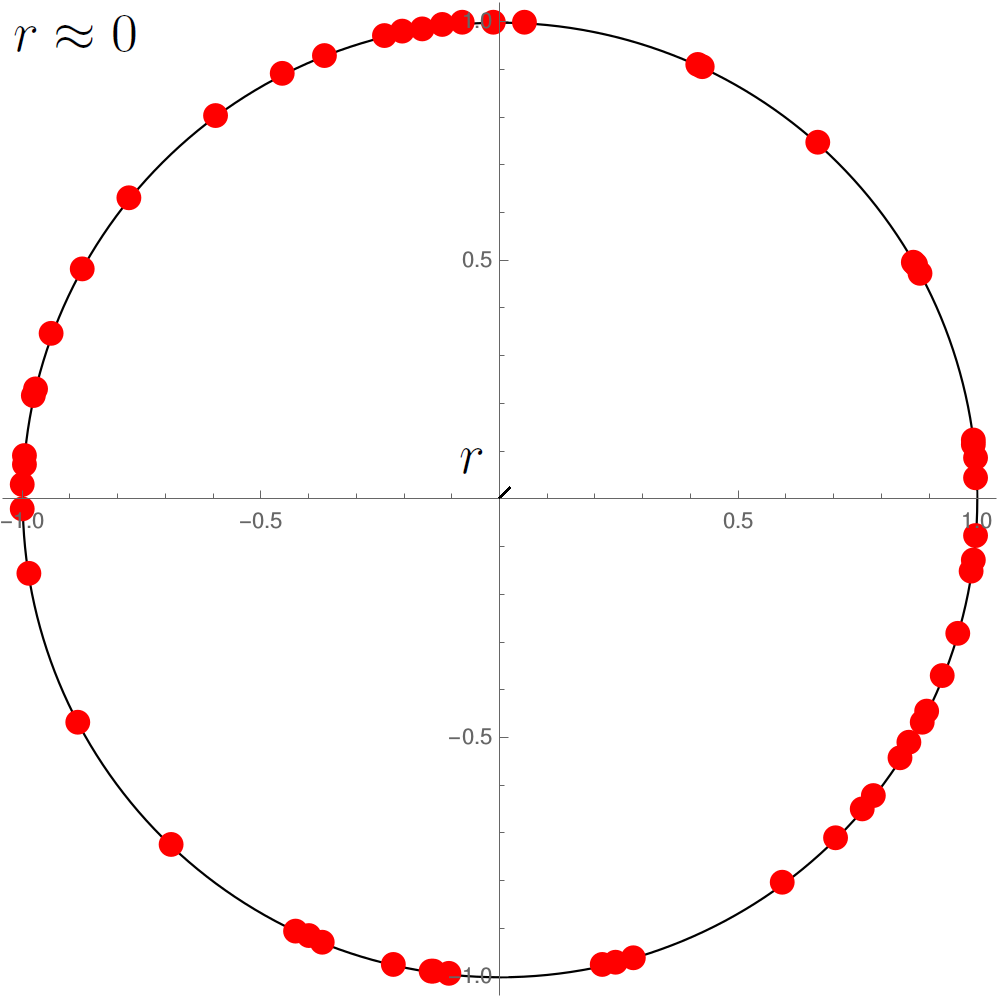}
\includegraphics[scale=0.16]{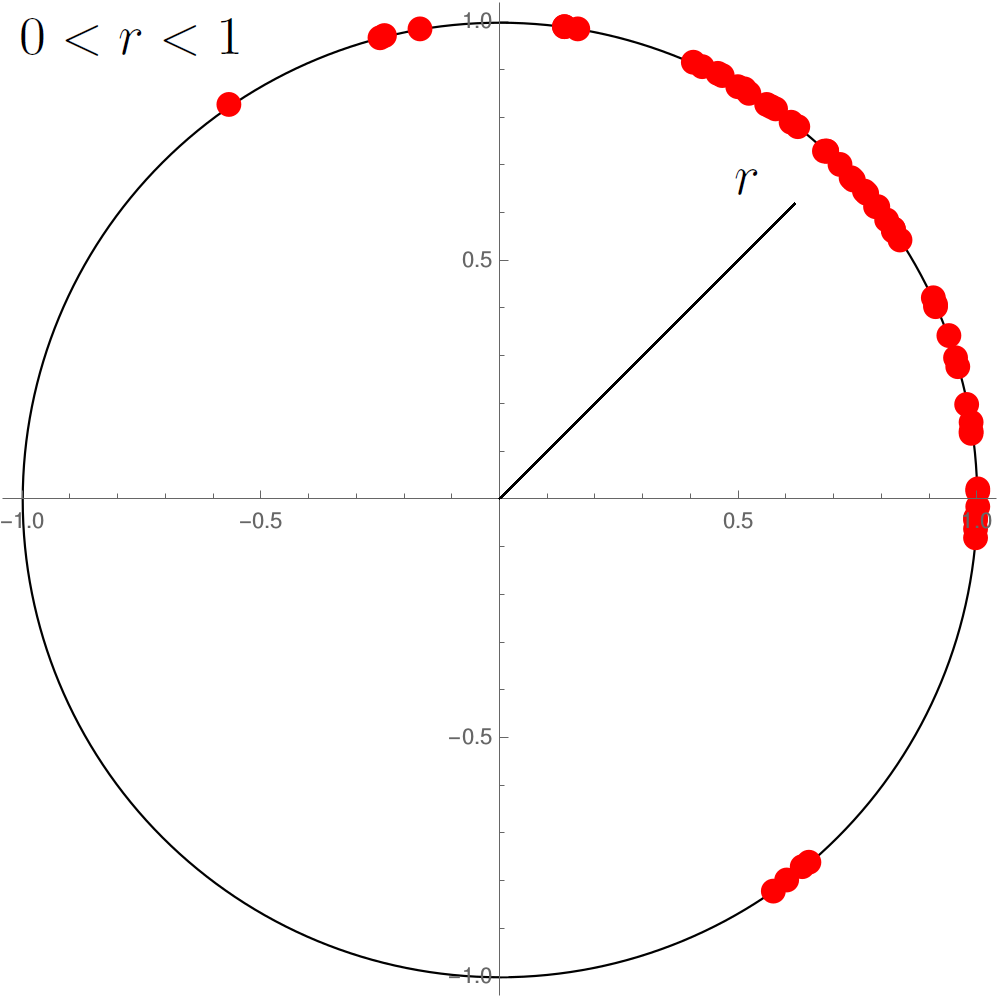}
\includegraphics[scale=0.16]{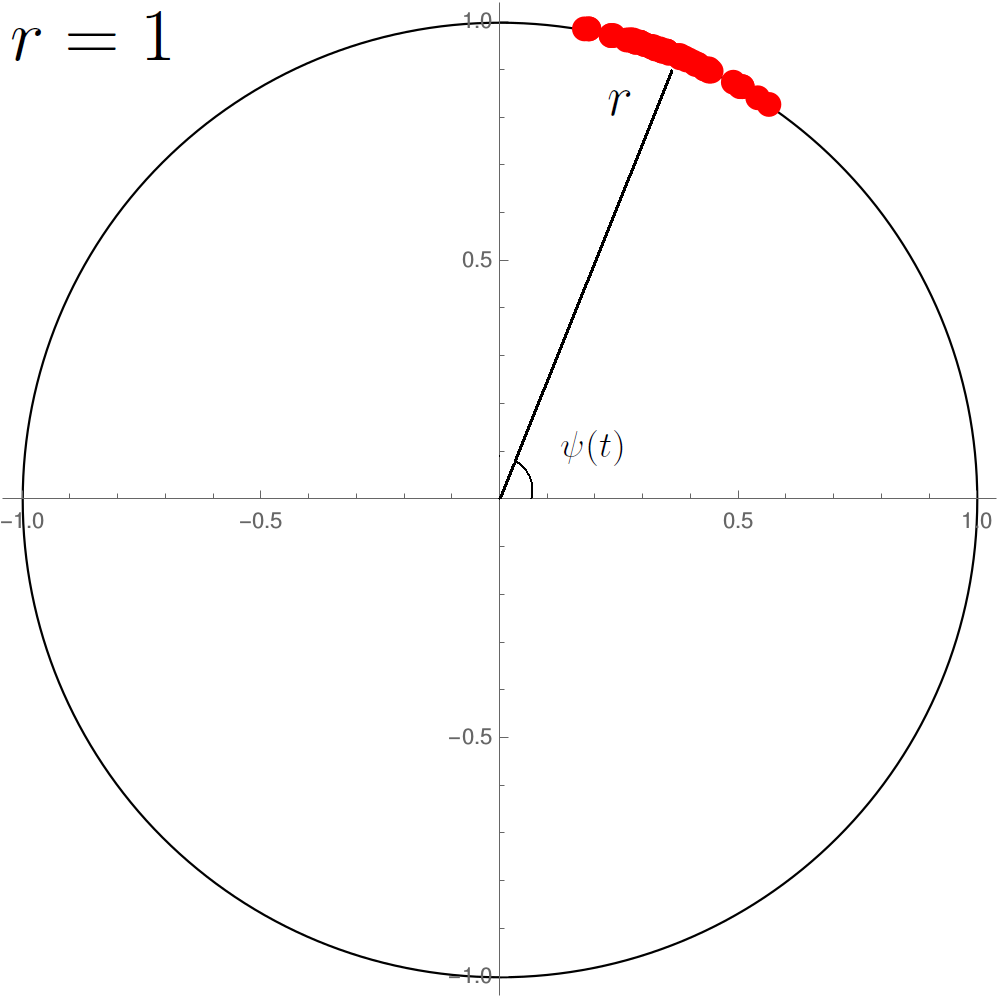}
\caption{Oscillators distribution on the unitary circle in non-synchronized configuration ($r \approx 0$), partial synchronization ($0 < r < 1$) and global synchronization ($r=1$). The complex number $z$ defined in equation (\ref{order}) is depicted as a black solid line, whose size (or module) is given by the order parameter $r$. In the full synchronized case, the oscillators move with velocity $\dot \psi$ as if they were a single element.}
\label{unitary_circle}
\end{figure}

The model of coupled oscillators introduced by Kuramoto consists of $N$ identical oscillators described by internal phases $\theta_i$ which rotate with natural frequencies $\omega_i$ typically selected from a symmetric distribution $g(\omega)$. In the original model all oscillators interact with each other according to the equations

\begin{equation}
\dot{\theta_i} = \omega_i + \frac{\lambda}{N} \sum_{j=1}^{N} \sin (\theta_j - \theta_i),
\label{original}
\end{equation}
where $\lambda$ is the coupling strength and $i = 1, ..., N$. The division by $N$ is necessary to avoid divergences on total interaction if the number of elements is too large. Although this problem does not involve a physical space, we can imagine the distribution of elements along a unitary circle, as depicted in figure \ref{unitary_circle}.

The frequency distribution $g(\omega)$ is responsible for the system disorder. If its mean value is $\bar{\omega}$ and its variance $\sigma^2$, we can infer that, the larger the variance, the larger the dispersion of natural frequencies and, therefore, it will be more difficult to synchronize the oscillators. The coupling parameter $\lambda$, on the other hand, has the role of bringing order to the system. For instance, if the oscillator $j$ is a little ahead of oscillator $i$, then $\sin (\theta_j - \theta_i) > 0$ and $\omega_i$ increases, so that $i$ can catch up with $j$. If $j$ is a little behind of $i$, then $\sin (\theta_j - \theta_i) < 0$ and $\omega_i$ decreases, so that $i$ can wait for $j$. It is worth noting that here the value $\lambda$ is constant for all oscillators and it determines the intensity of the coupling. Recent generalizations have also considered distributions of $\lambda$'s for each coupled pair \cite{Saa2019}.

\textbf{An example: two oscillators } In order to understand the behavior of the system, we consider a simple case of two coupled oscillators. In this example, equations (\ref{original}) become

\begin{equation}
\dot{\theta_1} = \omega_1 + \frac{\lambda}{2} \sin (\theta_2 - \theta_1),
\label{osc1}
\end{equation}

\begin{equation}
\dot{\theta_2} = \omega_2 + \frac{\lambda}{2} \sin (\theta_1 - \theta_2).
\label{osc2}
\end{equation}
Adding (\ref{osc1}) and (\ref{osc2}) we have

\begin{equation}
\dot{\theta_1} + \dot{\theta_2} = \omega_1 + \omega_2.
\end{equation}
Synchronization occurs when $\dot{\theta_1} = \dot{\theta_2}$. In this case, we obtain

\begin{equation}
\dot{\theta_1} = \dot{\theta_2} = \frac{\omega_1 + \omega_2}{2} \equiv \bar{\omega},
\end{equation}
which means that the synchronized state happens with average frequency. 

\textbf{General case } For any number $N$ of oscillators we can procedure as previously adding up all equations to obtain

\begin{equation}
\sum_{i=1}^{N} \dot{\theta_i} = \sum_{i=1}^N \omega_i + \frac{\lambda}{N} \sum_{i=1}^{N} \sum_{j=1}^{N} \sin(\theta_j - \theta_i).
\end{equation}
Since the sine function is odd, the double sum on the right is zero. It is simple to check for $N=3$:

\begin{equation}
      \begin{split}
        \sum_{i=1}^{3} \sum_{j=1}^{3} \sin (\theta_j - \theta_i) &= \sin (\theta_1 - \theta_1) + \sin (\theta_1 - \theta_2) + \sin (\theta_1 - \theta_3) \\
        &\quad {}+ \sin (\theta_2 - \theta_1) + \sin (\theta_2 - \theta_2) + \sin (\theta_2 - \theta_3) \\
        &\quad {}+ \sin (\theta_3 - \theta_1) + \sin (\theta_3 - \theta_2) + \sin (\theta_3 - \theta_3).
      \end{split}
\end{equation}

The terms where $i=j$ are null. The remaining cancel each other, since $\sin(\theta_j - \theta_i) = -\sin(\theta_i - \theta_j)$. The same occurs for any value of $N$.

If all oscillators synchronize, which characterizes a global synchronization state, all phase velocities are equal $\dot{\theta_1} = \dot{\theta_2} = ... = \dot{\theta_N}$, and then we have

\begin{equation}
\sum_{i=1}^{N} \dot{\theta_i} = N \dot{\theta_1} = \sum_{i=1}^{N} \omega_i.
\end{equation}
Thus
\begin{equation}
\dot{\theta_1} = \dot{\theta_2} = ... = \dot{\theta_N} = \frac{1}{N} \sum_{i=1}^{N} \omega_i \equiv \bar{\omega}.
\end{equation}
This result is identical to the case calculated for two oscillators. In these situations, we say that the system synchronizes spontaneously, since there is no external perturbation to cause the phenomenon.

In order to analyze the minimum value of the coupling strength that synchronize the system, Kuramoto introduced the complex number

\begin{equation}
z \equiv r e^{i \psi(t)} = \frac{1}{N} \sum_{j=1}^N e^{i \theta_j (t)},
\label{order}
\end{equation}
which is the phase average of oscillators. If all phases are equal, then $\theta_j = \psi$ and the sum is equal to $N$, which gives $r=1$. On the other hand, if $\theta_j$ are randomly distributed on unitary circle, $\theta_j = [0, 2 \pi]$, then $r \approx 0$, since the terms of the sum cancel each other. From this analysis we can see clearly that there is a change of behavior between these two extreme regimes. We say that $r$ is the \textit{order parameter} which delimits the transition between the disordered movement, $r \approx 0$, to the synchronized state, $r=1$. Figure \ref{unitary_circle} depicts the oscillators distribution and the order parameter on three configurations, $r \approx 0$ (non-synchronized state), $0<r<1$ (partial synchronization) and $r \approx 1$ (global synchronization).

To develop an analytical approach to study the transition, Kuramoto assumed that the frequency distribution $g(\omega)$ is centered at $\omega = \bar{\omega} = 0$, therefore, $g(\omega) = g(- \omega)$, which means that $g(\omega)$ is even and symmetric. He also took the limit of $N \rightarrow \infty$. In what follows, we will reproduce the analytical calculations made in his original work.

We start by reorganizing equation (\ref{order}) by multiplying both sides by $e^{-i \theta_j}$:
\begin{equation}
r e^{i \psi(t)} e^{-i \theta_j} = \frac{1}{N} \sum_{k=1}^N e^{i (\theta_k (t) - \theta_j (t))}.
\label{eq1}
\end{equation}
If we equal the imaginary parts of equation (\ref{eq1}) we obtain

\begin{equation}
r \sin(\psi - \theta_j) = \frac{1}{N} \sum_{k=1}^N \sin (\theta_k - \theta_j).
\label{order2}
\end{equation}
Comparing equations (\ref{original}) and (\ref{order2}) we eliminate the sum and we can write the dynamical equation as

\begin{equation}
\dot{\theta_j} = \omega_j + \lambda r \sin (\psi - \theta_j).
\label{dinamica}
\end{equation}
The interaction is defined by parameters $r$ and $\psi$. Besides, $\lambda$ appears multiplied by $r$, which gives a relationship between coupling and synchronization.

In order to take the limit of $N \rightarrow \infty$, we have to define a probability density, since we have to pass from discrete to continuous case. Then, we have to eliminate the index $j$ in the phase $\theta_j$ of each oscillator and develop a function that describes the phase $\theta$ of a group of oscillators in a given interval of unitary circle. Since each oscillator is in a position given by the phase $\theta_j$, we can imagine a phase distribution and, therefore, we are able to write it as a delta function: $\delta(\theta - \theta_1) + \delta(\theta - \theta_2) + ... + \delta(\theta - \theta_N)$. If the probability density $\rho$ gives the fraction of oscillators with phase between $\theta$ and $\theta + d \theta$ in a time $t$, then $\rho = \rho(\theta, t)$ must be normalized, that is,

\begin{equation}
\int \rho(\theta, t) d \theta = 1.
\end{equation}
In terms of delta function, we obtain

\begin{equation}
\frac{1}{N} \int \sum_{j=1}^N \delta(\theta - \theta_j) d \theta = 1.
\end{equation}
However, the construction of $\rho$ is still not complete. Each oscillator $\theta$ has a natural frequency that depends on distribution $g(\omega)$, and the position of oscillators in unitary circle depends on its natural frequency. Thus, the probability density must be rewrite as $\rho(\theta, g(\omega), t)$, or simply $\rho(\theta, \omega, t)$. This quantity gives the fraction of oscillators with phase in the interval $[\theta, \theta + d \theta]$ with natural frequency $\omega$ in time $t$, which is valid in the limit of $N \rightarrow \infty$. 

\begin{figure}[H]
\center
\includegraphics[scale=0.25]{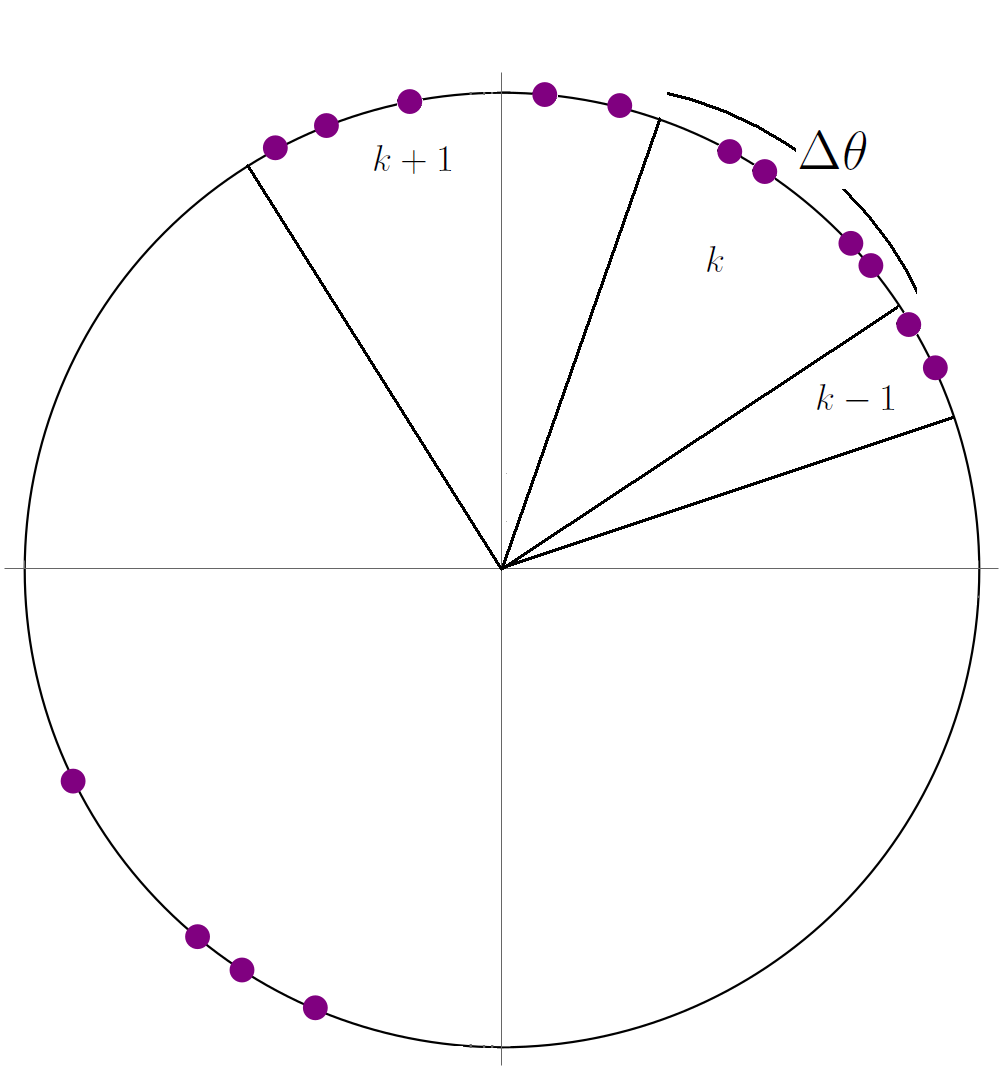}
\caption{Unitary circle divided in regions of size $\Delta \theta$. The region $k$ contains $N(k)$ oscillators.}
\label{box}
\end{figure}

As the number of oscillators is constant during dynamics, we can assert that $\rho$ must satisfy the equation with $J = \rho v$, where $J$ is the current, or flow, and $v$ is the angular velocity. In figure \ref{box} we depicted the unitary circle divided in regions of size $\Delta \theta$ each of them labeled by an index. The flow $J(k)$ is given by the number of oscillators, by per unit time, that leaves region indexed by $k$ and goes to region indexed by $k+1$. Let $N(k)$ be the number of oscillators in region $k$. Then, we can write the density of oscillators in $k$ as the ratio between the number of oscillators and the size of $k$, that is

\begin{equation}
\rho(k) = \frac{N(k)}{\Delta \theta}.
\end{equation}
Since the system is conservative we can assure that, if the quantity of oscillators on region $k$ changes, then there is a movement of elements on its boundaries, from $k -  1 \rightarrow k$ or from $k \rightarrow k + 1$, resulting on a increasing or decreasing of the number of oscillators in $k$, respectively. The variation of the number of oscillators on region $k$ can be written as 

\begin{equation}
\Delta N (k) = (- J (k) + J(k-1)) \Delta t.
\end{equation}
The negative sign on the first term reflects the movement of oscillators from $k$ to $k+1$, while the second term refers to the movement of elements from region $k-1$ to $k$. This results in a density variation,

\begin{equation}
\Delta \rho(k) = \frac{\Delta N(k)}{\Delta \theta} \rightarrow \Delta \rho(k) = \frac{(- J (k) + J(k-1)) \Delta t}{\Delta \theta}.
\end{equation}
Dividing by $\Delta t$ we obtain

\begin{equation}
\frac{\Delta \rho}{\Delta t} = - \left( \frac{\Delta J(k) - \Delta J(k-1)}{\Delta \theta} \right) \rightarrow \frac{\Delta \rho}{\Delta t} = - \frac{\Delta J}{\Delta \theta}.
\label{cont}
\end{equation}
The flow can now be computed as follows: all oscillators with angular velocity $v$ will traverse the interval $\Delta \theta$ in the time $\Delta t = \Delta  \theta / v$, passing to the next box. Therefore $J(k)$ is the number of oscillators with velocity $v$ in box $k$, that is, $\Delta \theta 
\rho(v, \theta, t)$, divided by $\Delta t = \Delta \theta / v$

\begin{equation}
J = \frac{\rho(v, \theta, t) \Delta \theta}{\Delta t} \rightarrow J = \rho v.
\end{equation}
If the number of elements which enter and leave the region $k$ is constant, then the density does not change. In the limit where the size of region goes to zero, $\Delta \theta \rightarrow 0$, equation (\ref{cont}) becomes

\begin{equation}
\frac{\partial{\rho}}{\partial {t}} + \frac{\partial{(\rho v)}}{\partial {\theta}} = 0.
\label{continui}
\end{equation}
Equation (\ref{continui}) is the \textit{continuity equation}.

Now, we must find a relationship between equation (\ref{continui}) and the order parameter $r$. Noting that $v$ is equal to $\dot{\theta}$ of equation (\ref{dinamica}), the angular velocity of the continuity equation can be written in terms of $\theta$, $\omega$ and $t$, that is

\begin{equation}
\dot{\theta} = v = \omega + \lambda r \sin (\psi - \theta),
\label{velo}
\end{equation}
where we took off the index $i$, since the system is now continuous. Thus, $v = v(\theta, \omega, t)$ is the angular velocity of an oscillator at coordinate $\theta$ with natural frequency $\omega$ in the time $t$. In terms of dynamical parameters, the continuity equation is rewritten as

\begin{equation}
\frac{\partial{\rho}}{\partial {t}} + \frac{\partial{[\rho \omega + \rho \lambda r \sin (\psi - \theta) ]}}{\partial {\theta}} = 0.
\end{equation}

Finally, equation (\ref{order}) in the limit of $N \rightarrow \infty$ as a function of density probability is given by 

\begin{equation}
r e^{i \psi} = \int_{- \pi}^{\pi} \int_{- \infty}^{+ \infty} e^{i \theta} \rho(\theta, \omega, t) d\omega d\theta.
\label{density}
\end{equation}
In equation (\ref{density}) we integrate over all phases and frequencies. It is worth noting that, by definition of $g(\omega)$ we have

\begin{equation}
 \int_{- \pi}^{\pi} \rho(\theta, \omega, t) d\theta = g(\omega).
 \label{gdef}
\end{equation}
As a consequence, the normalization of $\rho$ is written as

\begin{equation}
 \int_{- \infty}^{+ \infty} d \omega \int_{- \pi}^{\pi} \rho(\theta, \omega, t) d\theta = 1.
 \label{rhonorm}
\end{equation}

In what follows, we will study the dynamical behavior when the system is non-synchronized and partially synchronized, analysing the form of $\rho(\theta,\omega,t)$. Then, we will be able to calculate the order parameter $r$.

\subsection{Incoherent behavior - non-synchronized state}

In the non-synchronized state, the oscillators are distributed randomly on unitary circle. In this case, the density is uniform, $\rho = g(\omega) /2 \pi$, and equation (\ref{density}) becomes

\begin{equation}
r e^{i \psi} = \int_{- \pi}^{\pi}  e^{i \theta} d\theta \int_{- \infty}^{+ \infty} \frac{1}{2 \pi} g(\omega) d\omega.
\end{equation}
The integral on variable $\theta$ results in zero and, therefore, $r=0$. Besides, since $\partial{\rho} / \partial {t} = 0$, we can verify that equation (\ref{continui}) is satisfied,

\begin{equation}
\frac{\partial}{\partial \theta} (\rho \omega) - \rho \lambda r \sin (\psi - \theta) \rightarrow  \frac{\partial}{\partial \theta} (\rho \omega ) = 0,
\end{equation}
since $\rho$ is constant and $\omega$ does not depend on $\theta$.

\subsection{Partial synchronization}

We now assume that the system reached the steady state, $\partial \rho / \partial t = 0$, and that a fraction of oscillators is synchronized with $v = 0$, while the remaining are moving incoherently. In the case of $v=0$, equation (\ref{velo}) becomes

\begin{equation}
\omega = - \lambda r \sin (\psi - \theta) \rightarrow \omega = \lambda r \sin (\theta - \psi) \qquad \text{for} \qquad -\frac{\pi}{2} \leq \theta - \psi \leq \frac{\pi}{2}.
\label{sync}
\end{equation}
Because $\lvert \sin (\theta - \psi) \rvert < 1$, the synchronization only occurs for $\lvert \omega \rvert < \lambda r$. We can write expression (\ref{sync}) as

\begin{equation}
\theta = \psi + \arcsin \left(\frac{\omega}{\lambda r} \right).
\label{sync2}
\end{equation}
Equation (\ref{sync2}) provides the position where the oscillators with natural frequency $\omega$ stopped. As a consequence, oscillators with $\lvert \omega \rvert > \lambda r$ do not synchronize, since $\lambda$ is not strong enough to ``hold'' them together. To find the expression for density $\rho$ in the equilibrium, we have to divide our analysis in two cases: the synchronized and the non-synchronized parts.

\subsubsection{(A) The synchronized part}

From equation (\ref{sync2}) we write the density of the synchronized part as a delta function

\begin{equation}
\rho = \delta \left[\theta -  \psi + \arcsin \left(\frac{\omega}{\lambda r} \right) \right] g(\omega),
\end{equation}
indicating that the oscillators are centered close to $\psi$ with deviation given by $\arcsin (\omega / \lambda r)$. In order to rewrite the density conveniently, we can use the following property of the delta function,
\begin{equation}
\delta [\theta - \theta_0] = \lvert f'(\theta_0) \rvert \delta [f (\theta) - f (\theta_0)],
\label{prop}
\end{equation}
We can define properly the function $f(\theta)$ as

\begin{equation}
f(\theta) = \omega - \lambda r \sin (\theta - \psi).
\end{equation}
If we impose $f(\theta_0) = 0$ the condition for $\theta_0$ becomes

\begin{equation}
\sin(\theta_0 - \psi) = \frac{\omega}{\lambda r}.
\label{teta0}
\end{equation}
Thus

\begin{equation}
f'(\theta_0) = - \lambda r \cos (\theta_0 - \psi). 
\end{equation}
Using (\ref{teta0}), we obtain
\begin{equation}
f'(\theta_0) = - \sqrt{\lambda^2 r^2 - \omega^2}. 
\end{equation}
Finally, we can use (\ref{prop}) to rewrite the density as

\begin{equation}
\rho = \sqrt{\lambda^2 r^2 - \omega^2} \delta [\omega - \lambda r \sin (\theta - \psi)] g(\omega),
\label{syncpart}
\end{equation}
where $-\frac{\pi}{2} \leq \theta - \psi \leq \frac{\pi}{2}$. Equation (\ref{syncpart}) gives the density $\rho$ of the synchronized part.

\subsubsection{(B) The non-synchronized part}

In the steady state we must have $\partial(\rho v)/ \partial \theta = 0$. This condition implies that $\rho v$ is constant, independent of $\theta$. Thus, since $\lvert \omega \rvert > \lambda r$ on the non-synchronized part, we write the density as

\begin{equation}
\rho = \frac{C g(\omega)}{\lvert \omega - \lambda r \sin(\theta - \psi) \rvert},
\end{equation}
where $C$ is a normalization constant and we used $v$ from equation (\ref{velo}). To calculate $C$, we use the normalization condition of equation (\ref{gdef}), that is

\begin{equation}
\int_{-\pi}^{\pi} \frac{C  g(\omega)}{\lvert \omega - \lambda r \sin(\theta - \psi) \rvert} d \theta =  g(\omega).
\end{equation}
We cancel $ g(\omega)$ and can take off the modulus in the case of $\omega > 0$ and $\omega > \lambda r$. Performing a change of variables $\phi = \theta - \psi$, we obtain

\begin{equation}
C \int_{-\pi - \psi}^{\pi - \psi} \frac{d \phi}{\omega - \lambda r \sin \phi} = 1.
\end{equation}
We can verify that the integral above does not depend on $\psi$ and then we are able to integrate on the interval $[-\pi, \pi]$. Let $f(\sin \phi)$ be a function integrated on $[-\pi - \psi, \pi - \psi]$. We can rewrite as

\begin{equation}
\int_{-\pi - \psi}^{\pi - \psi} f(\sin \phi) d \phi = \int_{-\pi - \psi}^{- \pi} f(\sin \phi) d \phi +  \int_{-\pi}^{\pi - \psi} f(\sin \phi) d \phi.
\end{equation}
Now, let $\xi = \phi + 2 \pi$. It is simple to check that the first integral on the right hand side is integrated on $[\pi - \psi, \pi]$. Adding the two integrals we find  $\int_{-\pi}^{\pi} f(\sin \phi) d \phi$, that is independent of $\psi$.

The result of integration is, then

\begin{equation}
      \begin{split}
      &C \int_{-\pi}^{\pi} \frac{d \phi}{\omega - \lambda r \sin \phi} = 1  \\
      & \frac{2 C}{\sqrt{\omega^2 - \lambda^2 r^2 }} \arctan \left[\frac{\omega \tan \phi/2 - \lambda r}{\sqrt{\omega^2 - \lambda^2 r^2}}\right]_{-\pi}^{\pi} = 1.
      \end{split}
\end{equation}
For $\phi = \pm \pi$, $\tan \phi/2 \rightarrow \pm \infty$ and then, $\arctan(\pm \infty) \rightarrow \pm \pi/2$ that add up to $\pi$. Isolating the normalization constant we obtain $C = \sqrt{\omega^2 - \lambda^2 r^2} / 2 \pi$. The density $\rho$ for the non-synchronized part can be written as

\begin{equation}
\rho = \frac{g(\omega)}{2 \pi} \frac{\sqrt{\omega^2 - \lambda^2 r^2 }}{\lvert \omega - \lambda r \sin(\theta - \psi) \rvert}.
\label{nonsyncpart}
\end{equation}

\subsubsection{(C) The order parameter}

We developed the expressions of $\rho(\theta,\omega)$ for synchronized and non-synchronized oscillators. The final distribution is written using equations (\ref{syncpart}) and (\ref{nonsyncpart})

\begin{equation}
\rho(\theta,\omega) = g(\omega)
\begin{cases}
\sqrt{\lambda^2 r^2 - \omega^2} \delta [\omega - \lambda r \sin (\theta - \psi)], & \lvert \omega \rvert < \lambda r \\ \\
\frac{1}{2 \pi} \frac{\sqrt{\omega^2 - \lambda^2 r^2 }}{\lvert \omega - \lambda r \sin(\theta - \psi) \rvert}, & \lvert \omega \rvert > \lambda r.
\end{cases}
\end{equation}
Now, we are able to calculate the order parameter $r$. From equation (\ref{density})

\begin{equation}
r = \int_{- \pi}^{\pi} \int_{- \infty}^{+ \infty} e^{i (\theta - \psi)} \rho(\theta, \omega, t) d\omega d\theta \equiv r_s + r_{ns},
\end{equation}
where we divided equation (\ref{density}) by $e^{i \psi}$ and we separated the integral over $\omega$ into two cases, $r_s$ and $r_{ns}$ which refers to the synchronize part, where $\lvert \omega \rvert < \lambda r$, and to the non-synchronized part, where $\lvert \omega \rvert > \lambda r$, respectively. 

Since we assumed that $g(\omega)$ is symmetric, the non-synchronized integral is zero, $r_{ns} = 0$. We can verify this result by dividing the $\omega$ integral into two parts and performing a change of variables $\theta' = \theta - \psi$,

\begin{equation}
r_{ns} = \int_{- \pi}^{\pi} d\theta' \left[ \int_{- \infty}^{- \lambda r} e^{i \theta'} \rho(\theta', \omega, t) d\omega + \int_{+ \lambda r}^{+ \infty} e^{i \theta'} \rho(\theta', \omega, t) d\omega \right].
\end{equation}
We manipulate the first integral by changing $\omega \rightarrow -\omega$ and $\theta' \rightarrow \theta' + \pi$, whithout altering $\rho$, and we obtain

\begin{equation}
r_{ns} = \int_{- \pi}^{\pi} d\theta' \left[ \int_{- \infty}^{- \lambda r} e^{i (\theta' + \pi)} \rho(\theta' + \pi, -\omega, t) (- d\omega) + \int_{+ \lambda r}^{+ \infty} e^{i \theta'} \rho(\theta', \omega, t) d\omega \right].
\end{equation}
Since $e^{i \pi} = -1$, if we exchange the integration limits, we can see that the remaining term cancels the second integral, which leads to $r_{ns}=0$.

\begin{figure}[H]
\center
\includegraphics[scale=0.4]{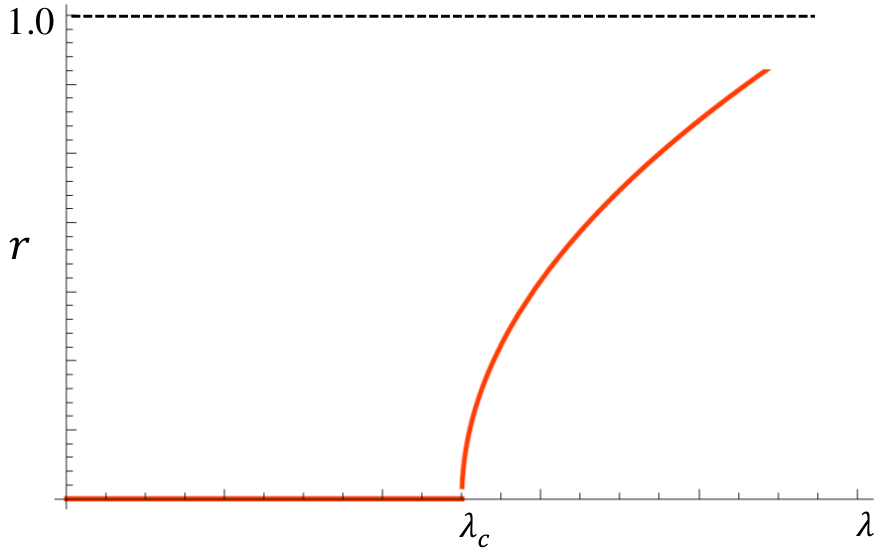}
\caption{Order parameter in function of coupling strength for the Kuramoto Model. The critical coupling $\lambda_c$ delimits the phase transition between the disordered (non-synchronization) and ordered (synchronization) phases.}
\label{curve_transition}
\label{space}
\end{figure}

In the synchronized integral we write $e^{ i \theta'} = \cos \theta' + i \sin \theta'$. We note that the imaginary part is zero, because the sine function is odd. We need to perform the integration over the real part,

\begin{equation}
r = \int_{- \pi/2}^{+ \pi/2} d \theta' \int_{-\lambda r}^{+ \lambda r} d \omega g(\omega) \cos \theta' \sqrt{\lambda^2 r^2 - \omega^2} \delta [\omega - \lambda r \sin \theta'].
\end{equation}
The $\omega$ integral is done using the delta function, which reduces to

\begin{align}
r &= \int_{- \pi/2}^{+ \pi/2} d \theta' g(\lambda r \sin \theta') \cos\theta' \sqrt{\lambda^2 r^2 - \lambda^2 r^2 \sin^2 \theta'}\\
&= \lambda r \int_{- \pi/2}^{+ \pi/2} d \theta'  g(\lambda r \sin \theta') \cos^2 \theta'.
\end{align}
This integral has two solutions: either $r=0$, which is trivial, or $r$ is given implicitly by

\begin{equation}
1 = \lambda \int_{- \pi/2}^{+ \pi/2} d \theta' g(\lambda r \sin \theta') \cos^2 \theta'.
\label{calcr}
\end{equation}
For $r = 0$ we can find the analytical expression of $\lambda$ for which the phase transition occurs, that is, the value that divides the synchronize and the non-synchronized states. This minimum value of coupling is denominated as \textit{critical parameter} and can be obtained by solving 

\begin{equation}
1 = \lambda \int_{-\pi/2}^{\pi/2} d \theta' \cos^2 \theta' g(0) \rightarrow 1 = \frac{\pi}{2} \lambda g(0),
\end{equation}
isolanting $\lambda = \lambda_c$, we obtain
\begin{equation}
\lambda_c = \frac{2}{\pi g(0)}.
\label{lambdac}
\end{equation}
The second solution occurs close to the phase transition for $r \approx 0$. We have to expand $g(\lambda r \sin \theta')$ in second order and calculate the behavior of the function for $\lambda > \lambda_c$,

\begin{equation*}
g(\lambda r \sin \theta') \approx g(0) + \lambda r \sin \theta' g'(0) + \frac{1}{2} \lambda^2 r^2 \sin^2 \theta' g''(0).
\end{equation*}
Substituting the expansion on the integral (\ref{calcr}), we have

\begin{equation}
1 = \lambda \int_{-\pi/2}^{\pi/2} d \theta' \cos^2 \theta' \left[g(0) + \lambda r \sin \theta' g'(0) + \frac{1}{2} \lambda^2 r^2 \sin^2 \theta' g''(0) \right].
\label{conta1}
\end{equation}
The first term inside the integral results in $\lambda_c ^{-1}$, and the second is zero, because of the sine function. The expression reduces to

\begin{equation}
1 = \frac{\lambda}{\lambda_c} + \frac{\lambda^3 r^2}{2} g''(0) \int_{-\pi/2}^{\pi/2} \cos^2 \theta' \sin^2 \theta' d \theta'.
\end{equation}
If we write $\cos^2 \theta' \sin^2 \theta'$ as $(\cos \theta' \sin \theta')^2 = (\frac{1}{2} \sin 2 \theta')^2$, we can perform a change of variables $u = 2 \theta$, and (\ref{conta1}) reduces to

\begin{equation}
1 = \frac{\lambda}{\lambda_c} + \frac{\lambda^3 r^2}{2} g''(0) \frac{\pi}{8}.
\end{equation}
Isolating $r$, we obtain

\begin{equation}
r^2 = \frac{16(\lambda_c - \lambda)}{\lambda^4 g''(0) \pi}.
\end{equation}
Because $r$ is in the vicinity of zero, we can consider that $\lambda \approx \lambda_c$ on the denominator. Thus,

\begin{equation}
r = \sqrt{-\frac{16(\lambda - \lambda_c)}{\lambda_c^4 g''(0) \pi}}.
\label{criticalr}
\end{equation}
Since $r \propto (\lambda-\lambda_c)^{\frac{1}{2}}$, the critical exponent of phase transition is 1/2. Figure \ref{curve_transition} shows the behavior of $r$ as a function of $\lambda$. We see that, for $\lambda < \lambda_c$, $r=0$ and the system is disordered (non-synchronization phase). On the other hand, for $\lambda > \lambda_c$, a small number of oscillators starts to move together and these fraction increases smoothly with coupling, until the system reaches global synchronization ($r=1$).

\section{The Kuramoto Model on Networks}

A natural extension of the Kuramoto system is to include the possibility that each oscillator interacts only with a subset of the other oscillators, which can be done by placing the system on a network whose topology defines the interactions. In this case, the system is described by the equations
\begin{equation}
\dot{\theta_i} = \omega_i  + \frac{\lambda}{k_i} \sum_{j=1}^{N} A_{ij} \sin (\theta_j - \theta_i),
\label{KMnet}
\end{equation}
where we added the adjacency matrix $A_{ij}$ and we replaced the division over $N$ by $k_i$, which is the number of terms in the sum. We made this choice because in heterogeneous networks, like the scale-free topology, where the degree distribution follows a power law function, the effect of the ``hubs'' (nodes with high degree) is considerably different from the remaining nodes. The same approach was develop in \cite{Oh2005}. We also note that the fully connected network is equivalent to the Kuramoto original model since each node (or oscillator) interacts with all the other nodes.

In order to verify the effect of network structure, we used the dynamical equations (\ref{KMnet}) on three different topologies: (i) fully connected with $N=200$ nodes (FC200); (ii) fully connected with $N=1000$ (FC1000) nodes; (iii) random Erdos-Renyi network with $N=200$ nodes (ER200) and average degree $\left\langle k \right\rangle = 10.51$; (iv) random Erdos-Renyi network with $N=1000$ nodes (FC1000) and average degree $\left\langle k \right\rangle = 19.87$; (v) scale-free Barabasi-Albert network with $N=200$ (BA200) computed starting with $m_0 = 11$ fully connected nodes and adding nodes with $m = 10$ links with  preferential attachment, so that $\langle k \rangle =$  9.83, and (vi) scale-free Barabasi-Albert network with $N=1000$ (BA1000) computed starting with $m_0 = 21$ fully connected nodes and adding nodes with $m = 20$ links with preferential attachment, so that $\langle k \rangle =$  39.56. In all simulations we have considered a Gaussian distribution of natural frequencies $g(\omega)$

\begin{equation}
g(\omega) = \frac{1}{\sqrt{2 \pi} a} e^{(\omega - \bar\omega)^2 / 2 a^2},
\label{gaussdist}
\end{equation}
with null mean $\bar\omega = 0$ and standard deviation $a=1.0$. Using equation (\ref{lambdac}) we can estimate the critical value $\lambda_c$. In this case, it is simple to verify that $g(0) = 1/\sqrt{2 \pi}$ and $\lambda_c = \sqrt{8/ \pi}$, that is, $\lambda_c \approx 1.6$.

\begin{figure}[H]
\center
\includegraphics[scale=0.7]{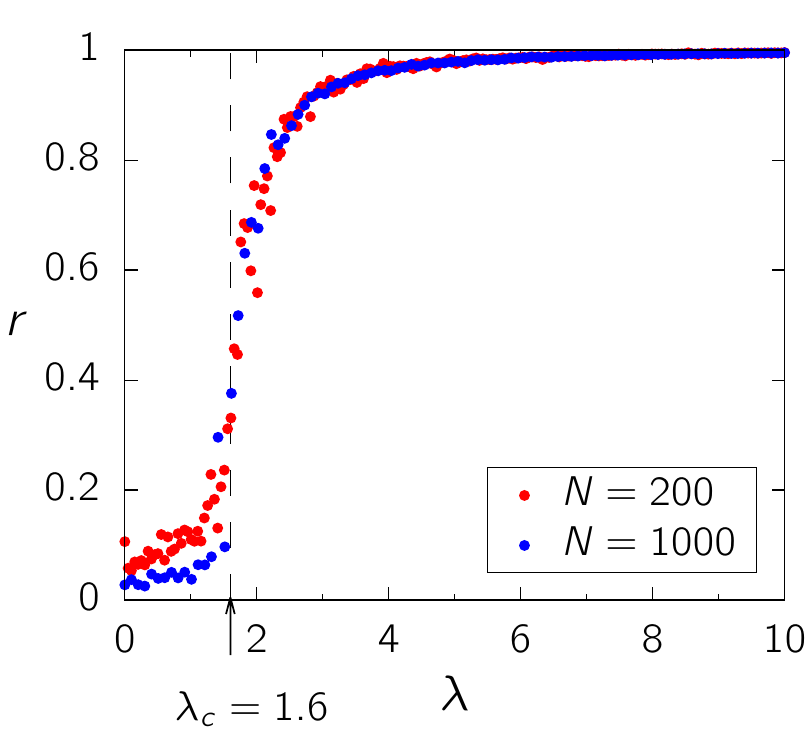}
\includegraphics[scale=0.7]{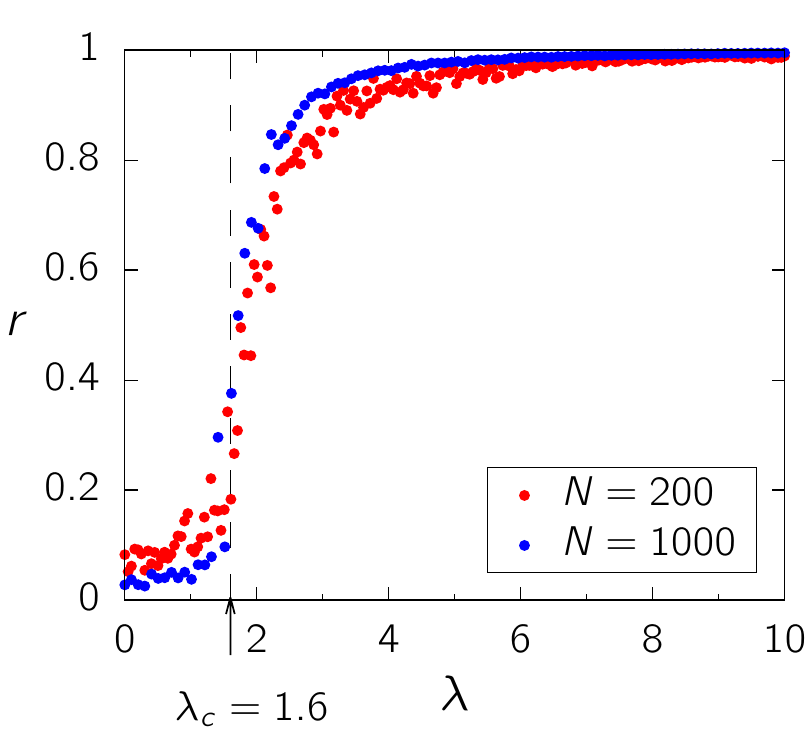}
\includegraphics[scale=0.7]{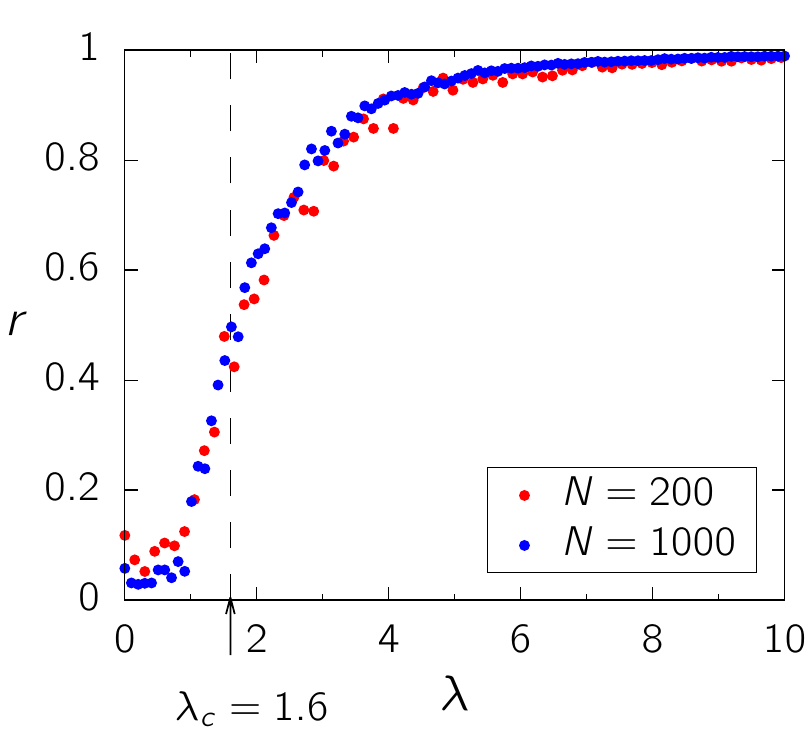}
\caption{Order parameter $r$ versus coupling $\lambda$ for three network topologies: fully connected (left), random (middle) and scale-free (right), for $N=200$ and $N=1000$ nodes. The critical coupling is independent of $N$. For a Gaussian distribution $g(\omega)$ the transition occurs at $\lambda_c $ = 1.6, delimited by the dashed line on each panel.}
\label{netsKM}
\end{figure}

Figure \ref{netsKM} computes the order parameter $r$ versus coupling $\lambda$ for all network configurations. In all cases we can verify that the theoretical curve behavior depicted on figure \ref{curve_transition} is satisfied and that the larger the number of nodes, the better is the result. This is expected once the theoretical development was made on the limit of $N \rightarrow \infty$.

\section{Explosive Synchronization}

So far we showed the original model proposed by Kuramoto and their applications on systems whose elements interact by a complex network structure. In all cases the transition from the non-synchronized to global synchronized states occurs smoothly, characterizing a second order phase transition with critical exponent $1/2$, as derived in equation (\ref{criticalr}). However, recent works have shown that the Kuramoto system, under specific conditions, has an abrupt change on order parameter, which is a first order phase transition. This behavior is termed explosive synchronization. In this section we will briefly present a review regarding first-order phase transitions using the Kuramoto model in complex networks.

The explosive synchronization phenomenon has been studied in several works \cite{Kim2017, Kim2016, Wang2016, Wang2017, Leyva2012, Deco2013, FARODRIGUES2012, SAARAFAEL2015} which demonstrate a relation between the natural frequency distribution $g(\omega)$ and the complex network structure. The applications range from waking from anesthesia (abrupt transition to conscious-unconscious states) \cite{Kim2017, Kim2016} to epileptic seizures \cite{Wang2016}.

One of the first works of explosive synchronization was reported in \cite{GomezGardenes}. In this paper, the authors showed the abrupt onset to global synchronization in scale-free networks using the Kuramoto model with the dynamical equations,

\begin{equation}
\dot{\theta_i} = \omega_i  + \lambda \sum_{j=1}^{N} A_{ij} \sin (\theta_j - \theta_i).
\label{ES-SF}
\end{equation}
The main difference between equation (\ref{ES-SF}) and our system, (\ref{KMnet}), is the lack of division by $k_i$. In order to study the behavior of phase transition, the internal frequency of each node is set as a function of its own degree, that is, $\omega_i = f(k_i)$. The correlation between degree and frequency introduces a relation between the network structure and the system dynamics. In particular, the authors used $k_i = \omega_i$. As a consequence, the frequency and the degree distributions are identical, $g(\omega) = P(k)$. In this situation, it is worth noting that in heterogeneous networks, as the scale-free topology, the hubs can synchronize more easily because of their large degree.

\begin{figure}[H]
\center
\includegraphics[scale=0.35]{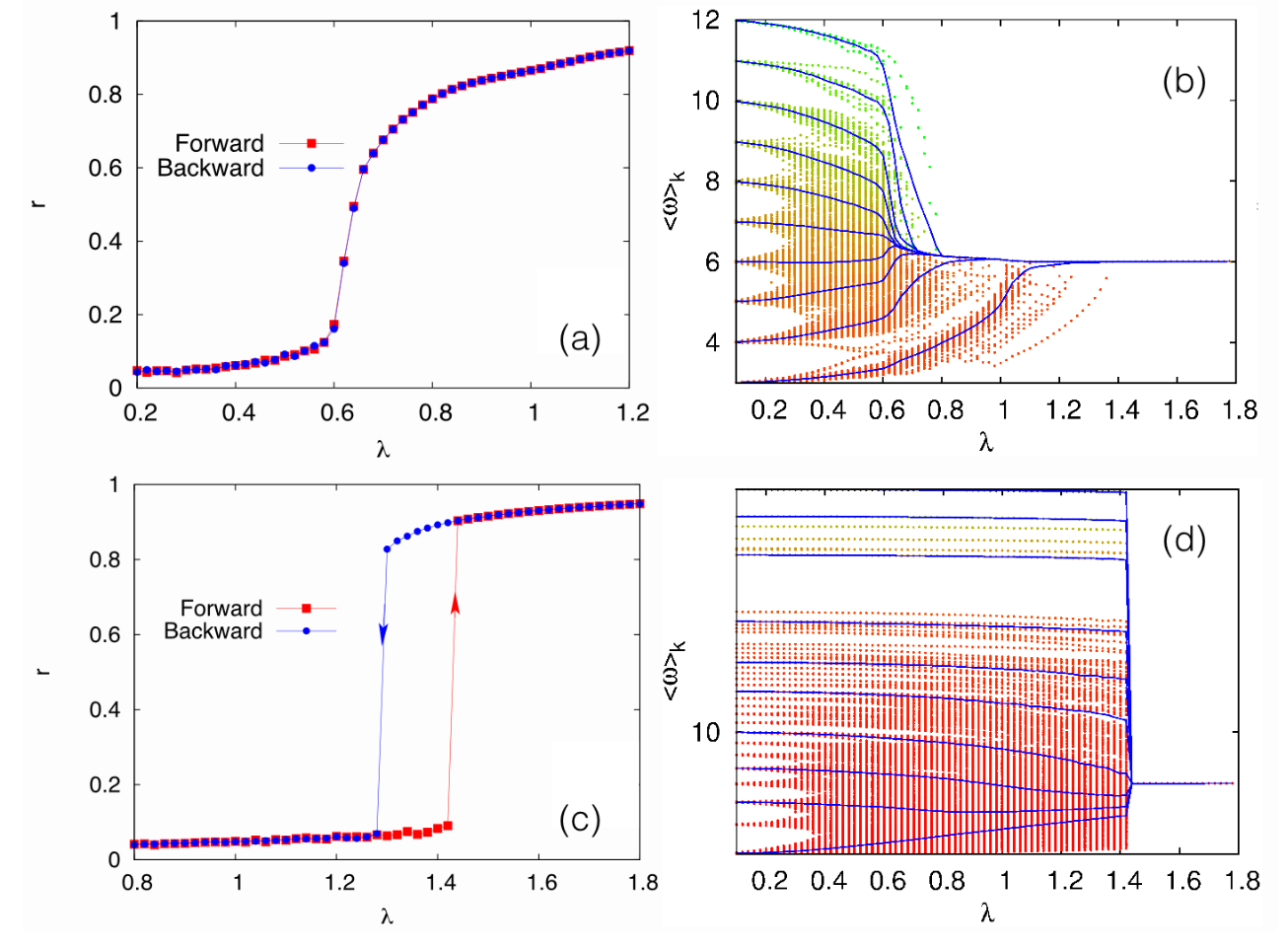}
\caption{Forward and backward curves $r(\lambda)$ for (a) random network and (c) scale-free network with $N = 10^3$ nodes and average degree $\left\langle k \right\rangle$ = 6. Panels (b) and (d) exhibit the results of equations (\ref{eff-freq}) and (\ref{eff-freq-avg}) depicted as dots and lines, respectively, as a function of coupling $\lambda$ in the forward continuation case for (b) random network and (d) scale-free network. Taken from \cite{GomezGardenes}.}
\label{paperPRL106}
\end{figure}

To analyze the role of network topology and the degree distribution $g(\omega)$ the equations (\ref{ES-SF}) were applied on random (Erdos-Renyi) and scale-free (Barabasi-Albert) networks. In both cases the adjacency matrix used is undirected, unweighted and the networks have the same number of nodes, $N = 10^3$, and average degree, $\left\langle k \right\rangle$ = 6. Figure \ref{paperPRL106} summarizes the results. Each panel shows two transition diagrams, labeled as \textit{forward}, where $\lambda$ is gradually increased, and \textit{backward}, where $\lambda$ is gradually decreased. We can see that, for the random network (panel \ref{paperPRL106} (a)), the $r(\lambda)$ curves are equal and smooth, indicating a second order phase transition, as usual. The opposite occurs for the scale-free network, (panel \ref{paperPRL106} (c)), where the forward and backward curves do not coincide and the phase transition is first order. By looking at both diagrams, the phase transition to synchronized state in each case occurs for different values of $r$, showing a strong hysteresis.

The authors also computed the effective frequency of each oscillator, defined as

\begin{equation}
 \omega_i^{eff} = \frac{1}{T} \int_{t}^{t+T} \dot\theta_i (\tau) d \tau,
\label{eff-freq}
\end{equation}
with $T >> 1$, as well as the average effective frequency of nodes with the same degree,

\begin{equation}
 \left\langle \omega \right\rangle_k = \frac{1}{N_k} \sum_{[i| k_i=k]} \omega_i^{eff},
 \label{eff-freq-avg}
\end{equation}
where $N_k = NP(k)$ is the total number of nodes with degree $k$.

Panels (b) and (d) exhibit the results of equations (\ref{eff-freq}) and (\ref{eff-freq-avg}) for the random and scale-free network, respectively. From panel (b) we can see that the oscillators with the largest degree converge firstly to the average frequency, $\Omega= \left\langle k \right\rangle$ = 6, contrary to what happens for nodes with small values of $k$. On the other hand, the explosive synchronization showed on diagrams $r(\lambda)$ of scale-free network is confirmed by the behavior of the effective frequencies on panel (d), since almost all nodes keep locked to their natural frequencies, $\omega_i = k_i$, until they reach the criticality at $\lambda \approx 1.42$, where they abruptly converge into one single value of average frequency,  $\Omega= \left\langle k \right\rangle$ = 6.

There are other mechanisms leading to explosive synchronization. A great description of these variety can be seen in reference \cite{Raissa2019}. The theoretical approach can be validate by constructing experimental setups that allows the observation of this phenomenon in real-world systems. As an example, in the work \cite{Leyva2012}, the authors demonstrated numerical and experimental evidence of the first-order synchronization transition in a network of phase coherent Rössler units (chaotic oscillators). The numerical results showed that these elements interacting in a scale-free topology on a chaotic regime exhibit explosive synchronization when there is a positive correlation between the network structure and the natural frequencies of oscillators. By constructing an electronic network device operating in the same regime of the theoretical system, the authors reported that the experimental diagram of synchronization have the same behavior of the numerical data, validating the appearance of explosive synchronization in real systems. Many other interesting experimental set up can be found in \cite{Raissa2019}.

The phenomenon of abrupt synchronization is also observed in neuroscience. As reported in \cite{Kim2017, Kim2016} the conscious-unconscious transition appears when the brain is awaking from anesthesia. The authors hypothesized that, in human brain networks, the conditions for explosive synchronization occur in the anesthetized brain just over the threshold of unconsciousness. In \cite{Deco2013} the authors reported that the unconsciousness and resting states are apparently related to a bifurcation point on the phase space where the dynamical system may lead to spontaneous synchrony. Another examples include the epileptic seizures, where the brain shows an abrupt dynamical behavior activity during an epileptic event \cite{Wang2017}, and the sensitive frequency detection of the cochlea \cite{Wang2016}, where the hair cells present in the structure are modeled as oscillators close to a Hopf bifurcation. In this paper, the authors studied a system composed of globally coupled units of the cochlea (the hair cells) which exhibits explosive synchronization in the absence of an external stimulus.

\chapter{The Forced Kuramoto Model}
\label{chapter3}

The original Kuramoto model studied on the last chapter exhibits spontaneous synchronization when the coupling strength is larger than a threshold, termed critical parameter. Synchronization in many biological systems, however, is not spontaneous, but frequently depends on external stimuli. A natural extension of the Kuramoto model, therefore, is to include the influence of an external periodic force acting on the system \cite{Sakaguchi1988,Ott2008,Childs2008,Hindes2015}.  In this chapter we review the dynamics of the forced Kuramoto model as studied in detail by Childs and Strogatz \cite{Childs2008}.

\section{Introduction}

The forced Kuramoto model is defined by the addition of a periodic external drive to the original equations (\ref{original}),

\begin{equation}
\dot{\theta_i} = \omega_i + \frac{\lambda}{N} \sum_{j=1}^{N} \sin (\theta_j - \theta_i) + F \sin(\sigma t - \theta_i),
\label{forcingchilds}
\end{equation}
where $F$ is the amplitude of forcing and $\sigma$ is the forcing frequency. As we have seen, the distribution $g(\omega)$ of natural frequencies tends to desynchronize the oscillators, while the coupling $\lambda$ is responsible for the spontaneous synchronization of the units. On the other hand, the role of external forcing is to drive the oscillators to the forcing frequency $\sigma$. The competition between these regimes (desynchronization, spontaneous and forced synchronization) can be analysed by varying the parameter space. 

In order to get rid of the explicit time dependence in equation (\ref{forcingchilds}) we can perfom a change of coordinates to analyse the dynamics in a reference frame corotating with the driving force:

\begin{equation}
\phi_i = \theta_i - \sigma t
\label{change}
\end{equation}
which leads to 
\begin{equation}
\dot{\phi_i} = (\omega_i -\sigma) + \frac{\lambda}{N} \sum_{j=1}^{N} \sin (\phi_j - \phi_i) - F \sin \phi_i.
\label{forcedd2}
\end{equation}

One of the first studies of the periodically forced Kuramoto model was made by Sakaguchi \cite{Sakaguchi1988}, where he analysed the dynamical behavior of equations (\ref{forcedd2}). The simulations showed that when $F$ or $\sigma$ are large, a fraction of oscillators synchronize with external force, a phenomenon called ``forced entrainment'', while the rest remained desynchronized. On the other hand, when $F$ and $\sigma$ are small enough, a fraction of oscillators becomes self-synchronized at a different frequency of external driving. This characterizes the ``mutual entrainment'' state. The competition between these two different regimes seems to meet on the phase diagram and could be a signature of a transition between them. The curves of the phase diagram correspond to different bifurcations, although Sakaguchi did not go any further in these analysis.

The work of Antonsen et al. \cite{Antonsen2008} showed an improvement in the analytical development made by Sakaguchi. Their numerical and linear stability analysis exhibit a set of bifurcation curves in a reduced dimensionality. In this sense, they described the transitions between the different regimes of synchronization in low dimensional picture, but the details of the parameter space were still unclear. However, using \cite{Antonsen2008} Ott and Antonsen \cite{Ott2008} made an important discovery. They showed that the forced Kuramoto model has an invariant manifold under the dynamics, using a specific family of states satisfying a set of conditions. In this sense, they found an exact closed form solution for the complex order parameter $z$ in a two-dimensional dynamical system in a particular case where the frequency distribution $g(\omega)$ is Lorentzian.

In this context, the work of Childs and Strogatz \cite{Childs2008} used the two-dimensional system derived in \cite{Ott2008}. Their work gives a complete analysis of the bifurcation structure for the forced Kuramoto model. The authors considered a system composed of infinitely many phase oscillators with random intrinsic natural frequencies, global sinusoidal coupling and external sinusoidal forcing, using equation (\ref{forcedd2}). In this chapter we will briefly rewiew the paper of Childs and Strogatz \cite{Childs2008}. We will derive the reduced equations by carrying out the continuum limit $N \rightarrow \infty$ in (\ref{forcedd2}), using similar techniques of Chapter \ref{chapter2}. In the next chapter we will extend this work for partially forced Kuramoto oscillators. Although our analysis is not so general, it will allow the possibility of complex networks, not just the fully connected cases considered before.

\section{Derivation of the reduced equations}

As we already did in the last chapter, to take the continuum limit we need to define the density function $\rho(\phi,\omega,t)$  which express the fraction of oscillators with phases in the interval $[\phi, \phi + d\phi]$ and natural frequencies between $\omega$ and $\omega + d\omega$ in time $t$. This quantity must obey the normalization condition

\begin{equation}
 \int_{-\infty}^{\infty} \int_{0}^{2 \pi} \rho(\omega,\phi,t) d\phi d\omega = 1,
\end{equation}
and by definition of $g(\omega)$ we have

\begin{equation}
 \int_{0}^{2 \pi} \rho(\omega,\phi,t) d\phi = g(\omega).
 \label{gomega}
\end{equation}
Following the derivation in Chapter \ref{chapter2} (see equation (\ref{continui})), the continuity equation is simply

\begin{equation}
 \frac{\partial \rho}{\partial t} + \frac{\partial (\rho v)}{\partial \phi} = 0.
\end{equation}
In this equation $v = v(\phi,\omega,t)$ corresponds to $\dot\phi_i$ in the limit of $N \rightarrow \infty$, that is,

\begin{equation}
 v(\phi,\omega,t) = \lim_{N \rightarrow \infty} \dot\phi_i = \omega - \sigma + \lambda \int_{-\infty}^{\infty} \int_{0}^{2 \pi} \sin(\phi' - \phi) \rho(\omega',\phi',t) d \phi' d \omega' - F \sin \phi.
 \label{veloc}
\end{equation}
By using the complex number defined in equation (\ref{order}) we can write the expression above in terms of $z(t)$ in the continuum limit,

\begin{equation}
 z(t) = \int_{-\infty}^{\infty} \int_{0}^{2 \pi} e^{i \phi'} \rho(\omega',\phi',t) d\phi' d\omega',
 \label{complexz}
\end{equation}
which is equivalent to equation (\ref{density}). Now, multiplying (\ref{complexz}) by $e^{-i \phi}$ in both sides, we obtain

\begin{equation}
 z(t) e^{-i \phi}  = \int_{-\infty}^{\infty} \int_{0}^{2 \pi} e^{i (\phi' - \phi)} \rho(\omega',\phi',t) d\phi' d\omega'.
 \label{order3}
\end{equation}
The imaginary part of (\ref{order3}) is

\begin{equation}
 \text{Im}(z e^{-i \phi}) = \int_{-\infty}^{\infty} \int_{0}^{2 \pi} \sin (\phi' - \phi) \rho(\omega',\phi',t) d\phi' d\omega'.
\end{equation}
If we use $\sin \phi = \text{Im}(e^{-i \phi})$ in the expression (\ref{veloc}) we obtain

\begin{equation}
 v = \omega - \sigma  + \lambda \text{Im}(z e^{-i \phi}) + F \text{Im}(e^{-i \phi}) \rightarrow v =  \omega - \sigma + \text{Im}[(\lambda z + F) e^{-i \phi}].
\end{equation}
Using the relation $\text{Im}(\xi) = \frac{1}{2 i} (\xi - \xi^*)$, the equation above becomes

\begin{equation}
 v = \omega - \sigma  + \frac{1}{2 i} [(\lambda z + F) e^{-i \phi} - (\lambda z + F)^* e^{i \phi}]
 \label{vimag}
\end{equation}
where $*$ denotes complex conjugation. Now, we are able to rewrite the continuity equation by using (\ref{vimag}), that is

\begin{equation}
 \frac{\partial \rho}{\partial t} + \frac{\partial}{\partial \phi} \left(\rho \left\{\omega - \sigma  + \frac{1}{2 i} [(\lambda z + F) e^{-i \phi} - (\lambda z + F)^* e^{i \phi}] \right \} \right) = 0.
 \label{eqcont}
\end{equation}

In order to solve the continuity equation we can expand $\rho$ as a Fourier series in $\phi$,

\begin{align}
 \rho(\omega,\phi,t) &= \frac{1}{2 \pi} \sum_{n = - \infty}^{+ \infty} \rho_n (\omega,t) e^{i n\phi} = \frac{1}{2 \pi} \rho_0(\omega,t) + \frac{1}{2 \pi} \sum_{n = 1}^{+ \infty} \rho_n (\omega,t) e^{i n\phi} + \text{c.c.}  \\
 &= \frac{g(\omega)}{2 \pi} \left[1 + \sum_{n = 1}^{+ \infty} \rho_n (\omega,t) e^{i n\phi} + \text{c.c.} \right],
 \label{series}
\end{align}
where c.c. denotes complex conjugate. We can verify that $\rho_0 \equiv g(\omega)$ by integrating $\rho$ in $\phi$ (see equation (\ref{gomega})). As pointed by \cite{Childs2008}, if we substitute equation (\ref{series}) into (\ref{complexz}) and (\ref{eqcont}) we would have an infinite set of coupled nonlinear ordinary differential equations, difficulting the analysis. However, using the Ott and Antonsen ansatz, we can restrict $\rho$ to a special family of densities, such that

\begin{equation}
 \rho_n(\omega,t) = [\alpha(\omega,t)]^n,
\end{equation}
for all $n \geq 1$ and $|\alpha(\omega,t)| \leq 1$ to avoid divergence of the series \cite{Ott2008}. Thereby, equation (\ref{series}) is rewritten as

\begin{equation}
 \rho(\omega,\phi,t) = \frac{g(\omega)}{2 \pi} \left[1 + \sum_{n = 1}^{+ \infty} [\alpha (\omega,t)]^n e^{i n\phi} + \text{c.c.} \right].
 \label{fourierf}
\end{equation}
Now, we have to perform the derivatives $\partial \rho / \partial t$ and $\partial (\rho v) / \partial \phi$ in order to rewrite the continuity equation. These calculations give

\begin{equation}
 \frac{\partial \rho}{\partial t} = \frac{g(\omega)}{2 \pi \alpha} \left[\sum_{n=1}^{\infty} n \alpha^n e^{in \phi} \frac{\partial \alpha}{\partial t} + c.c. \right],
 \label{part1}
\end{equation}

\begin{equation}
 \frac{\partial (\rho v)}{\partial \phi} = \left(\frac{\partial \rho}{\partial \phi} \right) v + \left(\frac{\partial v}{\partial \phi} \right) \rho,
\end{equation}
where

\begin{equation}
 \left(\frac{\partial \rho}{\partial \phi} \right) = \frac{g(\omega)}{2 \pi} \left[\sum_{n=1}^{\infty} \alpha^n (in) e^{in \phi} + \text{c.c}. \right] \qquad \text{and} \qquad \frac{\partial v}{\partial \phi} = - \frac{1}{2} [(\lambda z + F) e^{-i \phi} + \text{c.c}].
 \label{part2}
\end{equation}
Substituting (\ref{part1}) and (\ref{part2}) in equation (\ref{eqcont}) and filtering the terms of $e^{i n \phi}$ we obtain

\begin{equation}
 \frac{d \alpha}{d t} = \frac{1}{2} (\lambda z + F)^* - i(\omega - \sigma) \alpha - \frac{1}{2} (\lambda z + F) \alpha^2.
 \label{alfa}
\end{equation}
Since we need to evaluate the complex order parameter $z(t)$, we can rewrite expression (\ref{complexz}) in terms of $\alpha$:

\begin{equation*}
 z(t) = \int_{-\infty}^{\infty} d\omega \left [\int_{0}^{2 \pi} \frac{g(\omega) e^{i \phi}}{2 \pi} \left(1 + \sum_{n=1}^{+ \infty} [\alpha(\omega,t)]^n e^{in \phi} + \text{c.c.}  \right) d\phi \right].
 \end{equation*}
By perfoming the integral over $\phi$ we obtain

\begin{equation*}
 \frac{g(\omega)}{2 \pi} \int_{0}^{2 \pi} \left( e^{i \phi} + \sum_{n=1}^{+ \infty} [\alpha(\omega,t)]^n e^{i (n+1) \phi} + \sum_{n=2}^{+ \infty} [\alpha^*(\omega,t)]^n e^{-i(n-1) \phi} + \alpha^*(\omega,t) \right) d \phi = \alpha^*(\omega,t) g(\omega),
\end{equation*}
which reduces to

\begin{equation}
 z(t) = \int_{-\infty}^{\infty} \alpha^*(\omega,t) g(\omega) d\omega.
 \label{interz}
 \end{equation}
Now we can choose the frequency distribution $g(\omega)$ to be a Lorentzian,

\begin{equation}
 g(\omega) = \frac{\Delta}{\pi[(\omega - \omega_0)^2 + \Delta^2]},
 \label{lorentz}
\end{equation}
and the equation for $z(t)$ becomes

\begin{equation}
 z(t) = \frac{\Delta}{\pi} \int_{-\infty}^{\infty} \frac{\alpha^*(\omega,t) d\omega}{(\omega - \omega_0)^2 + \Delta^2} \rightarrow z(t) = \frac{\Delta}{\pi} \int_{-\infty}^{\infty} \frac{\alpha^*(\omega,t) d\omega}{[\omega -(\omega_0 + i \Delta)][\omega -(\omega_0 - i \Delta)]}.
 \label{intz}
 \end{equation}
 
\begin{figure}[H]
\center
\includegraphics[scale=0.2]{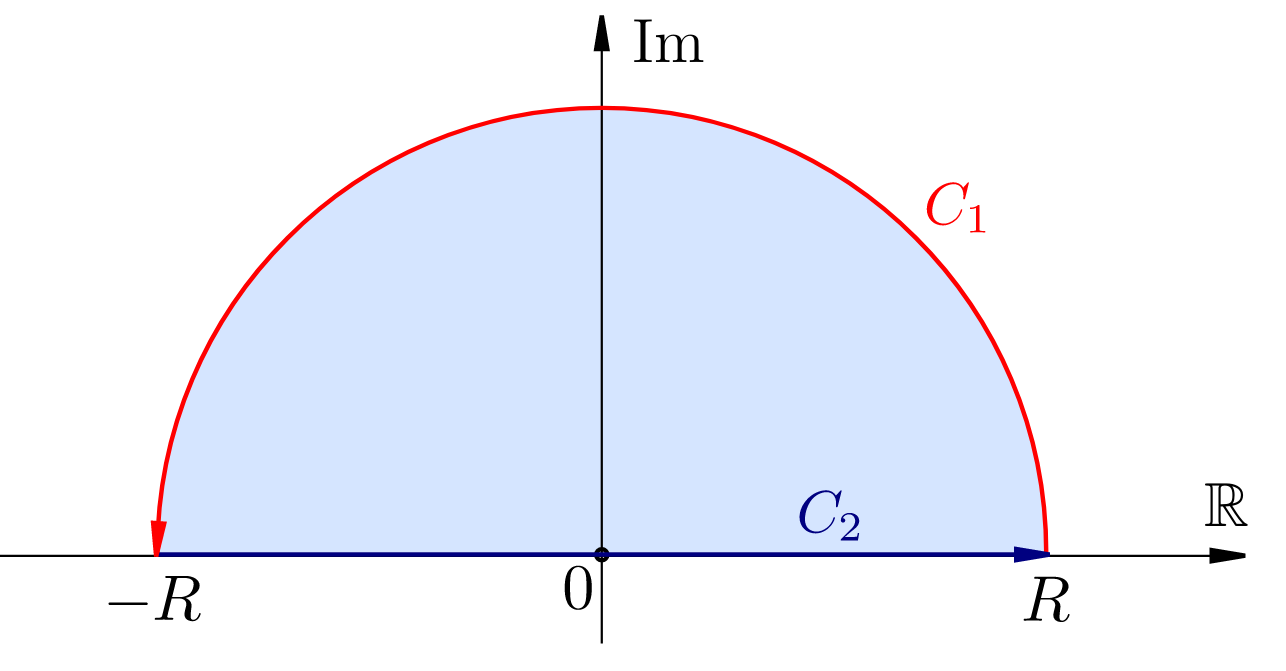}
\caption{Path integration for (\ref{intz}). The path $C$ is the concatenation of the paths $C_1$ and $C_2$.}
\label{circle}
\end{figure}

In order to perform the integration on the complex plane, the function $\alpha(\omega,t)$ has to obey some conditions, as noted in \cite{Ott2008}. First, $\alpha(\omega,t)$ must be analytically continued from real $\omega$-axis into the lower half $\omega-$plane for all $t \geq 0$ and, second, $|\alpha(\omega,t)| \rightarrow 0$ as Im$(\omega) \rightarrow - \infty$. The integral (\ref{intz}) diverges in two points: $\omega_1 = \omega_0 + i \Delta$ and $\omega_2 = \omega_0 - i \Delta$. To perform the calculation, let's define the contour $C$ that lies on the real axis from $-R$ to $R$ and then goes counterclockwise along a semicircle from $R$ to $-R$. This curve encloses the pole $z_0 = \omega_0 + i \Delta$ and the contour integral along $C$ is

\begin{equation}
 \int_C f(z)dz = \frac{\Delta}{\pi} \int_C \frac{\alpha^*(z,t) dz}{[z -(\omega_0 + i \Delta)][z -(\omega_0 - i \Delta)]}.
 \end{equation}
Using the residue theorem,

\begin{equation}
 \int_C f(z)dz = 2 \pi i \sum Res(f,z_0),
 \end{equation}
we have

\begin{equation*}
 \int_C f(z)dz = 2 \pi i \lim_{z \rightarrow z_0} (z - z_0) f(z_0) \rightarrow 2 \pi i \frac{\Delta}{\pi} \lim_{z \rightarrow z_0} \frac{(z - \omega_0 - i \Delta) \alpha^*(\omega_0 + i \Delta)}{(z - \omega_0 - i \Delta)(z - \omega_0 + i \Delta)},
\end{equation*} 
that is,

 \begin{equation}
 \int_C f(z)dz = \alpha^*(\omega_0 + i \Delta).
 \label{residu}
\end{equation}
We can split the contour $C$ in a curved arc $C_1$ and a straight part $C_2$, as depicted in figure \ref{circle}. Then, we have

\begin{equation*}
 \int_{C_1} f(z)dz + \int_{C_2} f(z)dz = \alpha^*(\omega_0 + i \Delta).
\end{equation*}
Since $C_2$ is contained on real axis, the integral over $C_2$ is real, that is,

\begin{equation*}
 \int_{C_2} f(z)dz = \int_{-R}^R f(z)dz 
\end{equation*}
If $f(z)$ is continuous on the semicircular contour $C_1$ for all large $R$, then by Jordan's lemma we have $\lim_{R \rightarrow \infty} \int_{C_1} f(z) dz = 0$, which means that the improper integral (\ref{intz}) is just the equation (\ref{residu}).

Finally, the result is $z(t) = \alpha^*(\omega_0 + i \Delta)$ and using the complex conjugate of equation (\ref{alfa}) we can compute the time evolution of $z$,

\begin{equation*}
 \frac{dz}{dt} = \frac{\partial \alpha^*}{\partial t} \biggr\rvert_{\omega = \omega_0 + i\Delta} \rightarrow \frac{dz}{dt} = \left[\frac{1}{2} (\lambda z + F) + i(\omega - \sigma) \alpha^* - \frac{1}{2} (\lambda z + F)^* (\alpha^*)^2  \right] \biggr\rvert_{\omega = \omega_0 + i\Delta},
\end{equation*}
then 
\begin{equation}
 \frac{dz}{dt} = \frac{1}{2} [(\lambda z + F) - z^2(\lambda z + F)^*] - [\Delta + i(\sigma - \omega_0)]z.
 \label{2dsystem}
\end{equation}

\section{Analysis of the reduced equations}

In this section we will analyse the reduced equations of the two-dimensional system of (\ref{2dsystem}). First we reduce the number of parameters by reescaling $\hat{t} = \Delta t$, $\hat{F} = F/\Delta$, $\hat{\lambda} = \lambda /\Delta$, $\hat{\sigma} = \sigma/\Delta$ and $\hat{\omega_0} = \omega_0/\Delta$. We also let $\Omega = \sigma - \omega_0$. In what follows, we will use $\Delta = 1$ and we will drop the hats for ease notation.

By introducing the polar coordinates, $z = r e^{i \psi}$, we can rewrite equation (\ref{2dsystem}) as

\begin{equation}
 \frac{d r }{dt} e^{i \psi} + i r e^{i \psi} \frac{d \psi}{dt} = \frac{1}{2} [(\lambda r e^{i \psi} + F) - r^2 e^{2i \psi}(\lambda r e^{i \psi} + F)^*] - (1 + i \Omega)r e^{i \psi}.
\end{equation}
Separating the expression above into real and imaginary parts we obtain the dimensionless evolution equations for $r$ and $\psi$,

\begin{equation}
 \dot r \equiv \frac{d r }{dt} = \frac{\lambda}{2} r (1 - r^2) - r + \frac{F}{2} (1 - r^2) \cos \psi,
 \label{rhoprime}
\end{equation}

\begin{equation}
 \dot \psi \equiv  \frac{d \psi}{dt} = - \left[\Omega + \frac{F}{2} \left(r + \frac{1}{r} \right) \sin \psi \right].
  \label{phiprime}
\end{equation}

\begin{figure}[!ht]
\center
\includegraphics[scale=0.26]{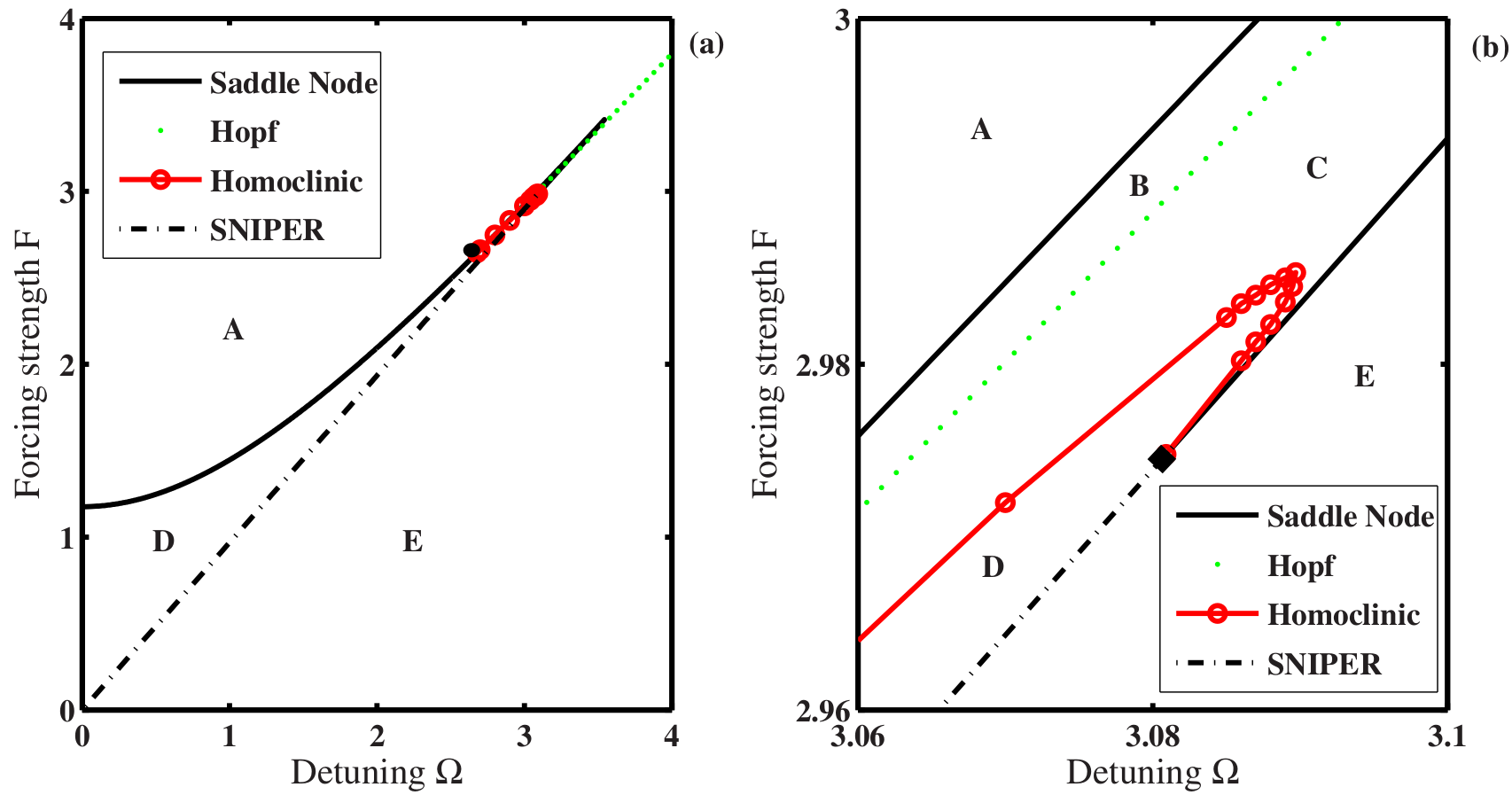}
\includegraphics[scale=0.26]{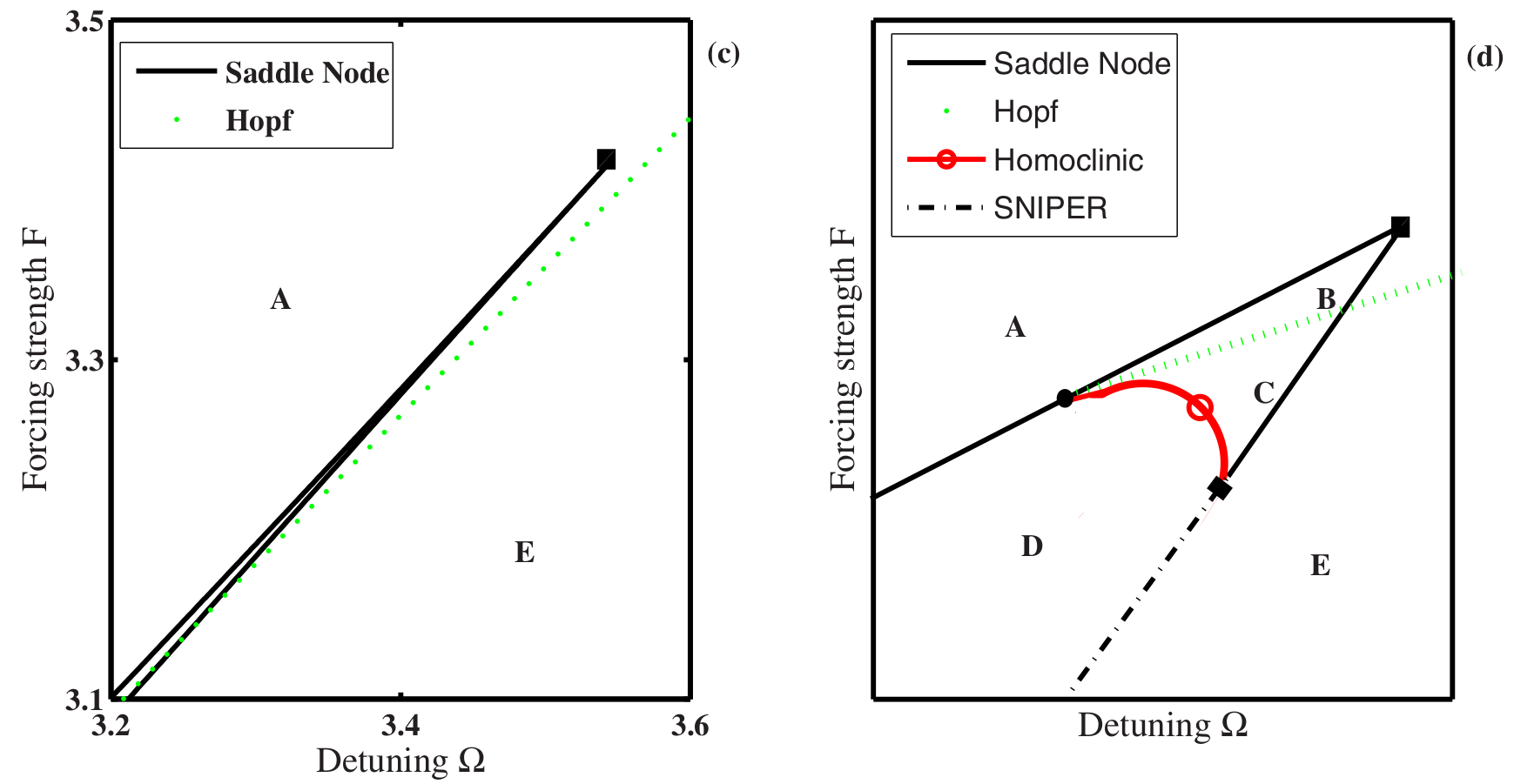}
\caption{Stability diagram for the forced Kuramoto model obtained from equations (\ref{rhoprime}) and (\ref{phiprime}). The bifurcation curves are represented as a function of intensity $F$ and frequency $\Omega$ of the external force. The coupling strenght is fixed at $\lambda = 5$. The system rich dynamics shows four types of bifurcations (panel (a)): the supercritical Hopf bifurcation, the homoclinic bifurcation, and two types of saddle-node bifurcations. The stability diagram is divided into 5 regions, each of them representing qualitatively different phase portraits. Panels (b) and (c) are enlargements of panel (a) (detailed explanation is on the text). Panel (d) represents a schematic version of the stability diagram in order to clarify the truncated crossover region. Figure taken from \cite{Childs2008}.}
\label{diagram}
\end{figure}

Before we reproduce the analytical results obtained in \cite{Childs2008}, we will first describe the resulting stability diagram, as depicted in figure \ref{diagram} extracted from \cite{Childs2008}. The rich dynamics exhibited by the system is essentially divided into two big regions: the one labeled ``A'' represents the forced entrainment, which means that a fraction of oscillators is moving in synchrony with the same frequency as the drive signal (induced synchronization). The other region, labeled ``E'', represents the mutual entrainment, where a fraction of the system is spontaneously synchronized. The remaining regions, ``B,'', ``C'' and ``D'', represent partial forced synchronization. We will briefly discuss each of them on the next section.

As we can see in figure \ref{diagram} the stability diagram is divided into 5 regions, each of them representing qualitatively different phase portraits. The system's rich dynamics shows the occurence of four types of bifurcation curves, namely saddle-node, Hopf, homoclinic and SNIPER (saddle-node infinite-periodic). Panels (b) and (c) are enlargements of the results of panel (a), in order to explore the details of the bifurcation curves. As pointed in \cite{Childs2008}, because all these figures are very hard to interpret, panel (d) proposes a schematic version of the stability diagram. In what follows, we will derive the analytical approach to find the parametric curves of the stability diagram and we will further discuss the results.

\subsection{Saddle-node bifurcations}

In bifurcation theory it is usual to find the fixed points in terms of the parameters of the problem and then study their stability. However, it is algebraically complicated to solve equations (\ref{rhoprime}) and (\ref{phiprime}) in terms of them. Because we are concerned with the bifurcation curves we can impose an appropriate condition for the bifurcation we want to analyse and solve for the parameters in terms of the fixed ponts. This technique will allow us to derive the bifurcation curve in closed form, either explicitly or parametrically.

We can start by using equations (\ref{rhoprime}) and (\ref{phiprime}) and defining the functions $\dot r = f(r, \psi)$, $\dot\psi = g(r,\psi)$ and $(r_0,\psi_0)$ as the fixed points. The Jacobian matrix is then written as

\begin{equation}
\mathcal{J} = 
\begin{pmatrix} 
\frac{\partial f}{\partial r} \biggr\rvert_{r_0,\psi_0} & \frac{\partial f}{\partial \psi} \biggr\rvert_{r_0,\psi_0} \\ \\
\frac{\partial g}{\partial r} \biggr\rvert_{r_0,\psi_0} & \frac{\partial g}{\partial \psi} \biggr\rvert_{r_0,\psi_0}
\end{pmatrix}
\end{equation}
By using equations (\ref{rhoprime}) and (\ref{phiprime}), $\mathcal{J}$ becomes

\begin{equation}
\mathcal{J} = 
\begin{pmatrix} 
\frac{\lambda}{2} (1- 3 r_0^2) - F r_0 \cos \psi_0 - 1 & -\frac{F}{2} \sin \psi_0(1 - r_0^2) \\ \\ \\
-\frac{F}{2} \sin \psi_0 \left(1 - \frac{1}{r_0^2} \right) & -\frac{F}{2} \cos \psi_0 \left(r_0 + \frac{1}{r_0} \right)
\end{pmatrix}
\end{equation}
To ease notation we will omit the index of the fixed points. The equilibrium condition imposes that $\dot r = 0$ and $\dot \psi = 0$. At a saddle-node bifurcation, one of the eigenvalues of $\mathcal{J}$ has to be 0, which means that the determinant of the Jacobian vanishes, $\det \mathcal{J} = 0$. These three conditions must be satisfied simultaneously:

\begin{equation}
 \dot r = 0 \rightarrow \lambda = \frac{2}{1 - r^2} - \frac{F \cos \psi}{r}, \qquad \dot \psi = 0 \rightarrow \Omega = - \frac{F}{2 r} (1 + r^2) \sin \psi,
 \label{phirhozero}
\end{equation}

\begin{equation}
 \det \mathcal{J} = 0 \rightarrow \left[\frac{\lambda}{2}(1 - 3 r^2)- F r \cos \psi - 1 \right] \left[-\frac{F}{2} \cos \psi \left(\frac{1 + r^2}{r} \right) \right] = \left(\frac{F}{2} \right)^2 \sin^2 \psi (1 - r^2) \left(\frac{r^2 - 1}{r^2} \right).
 \label{det0}
\end{equation}
Now we can substitute the expression of $\lambda$ (\ref{phirhozero}) in equation (\ref{det0}) and after some manipulations we obtain

\begin{equation}
 F = -\frac{4 r^3 (1 + r^2)\cos \psi }{(1 - r^2)^2(1 + r^2 \cos 2 \psi)}.
 \label{forceF}
\end{equation}
If we substitute (\ref{forceF}) into equations (\ref{phirhozero}) we can write $\lambda$ and $\Omega$ in terms of $r$ and $\psi$, that is

\begin{equation}
 \lambda = \frac{2(r^4 + 2 r^2 \cos 2 \psi + 1)}{(1 - r^2)^2(1 + r^2 \cos 2 \psi)}, \qquad 0 \leq r \leq 1,
 \label{lambda}
\end{equation}

\begin{equation}
 \Omega = \frac{(r^3 + r)^2 \sin 2\psi}{(1 - r^2)^2(1 + r^2 \cos 2 \psi)}, \qquad - \pi \leq \psi \leq \pi.
 \label{omega}
 \end{equation}
The resulting set of equations (\ref{forceF}, \ref{lambda}, \ref{omega}) gives the saddle-node surface and it is one of the various parametrizations possible. These results provide very important information. For instance, when $r = 0$ there is no external forcing ($F=0$). In this case, equation (\ref{lambda}) gives the minimum value of the coupling strenght, $\lambda = 2$, which is exactly the value of critical coupling in the original Kuramoto model with a Lorentzian $g(\omega)$, expression (\ref{lorentz}). We can confirm by using equation (\ref{lambdac}) where $g(0) = (\pi \Delta)^{-1}$ which gives $\lambda_c = 2 \Delta$, or just $\lambda_c = 2$, since we are using $\Delta = 1$. Hence $\lambda \geq 2$ increases monotonically with $r$ for fixed $\psi$.

In order to reduce the number of unknown parameters we can consider a slice through the saddle-node surface at a constant value of coupling strenght for $\lambda>2$ and then plot the respective saddle-node curves in the $(\Omega,F)-$ plane. To get this parametrization, we can eliminate the $\psi$ dependence by isolating $\sin 2\psi$ in equation (\ref{omega}) and $\cos 2\psi$ in equation (\ref{lambda}). The result is

\begin{equation}
 \sin 2\psi = \xi + \frac{\xi(1+ r^4 - \eta)}{\eta-2}, \qquad \xi \equiv \frac{\Omega(1-r^2)^2}{(r^3+r)^2}, \qquad \cos 2 \psi = \frac{1 + r^4 - \eta}{r^2(\eta-2)}, \qquad \eta \equiv \frac{\lambda(1-r^2)^2}{2}.
\end{equation}
Now, using $\cos^2 2 \psi + \sin^2 2 \psi = 1$, we obtain the parametrization of the saddle-node (SN) curve as a function of $\lambda$ and $r$:

\begin{equation}
\Omega_{SN}(\lambda,r) = \frac{(1+r^2)^{3/2} \sqrt{\lambda (r^2-1)[\lambda (r^2-1)^2-4] - 4(r^2+1)}}{2(r^2 -1)^2},
\label{eq24}
\end{equation}

\begin{equation}
F_{SN}(\lambda,r) = \frac{\sqrt{2} r^2 \sqrt{\lambda^2 (1-r^2)^3 + 2\lambda(r^4-4r^2+3) - 8} }{(r^2 -1)^2}.
\label{eq25}
\end{equation}
Figure \ref{diagram} shows the parametric plot for $\lambda=5$ fixed in the range $0 < r < 1$. It's worth noting that the two branches of the saddle-node curve intersect tangentially at a point named codimension-2 cusp and marked by the solid square in figure \ref{diagram}(d). The coordinates of the cusp can be found numerically by calculating $d F_{SN}/d r \rvert_{\lambda=5,r=r'} = 0$, which gives $r' \approx 0.7267$. Then, substituting this value on equations (\ref{eq24}) and (\ref{eq25}) we obtain $(\Omega_{cusp}, F_{cusp}) \approx (3.5445,3.4164)$. At the lower branch of the saddle-node curve, where $F \approx \Omega$, there is a large section of SNIPER bifurcations, which are responsible to create or destroy limit cycles in the phase portrait.

\subsection{Hopf bifurcation}

In order to find the Hopf bifurcation curve, we need to impose simultaneously that $\dot r = 0$ and $\dot \psi = 0$ (condition to equilibrium points) and tr $\mathcal{J} = 0$ and $\det \mathcal{J} > 0$ (condition to Hopf bifurcation, equivalent to require that the eigenvalues are pure imaginary), where ``tr'' denotes the trace of the Jacobian. We can start by isolating $F \cos \psi$ and $F \sin \psi$ of equations (\ref{phirhozero}),

\begin{equation}
 F \cos \psi = \frac{2 r - \lambda r (1 - r^2)}{1 - r^2}, \qquad F \sin \psi = -\frac{2\Omega r}{1 + r^2}.
 \label{fcosseno}
\end{equation}
The condition tr $\mathcal{J} = 0$ gives,

\begin{equation}
 \text{tr} \mathcal{J} = 0  \rightarrow \frac{\lambda (1 - 3 r^2)}{2} - F r \cos \psi - 1 = \frac{F \cos \psi}{2} \left(r + \frac{1}{r} \right).
 \label{tracej}
\end{equation}
Substituting $F \cos \psi$ of (\ref{fcosseno}) in (\ref{tracej}) and multiplying both sides by $r(1-r^2)$ we obtain, after some manipulations,

\begin{equation}
 (2r - \lambda r + \lambda r^3) (r^2 + 1) = r (1 - r^2) (\lambda - 3 \lambda r^2 - 2) - 2 r^2(2 r - \lambda r + \lambda r^3)
\end{equation}
Now, isolating $r$ we have

\begin{equation}
 r = \sqrt{\frac{\lambda - 2}{\lambda + 2}}.
 \label{eq28}
\end{equation}
Since the parameter $r$ depends only on $\lambda$, we can write an expression for $F(\lambda,\Omega)$ if we calculate $\cos^2 \psi + \sin^2 \psi = 1$ by using equations (\ref{fcosseno}) and then substituting into (\ref{eq28}). These manipulations lead to

\begin{equation}
 F_{Hopf} = \frac{1}{2 \lambda} \sqrt{\frac{(\lambda-2)[\lambda^4 - 4 \lambda^3 + 4(\Omega^2+1)\lambda^2 + 16 \Omega^2 \lambda + 16 \Omega^2]}{\lambda+2}},
\end{equation}
which reduces to

\begin{equation}
 F_{Hopf} = \frac{\sqrt{3}}{10 \sqrt{7}} \sqrt{225 + 196 \Omega^2},
\label{Fhopf}
 \end{equation}
for $\lambda = 5$. The curve $F_{Hopf}(\Omega)$ is depicted in figure \ref{diagram}.
  
To plot equation (\ref{Fhopf}) on the phase portrait with the saddle-node bifurcation, we still need to evaluate $\det \mathcal{J} > 0$, which is the remaining condition for the Hopf bifurcation to occur. The coordinates of this limit point can be found when we impose the conditions for the Takens-Bogdanov point, obtained when we calculate simultaneously $\dot r = 0$, $\dot \psi = 0$, tr $\mathcal{J} = 0$ and $\det \mathcal{J} = 0$. We can get the coordinates of the Takens-Bogdanov point analytically by substituting equation (\ref{eq28}) in expressions (\ref{eq24}) and (\ref{eq25}), that is,

\begin{equation}
 \Omega_{TB} = \frac{(\lambda-2)\lambda^2}{4(\lambda+2)}, \qquad F_{TB} = \frac{\lambda-2}{4} \sqrt{\frac{\lambda^3-2 \lambda^2 + 4\lambda - 8}{\lambda+2}}.
 \label{tbpoint}
\end{equation}
For $\lambda = 5$, we have $\Omega_{TB} = \approx 2.6786$ and $F_{TB} \approx 2.6441$. The Takens-Bogdanov (TB) point is represented by the filled circle on panels (a) and (d) of figure \ref{diagram}. As regards the dynamical behavior, the Takens-Bogdanov point separates the upper branch of the saddle-node bifurcation into two regions with distinct characteristics. Below the TB point, an unstable node collides with a saddle along the saddle-node curve, as can be seen by comparing regions D and A depicted on the phase portrais of figures \ref{childsfig2}, panels (d) and (a), respectively. On the other hand, above the TB point, a stable node collides with a saddle along the saddle-node curve, representing the transition between regions B and A, as shown in panels (b) and (a) in figure \ref{childsfig2}.

\subsection{Homoclinic bifurcation}

The homoclinic bifurcation occurs when a periodic orbit collides with a saddle-node. As a consequence, the limit cycle disappears after the collision. The theory of the Takens-Bogdanov bifurcation predicts that a curve of homoclinic bifurcation must occur from the codimension-2 point (black square on figures \ref{diagram}(c) and (d)), tangentially to the saddle-node and to the Hopf curves. The homoclinic curve can be computed numerically. As we can see on figure \ref{diagram} (b), the region where the homoclinic appears on the diagram is very narrow, which makes it almost indistinguishable from the Hopf curve. This produces a very small area between them. It is interesting to note that the homoclinic curve moves paralllel to the Hopf curve and then goes back until it ends on the codimension-2 ``saddle-node-loop'' point, marked as a black diamond on figures \ref{diagram}(b) and (d), where it meets at the lower branch of the saddle-node and SNIPER curves.

\section{Phase portraits and bifurcation scenarios}

\begin{figure}[H]
\center
\includegraphics[scale=0.23]{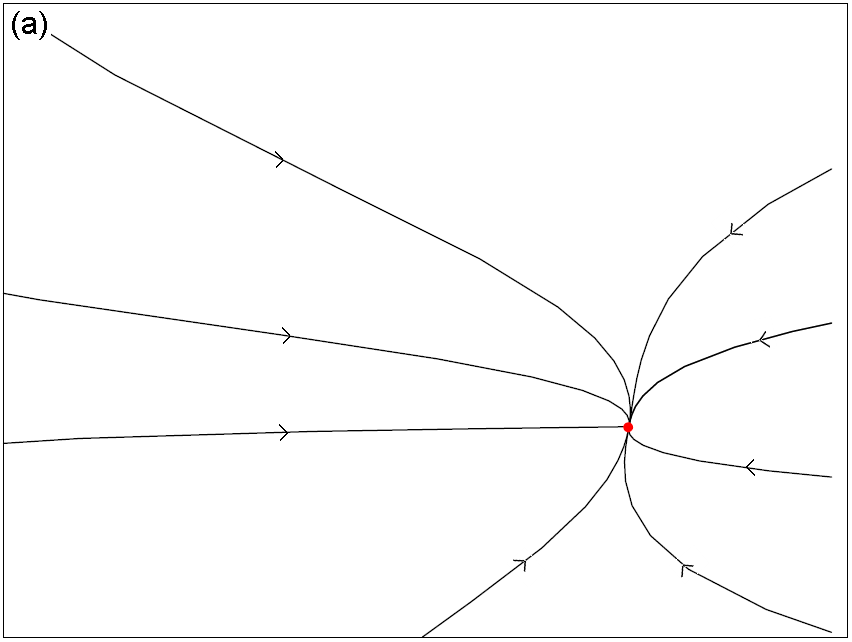}
\includegraphics[scale=0.23]{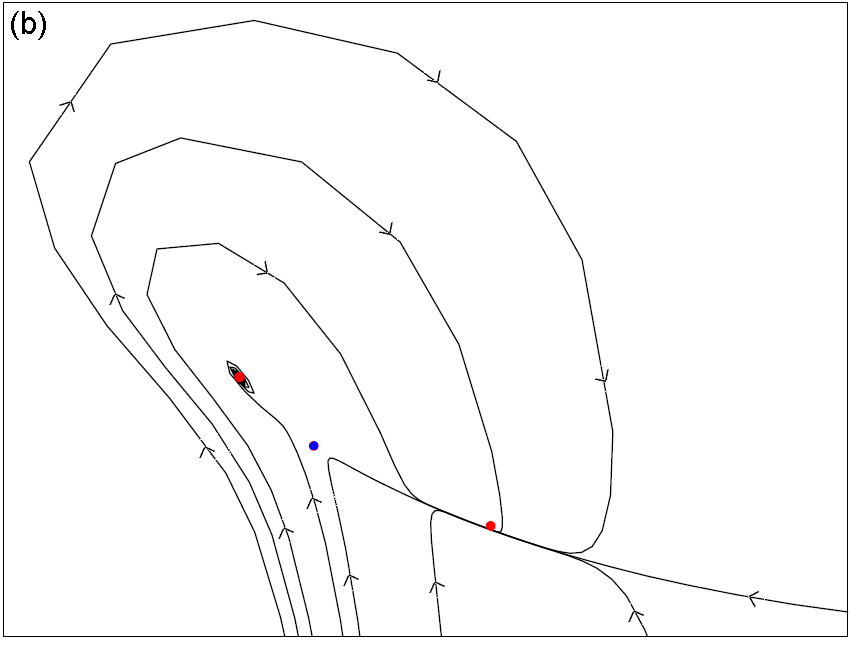}
\includegraphics[scale=0.23]{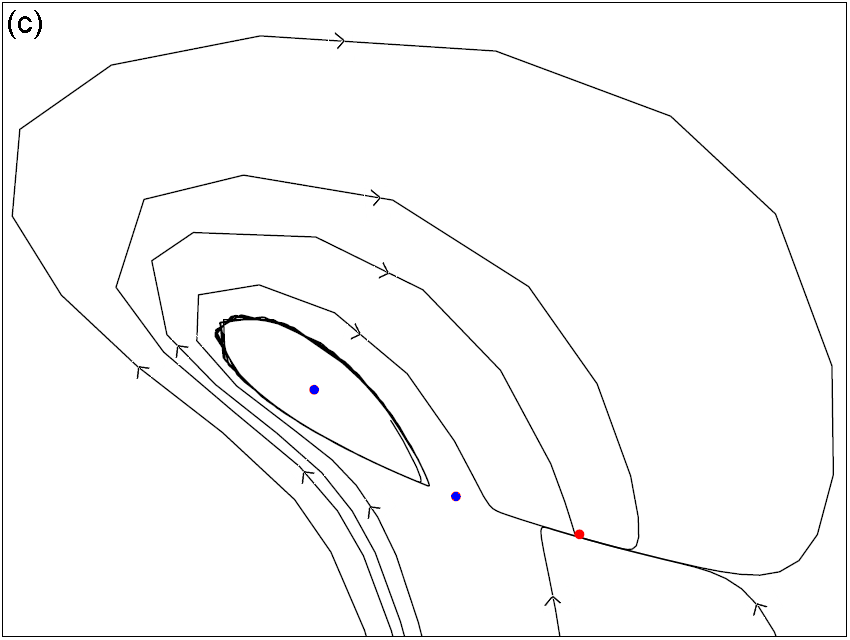}
\includegraphics[scale=0.23]{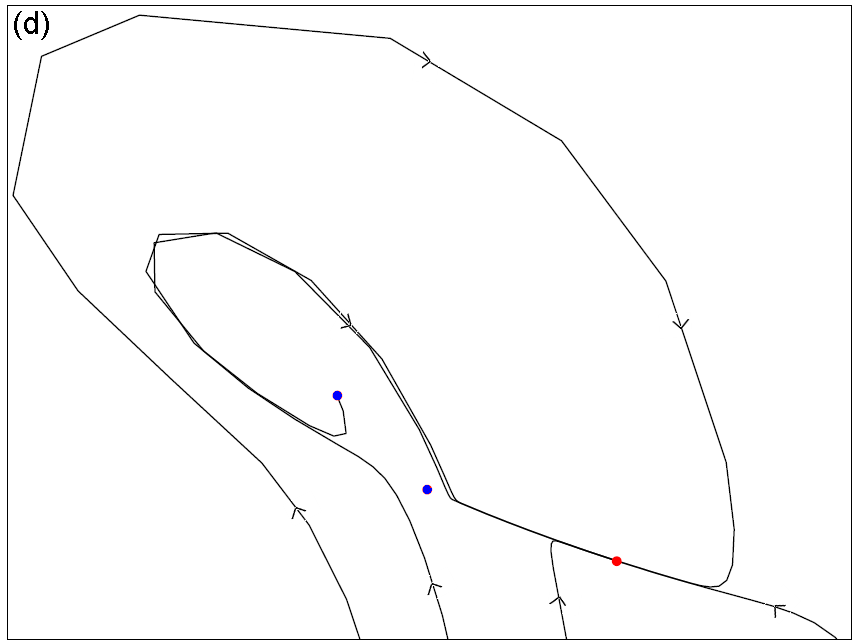}
\includegraphics[scale=0.23]{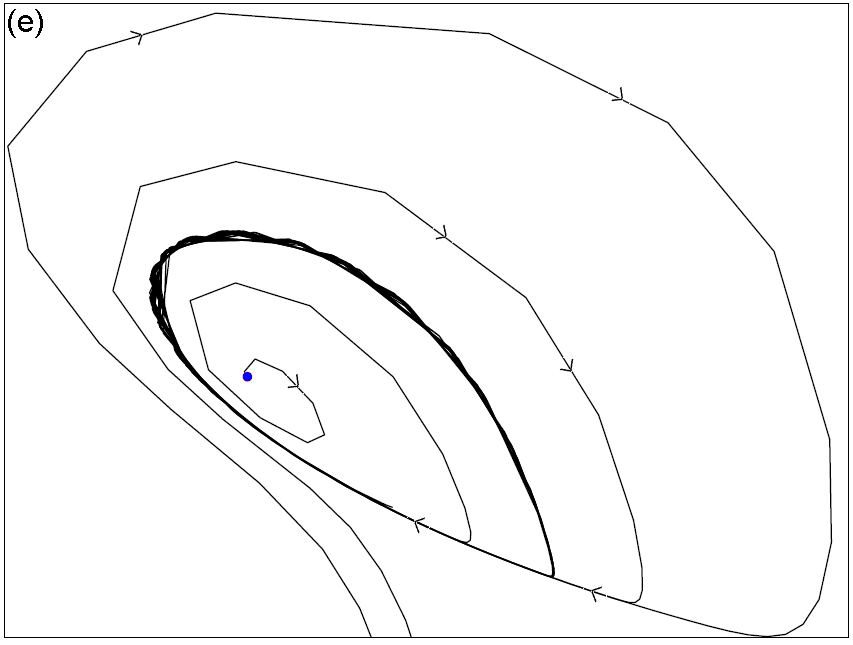}
\caption{Phase portraits for the variables $r$ and $\psi$ for the two-dimensional dynamics of $z$ written in polar coordinates representing the five existing regions. The fixed nodes are represented as red (stable) and blue (unstable) dots. The set of initial conditions to obtain all these curves can be found in appendix \ref{app1}.}
\label{childsfig2}
\end{figure}

So far we reproduced all the bifurcation curves analytically (saddle-node, SNIPER and Hopf) and numerically (homoclinic). Figure \ref{diagram} shows that these curves divide the phase diagram into five regions. We can now discuss the dynamical behavior and the transitions associated in each region, with the support of the phase portraits of figure \ref{childsfig2}. These figures can be computed by integrating numerically equations (\ref{rhoprime}) and (\ref{phiprime}) by varying the parameter space ($F$, $\Omega$ and $\lambda$) for different initial conditions ($r_0$, $\psi_0$).

\begin{enumerate}
 \item \textbf{Region A: forced entrainment}.
 
 In region A the order parameter $z$ converges to the stable fixed point for all initial conditions, as depicted on the phase diagram of figure \ref{childsfig2}(a). In the frame corotating with the drive, $z$ is phase-locked to the drive and moves periodically, which is represented by the fixed point. If we change to the original frame, a significant fraction of oscillators is moving in rigid synchrony with the same frequency of the external driving. In the case where $g(\omega)$ is centered in zero, that is $\omega_0 = 0$, the velocity of the oscillators in forced entrainment is $\dot \psi = 0$ on the corotating frame and $\dot \psi = -\Omega$ in the original frame.

 \item \textbf{Region B: bistability between two states of forced entrainment}

  As we can see on figure \ref{diagram}(b) the region B is very narrow. To understand the transition between regions A and B we can fix $\Omega$ and then decrease $F$. When we pass from A to B, a bifurcation occurs creating a pair of stable and unstable fixed nodes, coexisting with the stable node of the region A, as represented in the lower right part of figure \ref{childsfig2}(b). Region B depicts a bistability regime: for different initial conditions, the system goes to one of the two possible states, differing in the magnitude or in the argument of $z$.

  \item \textbf{Region C: bistability between forced entrainment and phase trapping}
  
  We can continue to analyse figure \ref{diagram}(b) for fixed $\Omega$ and decreasing $F$ until we reach region C, figure \ref{diagram}(c). In this case we pass through the curve of the Hopf bifurcation, where the stable fixed point created in region B loses stability and creates a small attracting limit cycle. On this cycle, $z$ remains running with the same average frequency of $F$, but now its amplitude and relative phase oscillate slightly, characterizing a phase trapping, that is, $z$ is frequency locking without phase locking. This behavior exists simultaneously with that seen on regions A and B, thus region C exhibits a bistability regime between forced entrainment and phase trapping.
  
  \item \textbf{Region D: forced entrainment}
  
  The transition between regions C and D occurs when we cross the homoclinic bifurcation curve. As we approximate from this curve, the limit cycle expands until it touches the saddle-node and forms a homoclinic orbit. Beyond the bifurcation, the limit cycle completly disappears, as depicted in figure \ref{childsfig2}(d). The consequence is a creation of an invariant loop, where the saddle and the original stable node of region A are connected by the branches of the saddle's unstable manifold. In region D, the stable node is the only attractor and the system converges into a state of forced entrainment.
  
  \item \textbf{Region E: mutual entrainment}
  
  In the region E the forced entrainment is completly lost. The transition to region E can occur in many ways. For example, we can pass from region D to E crossing the lower branch of the saddle-node curve, below and to the left of the saddle-node-loop point (black diamond of figure \ref{diagram}(b). In this case, as the bifurcation parameter is varied, the saddle and the stable node in region D of figure \ref{childsfig2}(d) collapse into a single stationary point on a closed orbit. In other words, the stable limit cycle is born with infinite period at the bifurcation point. This characterizes a SNIPER (saddle-node infinite-period) bifurcation, represented as the dashed line in figure \ref{diagram}. The result is a globally attracting limit cycle where the order parameter oscillates at a different frequency of the external driving, which means that a fraction of oscillators is dropped from the drive signal. 
  
  The other scenario possible to reach region E is to cross directly from C. In this case, we pass through the saddle-node bifurcation, where two fixed points collide and annihilate each other. We can imagine this phenomenon if we look at figure \ref{childsfig2}(c): the saddle-node in the middle and the stable node collides, and the limit cycle grows. After the bifurcation, the result is the phase portrait represented in figure \ref{childsfig2}(e). 
  
  The simpler case possible is to pass from region A to E. We can take, for example, a portion in the stability diagram where $F > F_{cusp}$ and $\Omega > \Omega_{cusp}$. For fixed $\Omega$ and decreasing $F$ we move directly from A to E crossing the Hopf bifurcation, leading to the birth of a periodic orbit. In all these cases the system has spontaneously synchronized or, in other words, has entered in a mutual entrainment state.  
   
\end{enumerate}

\section{Discussion}

In this chapter we studied the forced Kuramoto model by reviewing the Childs and Strogatz's work \cite{Childs2008}. We reproduced the analytical results showing the details of the main equations derived in their paper and we analysed the system's rich dynamics by plotting the stability diagram and the phase portrait with all bifurcation curves. To conclude this chapter, we will recapitulate the major ideas and results.

Inspired by several physical and biological systems, such as electrochemical oscillators, coupled metronomes, neutrino flavor oscillations, circadian rhythms and cardiac synchronization induced by heart cells, the Kuramoto model can be easily extended to allow the influence of external forcing. Mathematically, we introduced a periodic driving term on the Kuramoto original system, equation (\ref{forcedd2}).

Previous works on the forced Kuramoto model \cite{Sakaguchi1988, Ott2008} analysed the competition between two regimes: the induced synchronization, also called forced entrainment, where the system's average frequency is equal to the external driving, and the spontaneous synchronization, or mutual entrainment, where the external drive is not enough to drag the oscillators, recovering the Kuramoto original dynamics. Although these works present very relevant improvements on the analytical treatment of the model, they were not able to find the details of the bifurcation between these regimes.

In this sense, Childs and Strogatz, based on \cite{Ott2008}, derived a complete analysis of the forced Kuramoto model. They used equations (\ref{forcedd2}) and explored the reduced dimensionality of a infinite coupled differential equations into a two-dimensional system for a special family of functions proposed in \cite{Ott2008}, leading to a complete bifurcation analysis, where it was possible to derive exact results for the Hopf, saddle-node and Takens-Bogdanov bifurcations.

The stability diagram of figure \ref{diagram} (a) is the main result of the paper and it is substantially divided into two big regions, one concerning to the forced entrainment (region A), and the other to the mutual entrainment (region E). It's hard to see macroscopically the division between regions A and E, but when we zoom in in the parameter space it is possible to access the narrow bifurcation curves, \ref{diagram} (b).

In a zoom out scale, the stability diagram is essentially divided by the straight line $F \approx \Omega$. If we take the $\lambda \rightarrow \infty$ limit, equation (\ref{phiprime}) reduces to

\begin{equation}
 \dot\psi = - \Omega - F \sin \psi,
\end{equation}
which is the Adler equation, used to model systems like fireflies, lasers, and so forth. In the strong coupling $\lambda$ regime, $r \rightarrow 1$ is faster than $\psi$ and the oscillators behave as if they were a single giant element, with a very intensive attracting limit cycle. Analytically, we can study the Takens-Bogdanov point, which lies on the vicinity if the two big regions. Taking the $\lambda$-large limit on equations (\ref{tbpoint}), we obtain

\begin{equation}
 \lim_{\lambda \rightarrow \infty} \frac{F_{TB}}{\Omega_{TB}} \approx 1 - \frac{8}{\lambda^4} \rightarrow F \approx \Omega.
\end{equation}

In the next chapter we will study the forced Kuramoto model on networks, where the topology defines the interactions between the elements. We are going to consider the work of Childs and Strogatz in the regime where $F \approx \Omega$. We will also apply the external forcing only on a fraction of the oscillators. In this context, we are interested in the conditions for global synchronization with external force, if it exists.

\chapter{Global synchronization of partially forced Kuramoto oscillators on networks}
\label{chapter4}

In the last chapter we described a version of the forced Kuramoto model where an external stimulus, represented by a periodic force, was applied to all oscillators of the system. We reviewed the analytical and numerical results of the work \cite{Childs2008} and we showed the rich bifurcation structure of the system.

In this chapter we consider systems where the oscillators' interconnections form a network and where the force acts only on a fraction of the oscillators. We are interested in the conditions for global synchronization as a function of the fraction of nodes being forced and how it depends on network topology. The motivation for this study is to understand the response of a neural complex network to localized stimuli. We show that the minimum force $F_{crit}$ needed for global synchronization scales as $1/f$, where $f$ is the fraction of forced oscillators, and it is independent of the internal coupling strength $\lambda$. However, in order to reach synchronization with fraction $f$ a minimum internal strength is needed. The degree distribution of the network and the set of forced nodes modify the $1/f$ behavior in heterogeneous networks. We develop analytical approximations for $F_{crit}$ as a function of the fraction $f$ of forced oscillators and for the minimum fraction $f_{crit}$ for which synchronization occurs as a function of $\lambda$. 

This chapter was published in \cite{us}. We will follow its structure: in section \ref{model} we describe the partially forced Kuramoto model and present the results of numerical simulations in section \ref{numresults}. In section \ref{approx} we discuss the analytical calculations for $F_{crit}(f)$ and $f_{crit}(\lambda)$ that take into account network topology and explain most of the simulations. We summarize our conclusions in section \ref{conclusions}.

\section{The Forced Kuramoto Model on Networks}
\label{model}

In order to study the forced Kuramoto model on networks we need to consider two modifications on the system of equations (\ref{forcedd2}): first, to include the possibility that each oscillator interacts only with a subset of the other oscillators, the system will be placed on a network whose topology defines the interactions \cite{Arenas2008} and second we allow the external force to act only on a subset of the oscillators, representing the ``interface'' of the system that interacts with the ``outside'' world, like the photo-receptor cells in the eye \cite{gray1994}. 

The system is described by the equations
\begin{equation}
\dot{\phi_i} = \omega_i -\sigma - F \, \delta_{i,C} \sin \phi_i + \frac{\lambda}{k_i} \sum_{j=1}^{N} A_{ij} \sin (\phi_j - \phi_i),
\label{forced2}
\end{equation}
where $A_{ij}$ is the adjacency matrix defined by $A_{ij} = 1$ if oscillators $i$ and $j$ interact and zero if they do not; $k_i$ is the degree of node $i$, namely $k_i = \sum_j A_{ij}$; $F$ and $\sigma$ are respectively the amplitude and frequency of the external force; and $C$ is the subgroup of oscillators subjected to the external force. We have also defined $\delta_{i,C} = 1$ if $i \in C$ and zero otherwise and we shall call $N_C$ the number of nodes in the set $C$. In the next chapter we will consider cases where the network is weighted, i.e., where $A_{ij}$ can assume real values associated with the intensity of the coupling.

The behavior of the system depends now not only on the distribution of natural frequencies and coupling intensity $\lambda$, but also on the network properties, on the intensity and frequency of the external force and on the size and properties of the set $C$. The role of network characteristics in the absence of external forcing has been extensively studied in terms of clustering \cite{Mcgraw2005,Mcgraw2007,Mcgraw2008}, assortativity \cite{Restrepo2014} and modularity \cite{Oh2005,Arenas2006,Arenas2007}. 

The behavior of the system under an external force has also been considered for very large and fully connected networks when the force acts on all nodes equally, as we have seen on the last chapter \cite{Childs2008}. The system exhibits a rich behavior as a function of the intensity and frequency of the external force. In particular, it has been shown that if the force intensity is larger than a critical value $F_{crit}$ the system may fully synchronize with the external frequency. Among the questions we want to answer here are how synchronization with the external force changes as we make $N_C < N$ and how does that depend on the topology of the network and on the properties of the nodes in $C$. In particular we are interested
in studying how the critical intensity $F_{crit}$ of the external force increases as $N_C$ decreases
and if there is a minimum number of nodes that need to be excited by $F$ in order to trigger synchronization. In the next section we show the results of numerical simulations considering three
network topologies (random, scale-free and fully connected). Analytical calculations that describe
these results will be presented next.

\section{Numerical Results}
\label{numresults}

In order to get insight into the general behavior of the system we present a set of simulations for the following networks: (i) fully connected with $N=200$ nodes (FC200), (ii) fully connected with $N=500$ (FC500); (iii) random Erdos-Renyi network with $N=200$ and average degree $\left\langle k  \right\rangle = $  10.51 (ER200) and (iv) scale-free Barabasi-Albert network with $N=200$ (BA200) computed starting with $m_0 = 11$ fully connected nodes and adding nodes with $m = 10$ links with  preferential attachment, so that $\langle k \rangle =$  9.83. In all simulations we have considered a Gaussian distribution (equation (\ref{gaussdist})) of natural frequencies $g(\omega)$ with mean $\omega_0 = 0$ and standard deviation $a=1.0$ for the oscillators. In chapter \ref{chapter3}, $\Omega = \sigma - \omega_0$, thus $\Omega = \sigma$. In what follows, we will use $\sigma$ for the driving frequency.

For the fully connected networks the critical value $\lambda_c$ for the onset of synchronization can be estimated when $N \rightarrow \infty$ as $\lambda_c = 2a \sqrt{2/ \pi} \approx 1.6$ (see equation (\ref{gaussdist})). For finite networks the calculation of $\lambda_c$ can be performed numerically (see, for example, \cite{wang2015}) and we have checked that $\lambda_c=1.6$ is  a good approximation even for $N=100$ and for the other topologies we used. Full synchronization occurs only for larger values of $\lambda$ and we define $\lambda_f$ as the value where $r=0.95$ and $\dot{\psi}< 10^{-2}$. Here we are interested in scenarios where the system synchronizes spontaneously when $F=0$ and, therefore, we set $\lambda$ above $\lambda_f$ to assure full spontaneous synchronization. The coupling strength $\lambda$ has an important role in the synchronization process, as we discuss below. For each network type and fraction $f=N_C/N$ of nodes interacting with the external force we calculate the minimum (critical) force necessary for synchronization with the external frequency.

In order to characterize the dynamics we use the usual order parameter
\begin{equation}
z = r e^{i \psi} = \frac{1}{N}\sum_{i=1}^N e^{i \phi_i},
\end{equation}
where $r=1$ indicates full synchronization and $\dot{\psi}$ the frequency of the collective motion. We note that, since we are working on a rotating frame, synchronization with $\sigma$ will imply $\dot{\psi} = 0$ whereas spontaneous synchronization $\dot{\psi} = - \sigma$.

\begin{figure}[H]
\center
\includegraphics[scale=0.98]{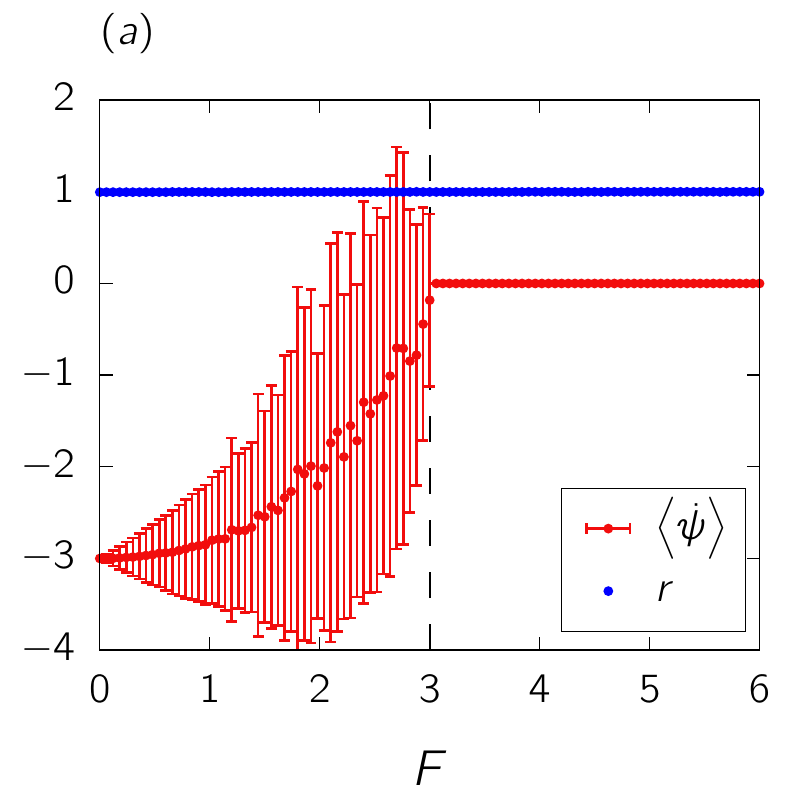}
\includegraphics[scale=0.98]{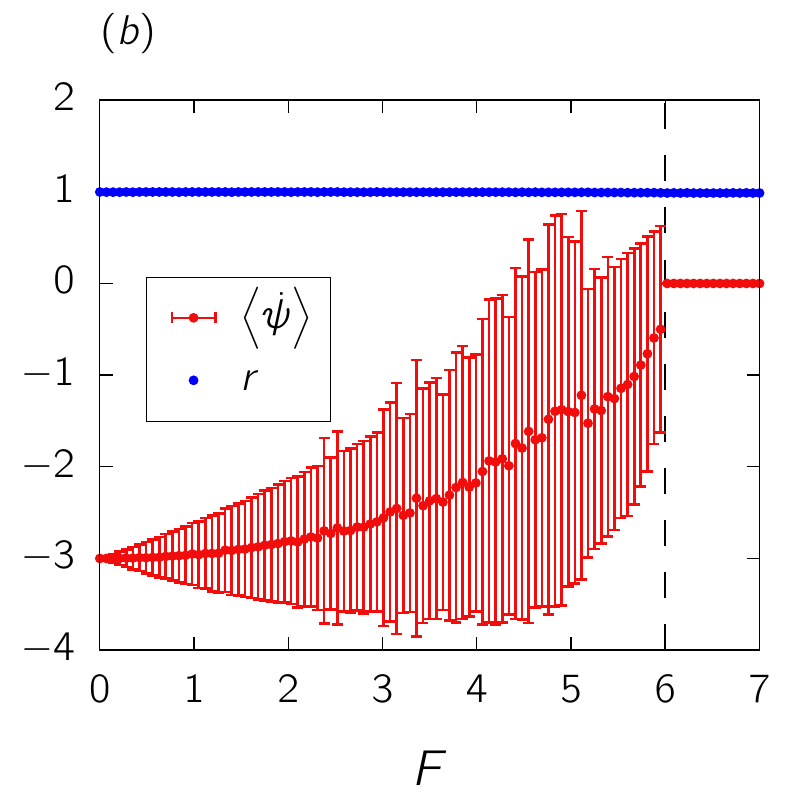}
\caption{Order parameter $r$ (blue) and the average of $\dot{\psi}$, $\langle \dot{\psi} \rangle$ (red), as a function of $F$ for a fully connected network with $N$ = 200, $\lambda$ = 20.0 and $\sigma$ = 3.0 for (a) $f = 1$ and (b) $f=0.5$. Red dots correspond to time averaged values calculated between $t=25$ to $t=50$. Error bars correspond to one standard deviation. The dashed lines indicate the critical force.} 
\label{errorbars}
\end{figure}

Fig. \ref{errorbars} shows $r$ and the average of $\dot{\psi}$, $\langle \dot{\psi} \rangle$, for FC200 as a function of $F$ for $\lambda = 20$ and $f=1$ and $f=0.5$. The system has been evolved up to $t=50$ starting with random phases, which was enough to overcome the transient period (see Fig.\ref{fig6}). Because the system is finite and there are fluctuations we computed time averages and standard deviations of $r$ and $\psi$ in the interval from time $25$ to $50$. The system remained fully synchronized for all values of $F$, first spontaneously ($F=0$) and later with the external frequency for $F > 3$ ($f=1$) and for $F>6$ ($f=0.5$). For intermediate values of the external force, $\dot{\psi}$ oscillates and the average and standard deviations are shown. In this regime the oscillators move together ($r=1$) but change directions constantly due to the competition between the couplings $\lambda$ and $F$. The critical force $F_{crit}$ was numerically computed as the value of $F$ where $\dot{\psi} < 10^{-2}$ and $r > $ 0.95.

\begin{figure}[H]
\center
\includegraphics[scale=0.98]{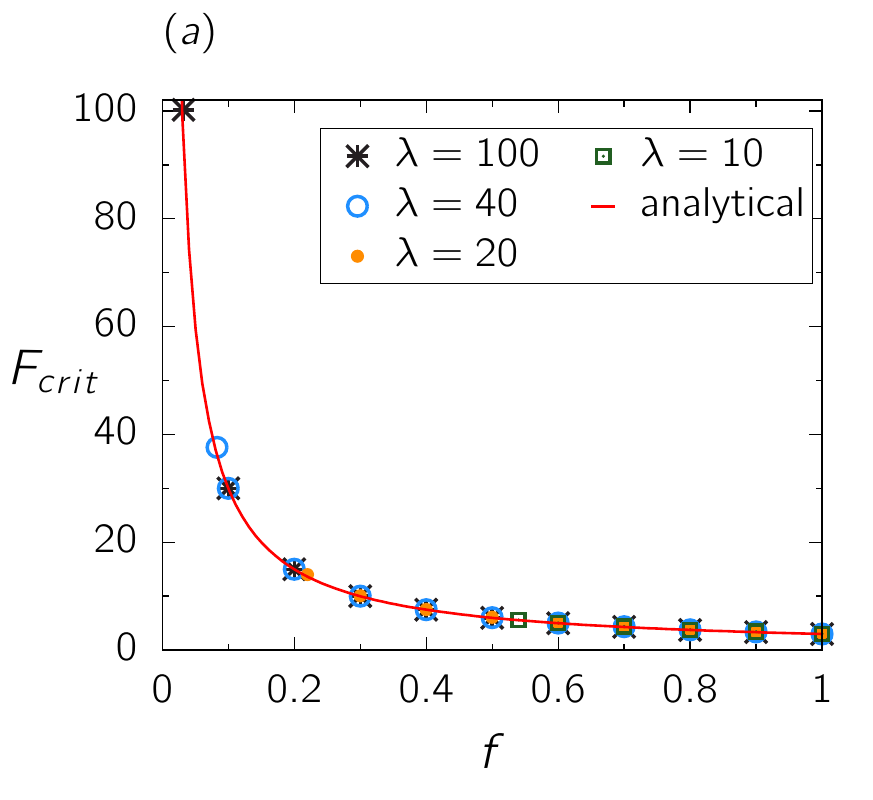}
\includegraphics[scale=0.98]{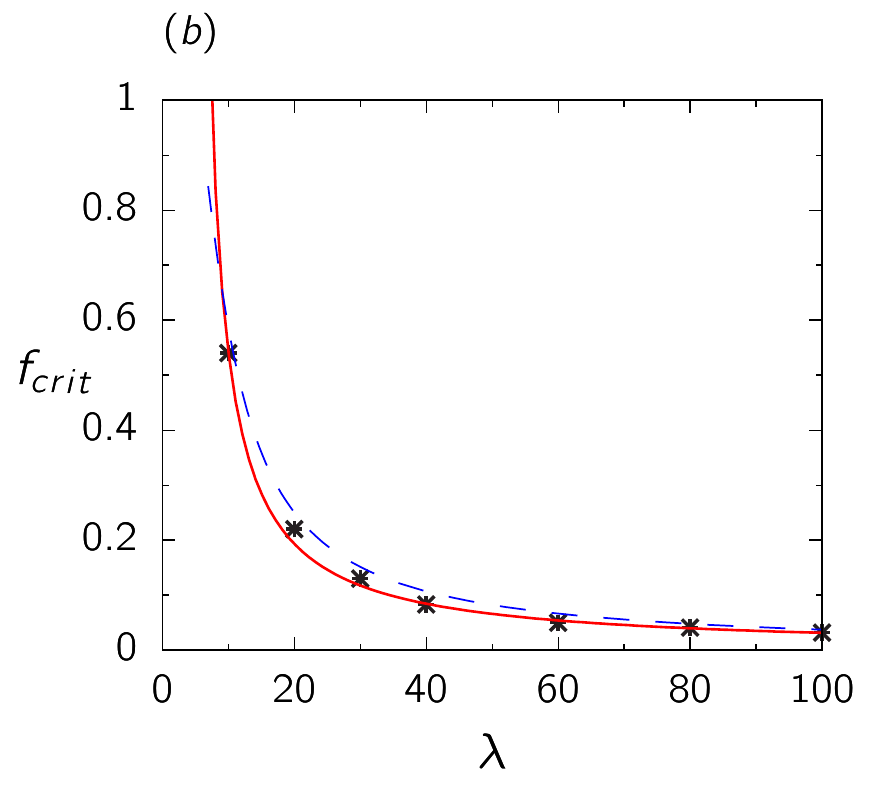}
\caption{(a) Critical force $F_{crit}$ versus fraction $f$ of forced nodes for the fully connected network FC200. The continuous red curve shows the analytical calculation and the symbols are the results of numerical simulations for different values of the coupling constant $\lambda$. The point with smallest $f$ for each $\lambda$ is defined as $f_{crit}$. (b)  $f_{crit}(\lambda)$ from numerical simulations (stars) and according to  Eq. (\ref{fcrit0}) (red curve). The dashed (blue) line was obtained from the parametric curve of Eq. (\ref{parametriclambda}).} 
\label{fig2}
\end{figure}

Fig. \ref{fig2}(a) shows $F_{crit}$ as a function of the fraction $f$ of excited nodes for FC200. It also shows that for a fixed value of the internal coupling $\lambda$ synchronization can only be achieved for $f$ larger than a critical value $ f_{crit}(\lambda)$. For example, for $\lambda=20$ (orange circles) synchronization is obtained only for $f > 0.22$. For $f < 0.22$ no synchronization is achieved for $\lambda=20$, no matter how large is the external force. The value of $f_{crit}$ is shown as the last point of the corresponding symbol on the plot.  Notice that the minimum value of $F$ for synchronization does not itself depend on $\lambda$, since the same value is obtained as long as $\lambda$ is large enough. Fig. \ref{fig2}(b) shows $f_{crit}$ as a function of $\lambda$. We have performed the same analysis for FC500 and both curves $F_{crit}(f)$ and $f_{crit}(\lambda)$ were essentially identical to the ones obtained for FC200, showing that these are independent of network size.

\begin{figure}[H]
\center
\includegraphics[scale=0.98]{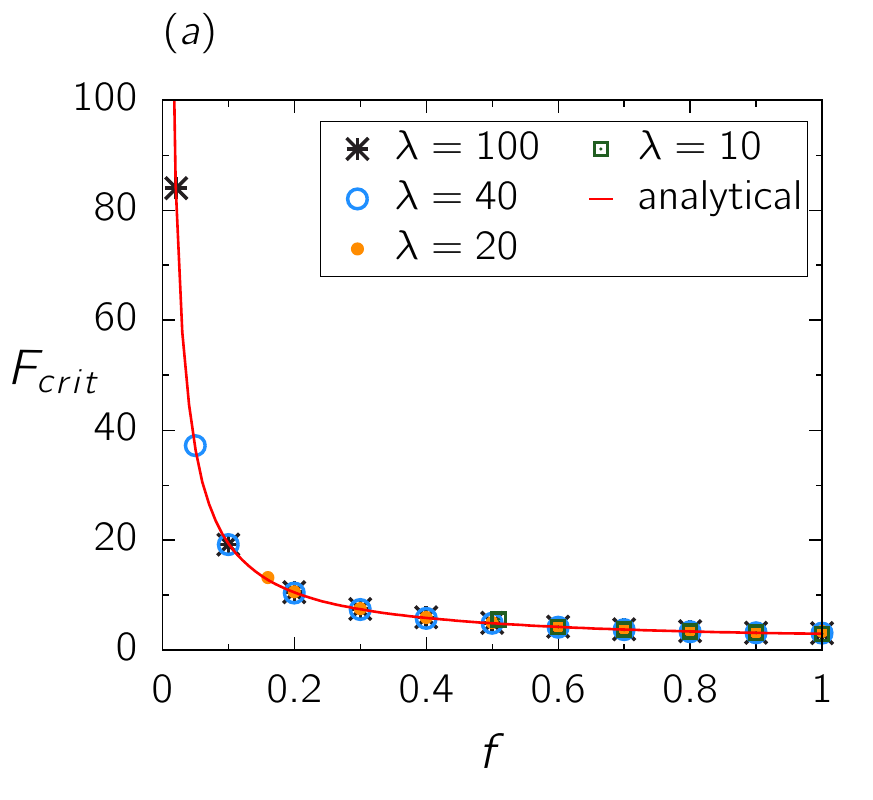}
\includegraphics[scale=0.98]{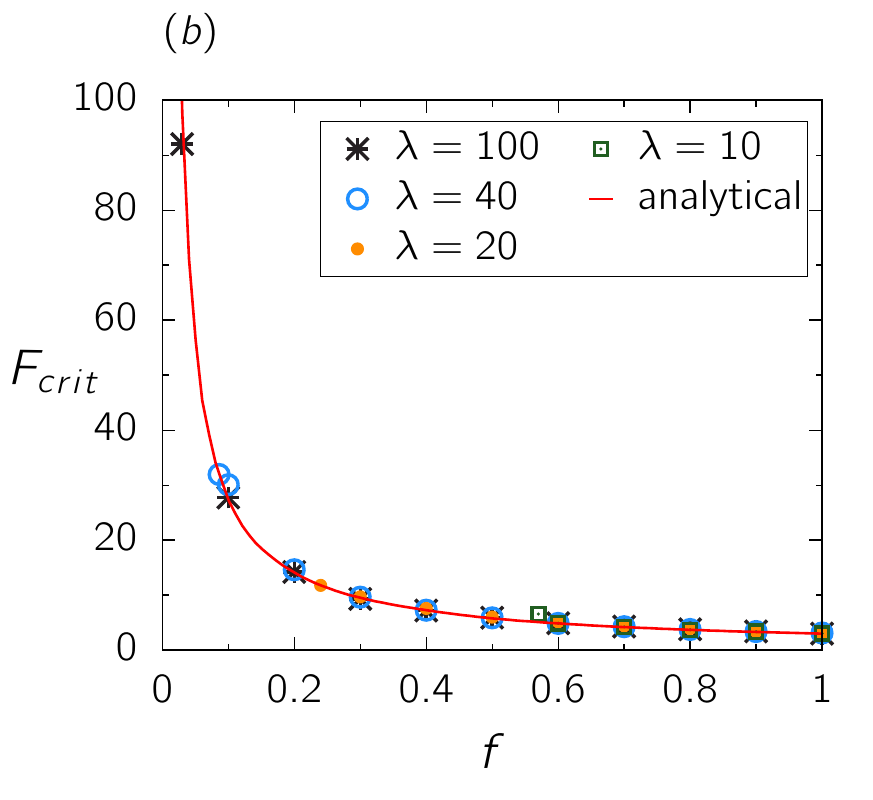}
\caption{Critical force $F_{crit}$ versus fraction $f$ of forced nodes for the random network ER200. The continuous red curve shows the analytical calculation and the symbols are the results of numerical simulations for different values of the coupling constant $\lambda$. The point with smallest $f$ for each $\lambda$ is defined as $f_{crit}$. Force is connected with nodes of (a) highest degrees; (b) random. For the red line on panel (b) we have computed the average degree $\langle k \rangle_C$ of forced set  over 10 simulations to eliminate fluctuations.}
\label{fig3}
\end{figure}

Fig. \ref{fig3} shows similar results for the ER200 random network. In this case the nodes have different degrees and it matters which nodes are selected to interact with the external force. For the results in panel (a) the nodes have been ordered from high to low degree and the $fN$ first (highly connected) nodes have been selected to interact with the force. In panel (b) the nodes were chosen at random.  The dependence of $f_{crit}$ on $\lambda$ is similar to the fully connected case and different values of $\lambda$ are shown with different symbols.

\begin{figure}[H]
\center
\includegraphics[scale=0.98]{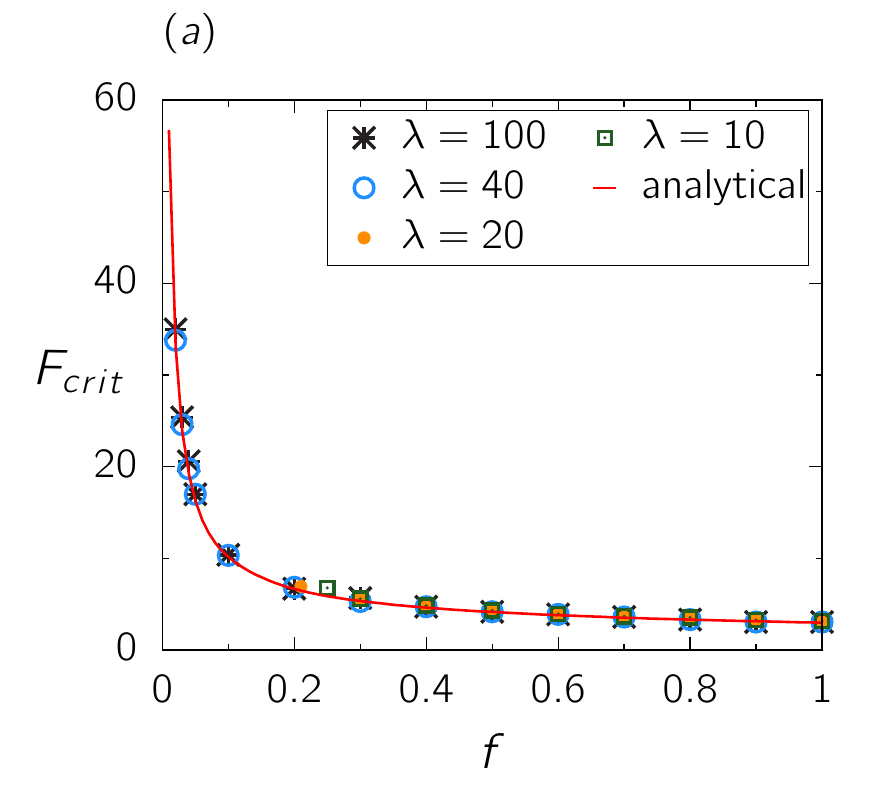}
\includegraphics[scale=0.98]{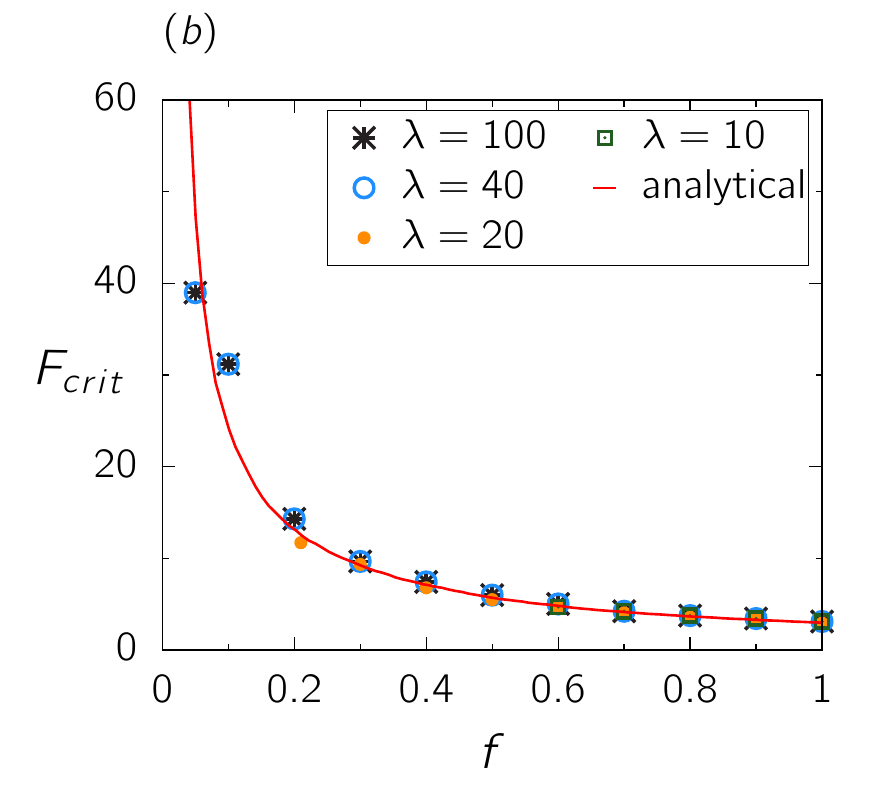}
\includegraphics[scale=0.98]{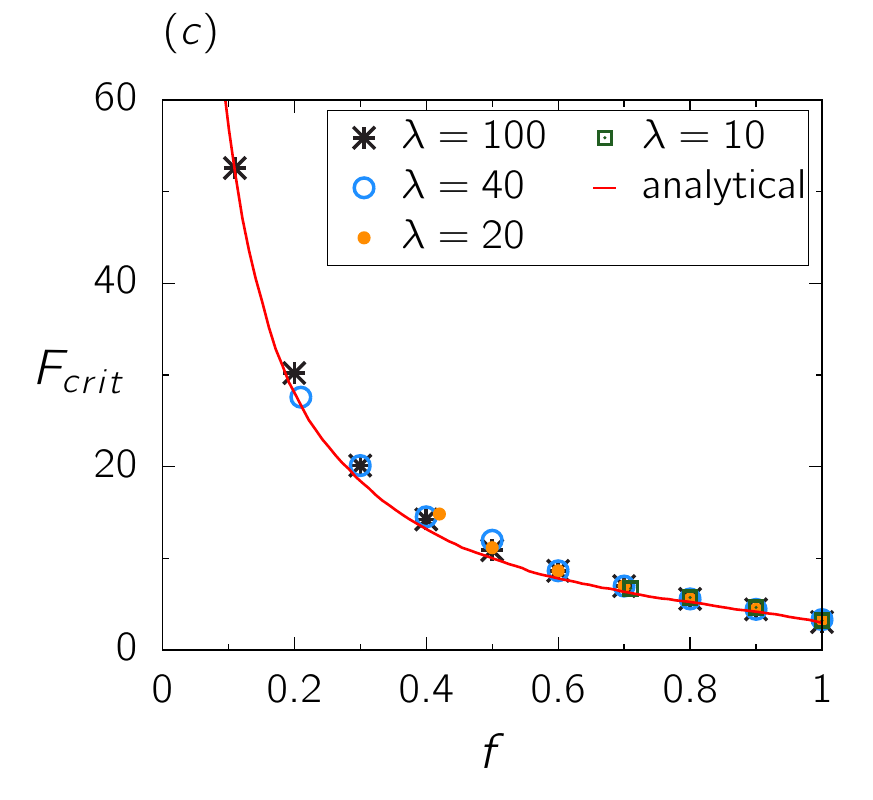}
\caption{Critical force $F_{crit}$ versus fraction $f$ of forced nodes for the scale free network BA200. The continuous red curve shows the analytical calculation and the symbols are the results of numerical simulations for different values of the coupling constant $\lambda$. The point with smallest $f$ for each $\lambda$ is defined as $f_{crit}$. Force is connected with nodes of (a) highest degrees; (b) random and (c) lowest degree. For the red line on panel (b) we have computed the average degree $\langle k \rangle_C$ of forced set over 10 simulations to eliminate fluctuations.}
\label{fig4}
\end{figure}

For the random network the differences between the two cases are not striking, since the distribution of nodes is quite homogeneous. This is not the case for the BA200 network, as shown in Fig. \ref{fig4}. When the  external source connects with nodes of highest degree, panel (a), the critical force for synchronization is smaller  than when connected randomly, panel (b), or with nodes of lowest degrees, panel (c), as expected. The analytical (red) curve for random connections shows an average over 10 simulations using the same network but different random choices of nodes.

\section{Analytical results}
\label{approx}

The numerical simulations show that: (i) $F_{crit}$ depends of $f$; (ii) for heterogeneous networks
it depends on the properties of the set $C$; (iii) there is a critical fraction $f_{crit}$, that depends on the network type, on $C$ and on $\lambda$, below which no synchronization is possible. In this section we derive a theory for $F_{crit}(f)$ and an approximation for $f_{crit}(\lambda)$.

\subsection{Critical Force}

In order to derive an expression for $F_{crit}(f)$ we use the fact that nodes directly affect all their neighbors and, therefore, their importance should be proportional to their degree. Defining $\bar{F} = fF$, we start by multiplying all terms of Eq.(\ref{forced2}) by $k_i/\langle k \rangle$, sum over $i$ and divide by $N$ to obtain 
\begin{equation}
\frac{d \langle \phi \rangle}{dt} = \langle \omega \rangle -\sigma  - \bar{F}  \langle \sin \phi \rangle_C
\label{ap3}
\end{equation}
where
\begin{equation}
\langle \phi \rangle = \frac{1}{N}  \sum_{i=1}^N  \frac{k_i}{\langle k \rangle} \phi_i ,
\end{equation}
\begin{equation}
\langle \omega \rangle = \frac{1}{N}  \sum_{i=1}^N  \frac{k_i}{\langle k \rangle} \omega_i 
\end{equation}
and
\begin{equation}
\langle \sin \phi \rangle_C = \frac{1}{N_c}  \sum_{i \in N_c}  \frac{k_i}{\langle k \rangle} \sin \phi_i .
\label{sinforc}
\end{equation}
The term proportional to $\lambda$, containing the coupling between the oscillators, cancel out exactly. When the oscillators synchronize with the external force Eq.(\ref{sinforc}) becomes
\begin{equation}
\langle \sin \phi \rangle_C = \sin \langle \phi \rangle \frac{\langle k \rangle_C}{\langle k \rangle},
\end{equation}
where $\langle k \rangle_C = 1/N_c \sum_{i \in N_c} k_i$ is the average degree of the set C. 

Since $\langle \phi \rangle$ is constant in the synchronized state Eq.(\ref{ap3}) implies
\begin{equation}
\sin \langle \phi \rangle = \frac{\langle \omega \rangle -\sigma}{ \bar{F_k}} 
\label{apc}
\end{equation}
where we have defined
\begin{equation}
\bar{F_k} = f \frac{\langle k \rangle_C}{\langle k \rangle} F.
\label{app1b}
\end{equation}
Because the $\omega_i$ are randomly distributed with zero average, $\langle \omega \rangle$ is generally small for large networks (although not zero in a single realization of the frequency distribution). Since $| \sin \langle \phi \rangle| \leq 1$, Eq.(\ref{apc}) holds only if $\bar{F_k} \geq |\sigma - \langle \omega \rangle|$ so that the critical force can be estimated as $\bar{F}_c =\sigma -\langle \omega \rangle$, or 
\begin{equation}
F_{crit} = \frac{\sigma - \langle \omega \rangle}{f} \, \frac{\langle k \rangle}{\langle k \rangle_C} 
        \approx \frac{\sigma}{f} \, \frac{\langle k \rangle}{\langle k \rangle_C} .
\label{force_crit}
\end{equation}
For regular networks, in particular,  where all nodes have the same degree, $\langle k \rangle = \langle k \rangle_C$, the critical force is reduced to
\begin{equation}
F_{crit} = \frac{\sigma}{f}.
\label{apd}
\end{equation}

Eq. (\ref{force_crit}) shows that when nodes with high degree are being forced, $\left\langle k \right\rangle_C > \left\langle k \right\rangle$,  the critical force for synchronization is smaller than the value obtained by equation (\ref{apd}), since the external force is directly transmitted to a large number of neighbors.  On the other hand, if $\left\langle k \right\rangle_C < \left\langle k \right\rangle$ (nodes with low degree are being forced) the critical force must be higher than that estimated by (\ref{apd}), since these nodes have few neighbors. This agrees with the results shown in Figs. \ref{fig2}-\ref{fig4}  where the continuous (red) line shows the approximation Eq (\ref{force_crit}). For the scalefree network, in particular, when the force acts on nodes of highest degree, Fig. \ref{fig4}(a), $F_{crit} \approx 5$ for $f = 0.4$, whereas $F_{crit} \approx 15$ for the same value of $f$ when the force acts on the nodes with smallest degree Fig. \ref{fig4}(c).

\subsection{The critical fraction}

Eq.(\ref{ap3}) is exact and it might appear to be completely independent of $\lambda$. This, however, is not true, since the dynamics of the angles $\phi$ are implicitly coupled by $\lambda$ and synchronization is only possible if $\lambda$ is large enough. As $f$ decreases the amplitude of the external force needed for synchronization increases and if it gets too much larger than $\lambda$ the oscillators start to move almost independently and synchronization is hindered.

An approximation for minimum value of $f$ that can lead to synchronization for a given $\lambda$ can be obtained by setting the internal coupling strength per node to the intensity of the external force, i.e., $\lambda \simeq F$. Along the curve $F=F_{crit}$ this becomes  $\lambda \simeq \sigma \langle k \rangle/ (f \langle k \rangle_C)$ (see Eq.(\ref{force_crit})). However, since complete spontaneous synchronization only happens for $\lambda$ sufficiently large (of the order of $\lambda_f$) we propose that $f_{crit}$ can be estimated  from the relation $\lambda - \lambda_0 = F_{crit}$, or
\begin{equation}
f_{crit} (\lambda) = \frac{\sigma}{\lambda-\lambda_0} \; \frac{\langle k \rangle}{\langle k \rangle_C},
\label{fcrit0}
\end{equation}
where $\lambda_0$ is a fit parameter, whose value has to be at least $\lambda_c$. For fully connected networks $\langle k \rangle = \langle k \rangle_C$ and Eq. (\ref{fcrit0}) reduces to $f_{crit} (\lambda) = \sigma/(\lambda-\lambda_0)$. For the red curve in Fig.\ref{fig2}(b) we obtained $\lambda_0=4.48 \pm 0.12$ which fits very well the numerical results (black stars). Note that the value of $\lambda$ for $f=1$ is $\lambda_0+\sigma = 7.48$ for which we find $r=0.99$ for $F=0$ although $\dot{\psi}$ is still fluctuating. Full spontaneous synchronization ($r>0.95$ and $\dot{\psi} < 10^{-2}$) only occurs for $\lambda_f=11.3$.

The heuristic approximation given by Eq.(\ref{fcrit0}) can be made more precise using the bifurcation surfaces derived by Childs and Strogatz \cite{Childs2008} for the case where the external force acts on all nodes. The derivation assumed a Lorentzian distribution for the oscillator's natural frequencies, but is believed to be valid for a larger class of such distributions. The full bifurcation diagram is divided into five regions but is dominated by only two: one where the oscillators are locked to the same frequency as the external force and one with mutual, spontaneous, synchronization. These two main regions are separated by saddle-node bifurcations given in the $F$ versus $\sigma$ plane, for $\lambda$ fixed, by the parametric equations
\begin{equation}
\sigma(\lambda,r) = \frac{(1+r^2)^{3/2}}{2(1-r^2)^2} \sqrt{\lambda (r^2-1)[\lambda (r^2-1)^2-4] - 4(r^2+1)},
\end{equation}

\begin{equation}
F(\lambda,r) = \frac{\sqrt{2} r^2}{(1-r^2)^2} \sqrt{\lambda^2 (1-r^2)^3 + 2\lambda(r^4-4r^2+3) - 8},
\end{equation}
where $r$ varies from approximately $0.66$ to $1.0$. These expressions were derived on the last chapter; see equations (\ref{eq24}) and (\ref{eq25}), respectively. The resulting curve $F=F(\sigma)$ can be
approximated by the simple relation $F = \sigma$, as predicted by eq.({\ref{apd}}). This
approximation becomes exact as $\lambda$ goes to infinity, or when $r=1$ and $\dot{\psi}=0$.

Solving these equations for $F$ and $\lambda$ we obtain
\begin{equation}
\lambda(\sigma,r) = \frac{2}{(r^2-1)^2} + 2 \sqrt{\frac{r^4}{(r^2-1)^4}+\frac{\sigma^2(r^2-1)}{(r^2+1)^3}}
\label{parametriclambda}
\end{equation}
and $F(\sigma,r) = F(\lambda(\sigma,r) ,r)$. This new set of parametric equations results in the critical curve $F=F(\lambda)$, for fixed $\sigma$. Finally, using eq.(\ref{apd}) $F=\sigma/f$ we can compute $f=f(\lambda)$ with the parametric functions $(\lambda(\sigma,r),\sigma/F(\sigma,r))$). This curve is shown as dashed (blue) line in Fig. \ref{fig2}(b) and differs from the heuristic approximation only for small values of $\lambda$.

\subsection{Transition from forced to mixed dynamics}

Synchronization with the external force is possible only if $F > F_{crit}$, estimated by Eq. (\ref{force_crit}). If $F < F_{crit}$ the system's behavior is determined by the competition between spontaneous and forced motion. The transition between these two regimes was studied in detail in ref. \cite{Childs2008} for the case of infinitely many oscillators, all of which coupled to the external drive. Here we present a simplified description of the transition using the analytical approach developed above. 

Making the approximations $\langle \omega \rangle_k = 0$ and $\langle \sin \phi \rangle_{k,C} = \sin \langle \phi \rangle$, Eq. (\ref{ap3}) simplifies to the Adler equation \cite{Adler1973}
\begin{equation}
\frac{d \phi }{dt} = -\sigma  - \bar{F}  \sin \phi 
\label{ap6}
\end{equation}
where we are omitting the average symbol and considering regular networks to simplify the notation. For general networks we only need to make $\bar{F} \rightarrow \bar{F}_k$. This equation, which has been used to model fireflies \cite{Ermentrout1984} among other systems \cite{Childs2008}, can be solved exactly to give
\begin{equation}
\sigma \tan{\phi/2} = \bar{F}  + \sqrt{\bar{F}^2 - \sigma^2} \tanh{\left[\frac{1}{2}\sqrt{\bar{F}^2 - \sigma^2}(t-t_0)\right]}
\end{equation}
for $\bar{F} > \sigma$. In this case $\phi$ converges to a constant value and the system stops (synchronizes with $F$). For $\bar{F} < \sigma$, on the other hand, the solution is oscillatory, 
\begin{equation}
\sigma \tan{\phi/2} = \bar{F}  - \sqrt{\sigma^2-\bar{F}^2} \tan{\left[\frac{1}{2}\sqrt{\sigma^2 - \bar{F}^2}(t-t_0)\right]}
\end{equation}
with period \cite{Jensen2002}
\begin{equation}
\tau = \frac{2\pi}{\sqrt{\sigma^2 - \bar{F}^2}}.
\label{period}
\end{equation}
\begin{figure}[H]
\center
\includegraphics[scale=0.91]{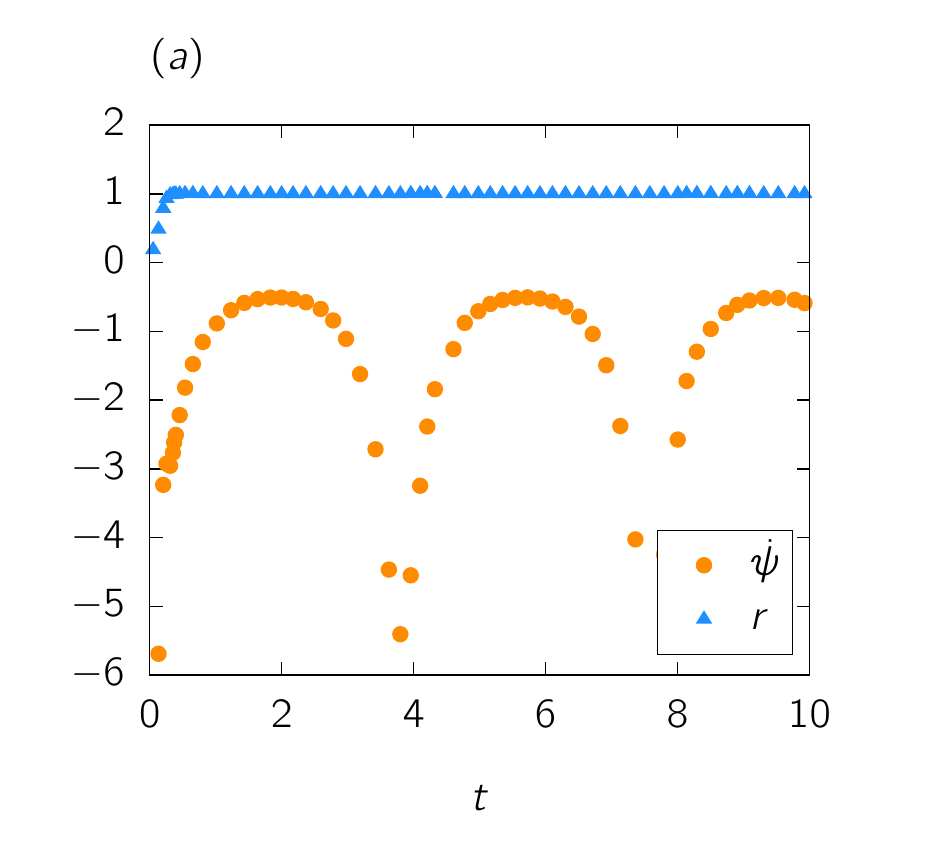}
\includegraphics[scale=0.91]{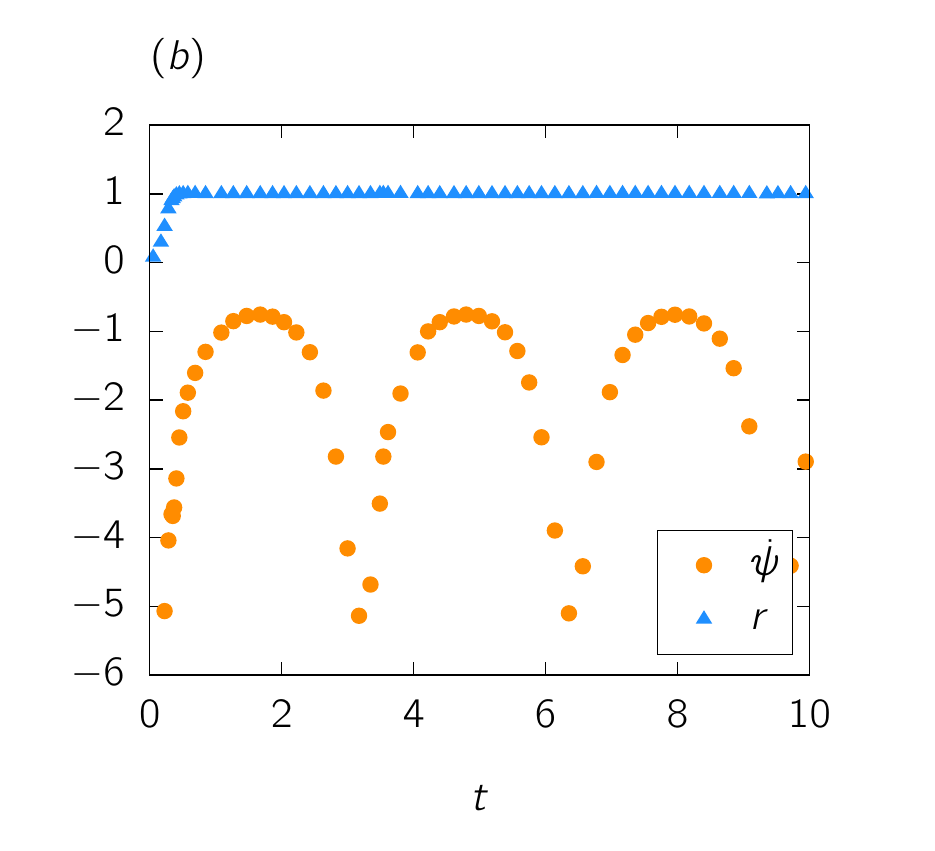}
\includegraphics[scale=0.91]{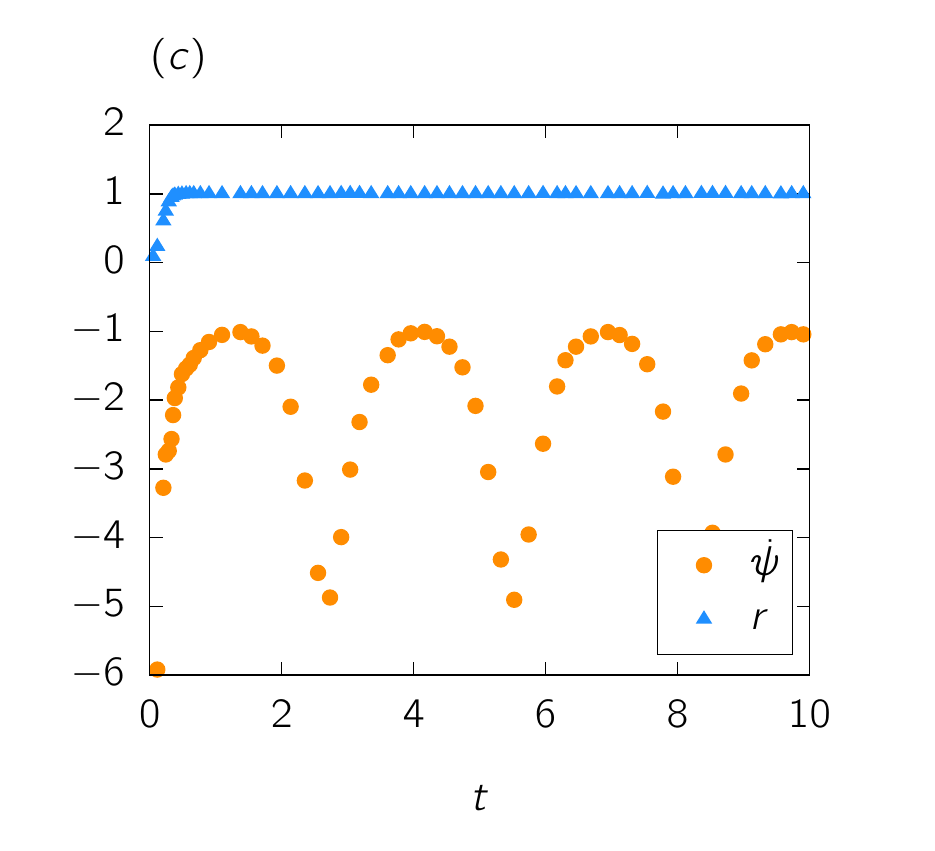}
\includegraphics[scale=0.91]{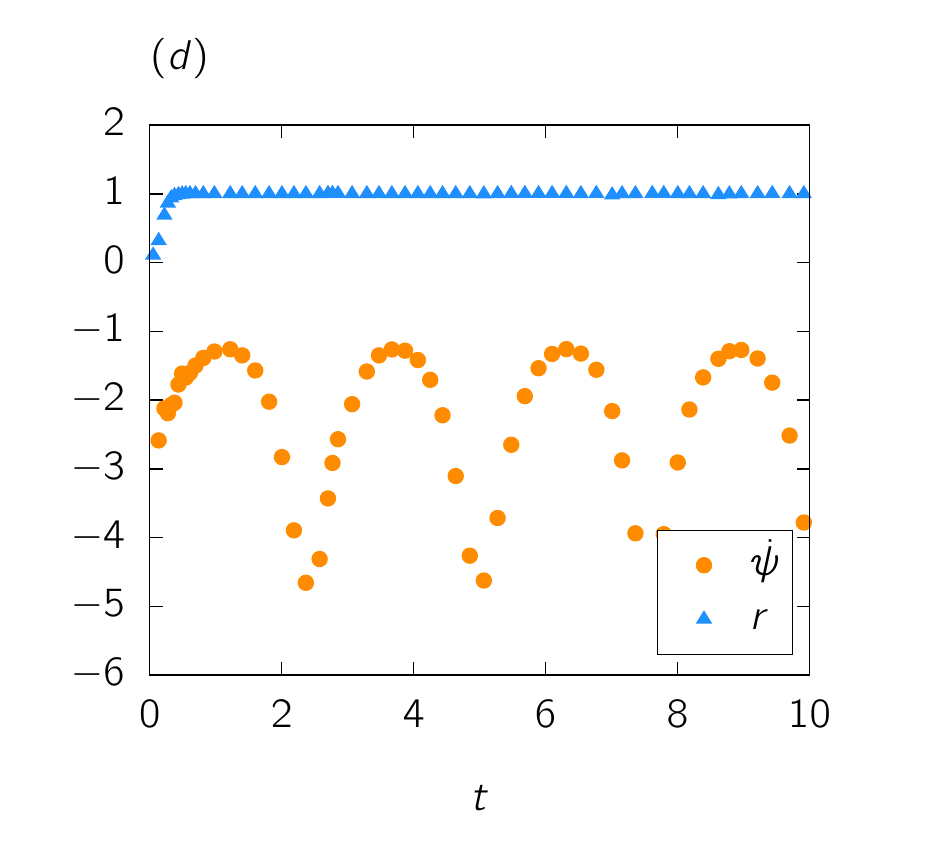}
\caption{Frequency of oscillations for the fully connected network with 200 nodes for $F=2.5$ fixed and fraction (a) $f=100 \%$; (b) $f=90 \%$; (c) $f=80 \%$ and (d) $f=70 \%$. The points show $r$ (blue triangles) and $\dot{\psi}$ (orange circles). The periods estimated from Eq. (\ref{period}) are (a) $\tau = 3.8$; (b) $\tau = 3.4$; (c) $\tau = 3.1$ and (d) $\tau = 2.9$.}
\label{fig5}
\end{figure}

Figure \ref{fig5} illustrates the frequency of oscillations for $F < F_{crit} = 3$ fixed and different number of nodes that receive the external drive, showing $r$ and $\dot{\psi}$ as a function of $t$, for a fully connected network. Although $r$ approaches 1 quickly (i.e., the system does synchronize), $\dot{\psi}$ oscillates with growing periods as the number of nodes on $C$ increases, remaining always negative. This means that $\psi$ decreases monotonically and the order parameter $z(t)$ oscillates, implying that a finite fraction of the oscillators has synchronized spontaneously, due to their mutual interactions and not to the drive. The approximation (\ref{period}) for the periods of oscillation matches very well the results of the simulations.

\subsection{Time to equilibrium}

The time scale of dynamical processes also changes with the fraction of forced nodes. The time to equilibrium should increase when $f$ decreases, but no simple relation seems to exist. When $F$ is large,  we can approximate Eq.(\ref{ap3}) by 
\begin{equation}
\frac{d \langle \phi \rangle_k}{dt} = - F f \frac{\langle k \rangle_C}{ \langle k \rangle}  \sin \langle \phi \rangle_k.
\label{forced4}
\end{equation}
\begin{figure}[H]
\center 
\includegraphics[scale=0.55]{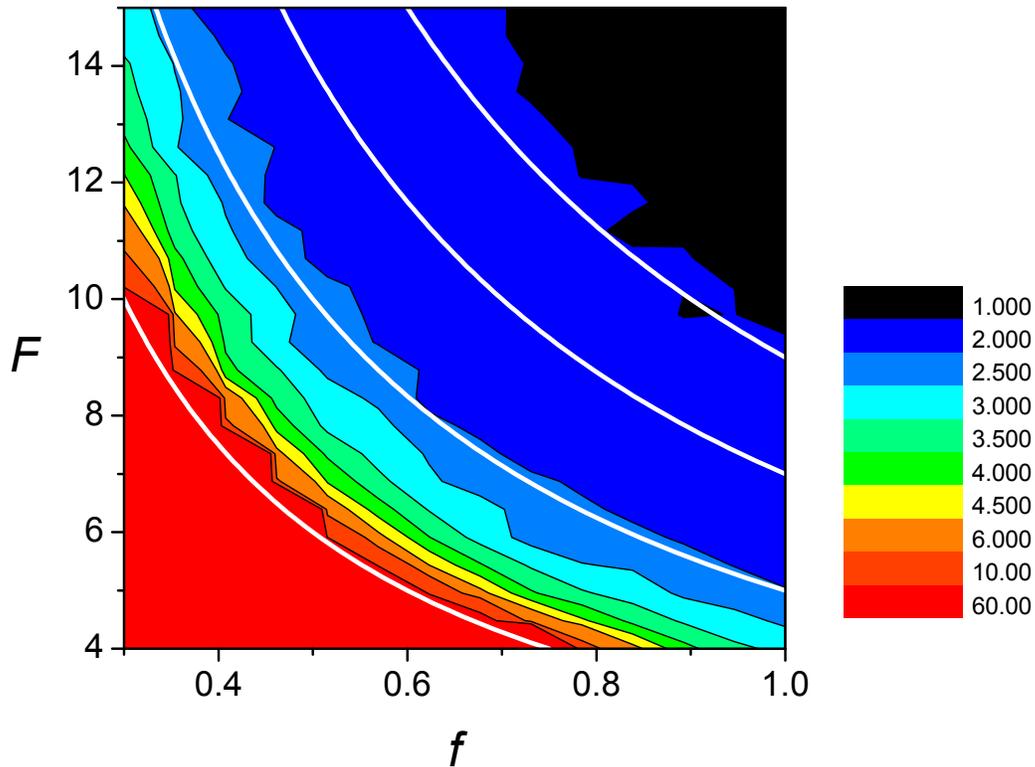}
\caption{Contour plot of time to equilibration for different values of $F$ and $f$ and fixed $\lambda = 40$. Thick lines correspond to constant times according to the approximation $F=F_0/f$ for $F_0=3$, 5, 7 and 9.}
\label{fig6}
\end{figure}

Defining $t' = t F f \langle k \rangle_C/ \langle k \rangle$ this equation becomes identical to that of a system where the force acts on all nodes. Therefore, within this crude approximation we expect that: (i) for fixed $F$, the time to equilibration should scale as $\tau(f) = \tau_0 \, \langle k \rangle/ [f \langle k \rangle_C]$, where $\tau_0$ is the equilibration time at $f=1$ and; (ii) along the curve $F_{crit} (f) = F \, { \langle k \rangle} / [ f \langle k \rangle_C]$ the time to equilibration remains constant, since the factors multiplying $F$ in Eq.(\ref{forced4}) cancel out. Fig.(\ref{fig6}) shows contour levels of numerically computed equilibration times in the $F \times f$ plane. Thick white lines shows predicted curves of constant times, which indeed provide a somewhat poor approximation to the computed values. 

\section{Conclusions}
\label{conclusions}

 In this chapter we considered the problem of periodically forced oscillators where the external drive acts only on a fraction of them \cite{Childs2008}.  When the periodic drive acts on all oscillators, the system always synchronize with the forced period if the force intensity is sufficiently large \cite{Childs2008}. Using numerical simulations and analytical calculations we have shown that the force required to synchronize the entire set of oscillators increases roughly as the inverse of the fraction of forced nodes. The degree distribution of the complete network of interactions and of the set of forced nodes also affect the critical force for synchronization. Forcing oscillators with large number of links facilitates global synchronization in proportion to the average degree of the forced set to the total network.

We have also shown that below a critical fraction, that depends on $\lambda$, no synchronization occurs, no matter how large the force. We believe this is an interesting result of this study that might have consequences for adaptive systems relying on synchronization. The set of $N_C=fN$ nodes that directly receives the external drive can be interpreted as the interface of a system where the remaining $(1-f)N$ nodes are the ``processing unit'', that needs to synchronize with the external signal $F$ to perform a function. In this case it would be desirable to have $N_C$ as small as possible to increase the processing power. However, synchronization with small $f$ requires large couplings between the units, which can be costly. An example is neural network of {\it C. elegans} where multiple links can connect the same two nodes and the cost of a connection is proportional to the number of such links (synaptic connections) that make it \cite{Latora2003}. In these cases it is expected that a balance between interface size and network cost is attained, and natural systems should evolve toward this condition. The final balance will, of course, depend on the cost. If the cost is zero the system should evolve to the minimum possible interface size, given by $f=\sigma/F$ and $\lambda = \lambda_c + \sigma/f = \lambda_0+ F$ (for a fully connected network). If there is a cost it might be advantageous to work with a smaller processing unit (and larger interface) that requires smaller values of $\lambda$.

The theory developed here for $F_{crit}(f)$ considered only constant values of the coupling strength $\lambda$. In this case the term containing $\lambda$ in Eq.(\ref{forced2}) disappears from the averaged Eq.(\ref{ap3}). This equation, however, remains valid for arbitrary symmetric couplings $\lambda_{ij}=\lambda_{ji}$, as can be easily verified by inspection. For asymmetric couplings,  $\lambda_{ij} \neq \lambda_{ji}$, this is not true and an extra term has to be included in Eq. (\ref{ap3}). However, since this term is proportional to $\sin(\theta_j - \theta_i)$ it vanishes when the system synchronizes and Eq.(\ref{force_crit}) still holds, being, therefore, a very robust result.

As a final comment we note that here we have picked nodes for the set $C$ at random or based on their degree. Another interesting choice would be to pick them according to their natural frequencies $\omega_i$. For finite systems the oscillator with the largest frequency determines the spontaneous synchronization of the system \cite{wang2015} and forcing the fastest nodes might also result in interesting dynamics.

\chapter{Modular structure in \textit{C. elegans} neural network and its response to external localized stimuli}
\label{chapter5}

In this chapter we probe the community structure of the neural electrical junction network of the \textit{C. elegans} using the partially forced Kuramoto model of synchronization \cite{us}. We aim to understand how the network responds to external localized stimuli and which modules are more affected when a specific group of neurons, that can be a functional group or a physically arranged module, is stimulated. We use two different metrics to characterize the overall behavior of the network under a localized stimulus: the synchronization of neurons within and between modules, as measured by the usual Kuramoto order parameter, and the phase-velocity inter-neuron correlation. We want to investigate the behavior of the system as a function of parameters such as stimulus intensity and inter-neuron connection strength. In particular we are interested in cases leading to global induced synchronization and highly correlated behavior, where the network responds as a whole, or to uncorrelated states, where neurons do not react to each other. Our simulations are guided by the results of chapter 4 where we studied the partially forced Kuramoto model on synthetic networks, using the external force to simulate a localized stimulus.

\textit{C. elegans} is a nematode animal, unsegmented and with bilateral symmetry, exhibiting physiological similarity to mammals as regards the nerves and neurotransmitters morphologies. The \textit{C. elegans} is considered a model organism in studies of disorders related to human nervous system, such as epilepsy \cite{epi1,epi2} and Parkinson's disease \cite{parkinson1, parkinson2}. It was the first multicelular animal to have its whole nervous system mapped, containing only 302 neurons. Its neural network is available in open source data centers, such as the WormAtlas \cite{wormatlas} and the OpenWorm \cite{openworm}. Here we extracted all necessary data from WormAtlas.

The 248 neurons of the electrical junction network are anatomically classified as belonging to head, body or tail, and neuron types are divided into motoneurons, interneurons and sensory neurons. We have also performed a classification into 10 ganglia (A: anterior ganglion, B: dorsal ganglion, C: lateral ganglion, D: ventral ganglion, E: retrovesicular ganglion, F: posterolateral ganglion, G: ventral cord neuron group, H: pre-anal ganglion, J: dorsorectal ganglion, K: lumbar ganglion \cite{wormatlas}) which is a finer division of the anatomical classification into head, body and tail. 

We also decomposed the network into three modules based on topological properties and numbered by $M_1$, $M_2$ and $M_3$ from largest to smallest. This modularization procedure was made with the software Cytoscape using the app ModuLand \cite{moduland1,moduland2}. Each module contains neurons from the three anatomical parts, and consequently the 10 ganglia, and of the three types. We applied the stimulus to the largest module $M_1$, then to the ganglion C and finally to the sensory neurons and we observed the response of other neurons. We will show that no single partition of the brain into communities can account for its behavior under stimuli. All partitions analyzed here, topological, anatomical and functional, play a role in the response to external localized stimuli, revealing the complexity of the brain's wiring and function.

This chapter was published in \cite{Arruda2019}. We will follow its structure: in section \ref{mm} we describe the materials and methods, showing the partially forced Kuramoto model, the \textit{C. elegans} neural connectome and the order parameters used to measure the state of the network. The results of numerical calculations and its analysis are in section \ref{elegansresults}. Finally, we summarize our discussion in section \ref{elegansconclusion}. The supplementary material is in section \ref{sp}.

\section{Materials and Methods}
\label{mm}

\subsection{Partially forced Kuramoto model}
\label{pfk}

The Kuramoto  model of coupled oscillators \cite{Kuramoto1975} is a paradigm in the study of synchronization and has been explored in connection with biological systems, neural networks and the social sciences \cite{Rodrigues2016, Acebron2005}. Here we consider a modified version of the original Kuramoto model where each oscillator interacts only with a subset of the other oscillators, as specified by a network of connections \cite{Arenas2008}. Moreover, part of the oscillators also interacts with an external periodic force \cite{Sakaguchi1988,Ott2008,Childs2008,us}. The oscillators are described by their phase $\theta$ and system is governed by the equations

\begin{equation}
\dot{\phi_i} = \omega_i -\sigma - F \, \delta_{i,C} \sin \phi_i + 
\frac{1}{s_i} \sum_{j=1}^{N} \lambda_{ij} \sin (\phi_j - \phi_i),
\label{forcedworm}
\end{equation}
where $\lambda_{ij} = \lambda A_{ij}$. The adjacency matrix $A_{ij}$ gives the strength of interaction between oscillators $i$ and $j$. For unweighted networks $A_{ij}$ assumed the value 1 if they interact and 0 otherwise, but weighted networks like that of the {\it C. elegans}, might have very inhomogeneous distributions of weights. To distinguish this case from the unweighted networks we define $s_i = \sum_j A_{ij}$ as the weighted degree of neuron $i$. For networks that can be divided into anatomical or functional communities, the external force can be applied to one of the communities as a way to probe its influence on the others. Thus, we will investigate how the control parameters, $\lambda$ and $F$, affect the spontaneous and induced synchronization of the focal community (where the force is applied) and how it spreads to the other communities of the system.

As we already discussed on Chapter \ref{chapter4} if there is no external force and if  the internal coupling constant $\lambda$ is sufficiently large the oscillators synchronize spontaneously with frequency $\bar\omega = \sum \omega_i/N$ in the original coordinates $\theta$ or with frequency $\bar\omega - \sigma$ in the rotating frame $\phi$. On the other hand, if both $\lambda$ and $F$ are large the system synchronizes with the external frequency $\sigma$ in the original frame or 0 in the rotating frame. In our simulations, since the Gaussian distribution is symmetric, $\bar{\omega}=0$, the spontaneous synchronization corresponds to global frequency $ \dot{\psi} =  - \sigma$ and forced synchronization to frequency $ \dot{\psi} =   0$ \cite{us}.

We can estimate the minimum intensity of the external force, $F_c$, required to induce global synchronization using equation (\ref{force_crit})
\begin{equation}
F_{c} = \frac{\sigma}{f} \frac{\langle k \rangle}{\langle k \rangle_C}.
\label{Fcrit}
\end{equation}
In the context of this work, $f=N_C/N$ is the fraction of forced neurons; $\langle k \rangle$ and $\langle k \rangle_C$ are the average degree of the network and the forced module, respectively.

\subsection{Modularization}
\label{mod}

Understanding how connections are arranged in neural networks is key to understand how the brain functions and transmits information \cite{Baptista2010,Antonopoulos2015,Borges2017}.  Neurons can be grouped by their location in the brain, by their function and also by their connectivity with other neurons, independently of their position or function. Networks can generally be decomposed into these {\it topological} modules, also called clusters or communities, where a large number of links join nodes of the same module and comparatively few links join nodes belonging to different modules. Several methods have been recently proposed to detect modules, each providing a different decomposition, and no optimal algorithm has yet been devised \cite{Fortunato2010}. The strength of each decomposition, however, can be measure by the modularity coefficient \cite{newman2004,newman2006} 
\begin{equation}
 Q_w = \frac{1}{2m} \sum_{i,j} \left(A_{ij} - \frac{s_i s_j}{2m} \right)  \delta(c_i,c_j),
\label{modularity}
 \end{equation}
where $A_{ij}$ is the (weighted) adjacency matrix, $s_i$ is the sum of the weights of all links attached to node $i$, $c_i$ is the module of node $i$, $\delta(c_i,c_j) = 1$ if nodes $i$ and $j$ belong the same module and $2m$ is the sum of all of the link weights in the network. Eq. (\ref{modularity}) gives $Q_w = [-0.5, 1.0]$, where positive values indicate that there exist a larger number of connections between nodes of the same community than if connections were made randomly, and negative values mean less intra-module connections. For real complex networks, $0.3 < Q_w < 0.7$ \cite{newman2004,newman2006}. For unweighted networks the same formula can be used, replacing $s_i$ by the degree $k_i$ of node $i$ and $m$ by the total number of links in the network.

Previous analysis of neural networks have shown that they do exhibit modular organization \cite{Fortunato2010, newman2006}. The most common algorithms for module detection in biological network analysis are the so called hierarchical clustering \cite{Fortunato2010}. This technique is classified in two types: in the first, individual neurons are initially grouped if they have high similarity; then these groups are further clustered together and so on until the desired number of modules is formed (bottom up, agglomerative algorithm). In the second type of algorithm the network is divided in groups by the removing links that connect nodes with low similarity (top down, divisive algorithm).

In this work we used the ModuLand plug-in \cite{moduland1,moduland2} of the software Cytoscape \cite{cytoscape} to study the modularization of the \textit{C. elegans} network. This tool uses a hierarchical algorithm which detects multiple layers of communities, where nodes of the higher hierarchical step are modules of the lower step. ModuLand was tested in biological systems, such as protein structure and metabolic networks, providing modules which correspond to relevant biological communities \cite{moduland1,moduland2}.  For the present case of the  \textit{C. elegans} electric junction network ModuLand divided the network into 3 modules with $Q_w=0.47$.  Other procedures result in different partitions; for example \cite{antono2015} obtained 6 modules with $Q_w=0.375$, \cite{sohn} found 5 modules with $Q_w=0.49$ and \cite{pan2010} divided the network into 11 and 15 modules with $Q=0.63$ and $Q_w=0.66$, respectively. They all used slightly different versions of the  \textit{C. elegans} network, including or excluding some neurons. Here we focused on the case of 3 modules to compare with the 3 functional categories (motor, sensory, interneurons) and 3 major anatomical classes (head, midbody, tail). Further information is summarized in table \ref{table-modularity} of section \ref{sp}. We have also considered two other partitions, containing 5 and 10 modules respectively. The details are described on section \ref{sp-2}.

\subsection{\textit{C. elegans} neural connectome}
\label{cenc}

Based on structural and functional properties of the neural network of \textit{C. elegans}, Varshney et al \cite{chen2011} and Yan et al \cite{nature} presented a division of neuronal classes, totalizing 118, in three categories: sensory neurons (SN), which respond to environmental variations, motoneurons (MN), recognized by  the presence of neuromuscular junctions and responsible by locomotion, and the interneurons (IN), which cover all of other classes. The adjacency weighted matrix is defined as follows: the element $w_{ij}$ represents the total number of synapses interchange between the pair of neurons $ij$. In \cite{chen2011} the authors also divide the set into the gap junction network, which refers to the electrical synapses, and  the chemical synapses network.

Gap junctions are a medium for electrical coupling between neurons and, since the electric signal can be made in both directions, the electrical junction network is considered undirected and, consequently, its adjacency matrix symmetric. On the other hand, the chemical synapses network is a directed and weighted network, whose adjacency matrix is assymetric. Here we will concentrate on the electrical junction network only.

We analyzed the gap junctions neural network of nematode \textit{C. elegans} extracting the data from WormAtlas \cite{wormatlas}. The full connectome has 279 neurons (nodes) and 514 gap junctions (connections) divided  into a giant component with 248 neurons plus 31 neurons not connected with it. Here we will study the dynamics on the giant component. Thereby, we built the weighted electrical junction (EJ) network of the \textit{C. elegans} with 248 neurons and 511 gap junctions.  We also used a hierarchical algorithm to detect communities on the EJ network. For that, we used the package ModuLand \cite{moduland1,moduland2} available on the free software Cytoscape \cite{cytoscape}. The algorithm provided three modules ($M_1$, $M_2$, $M_3$) with modularity $Q_w = 0.47$.

\begin{figure}[H]
\center
\includegraphics[scale=0.3]{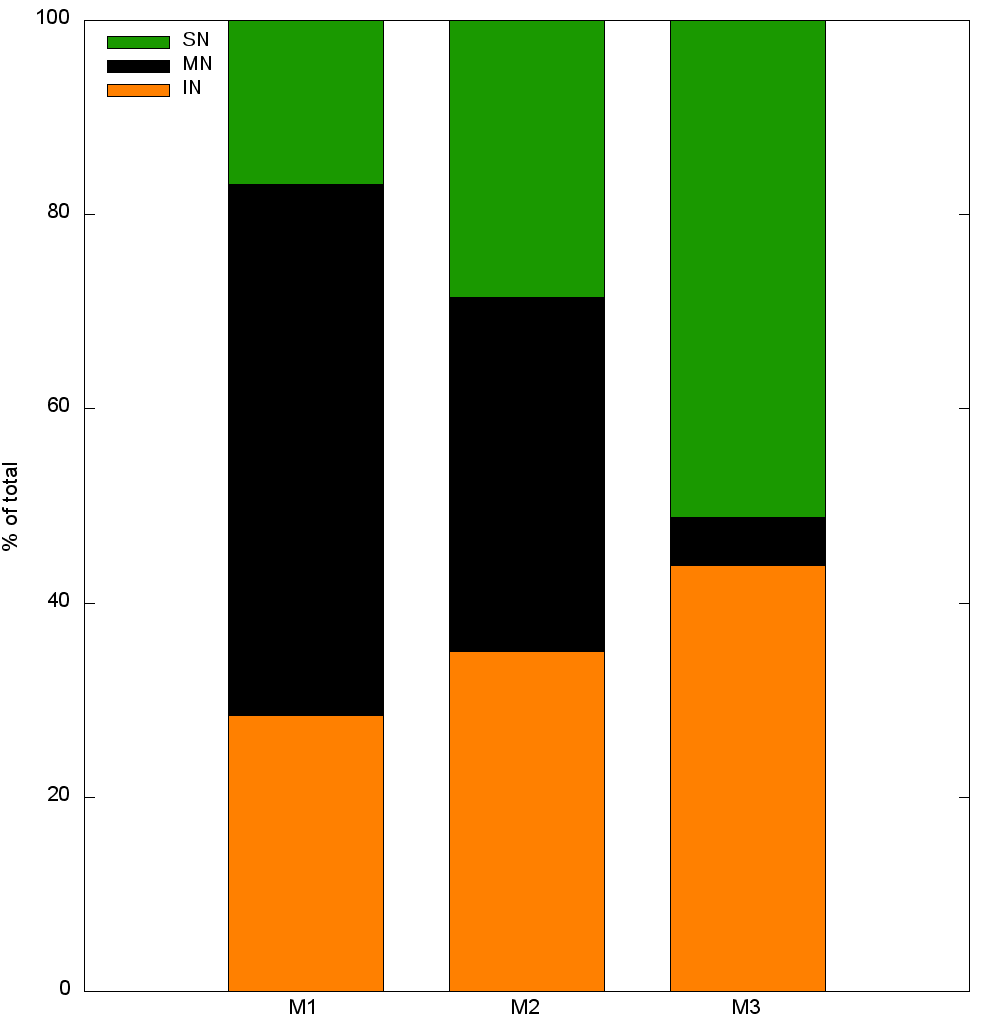}
\includegraphics[scale=0.3]{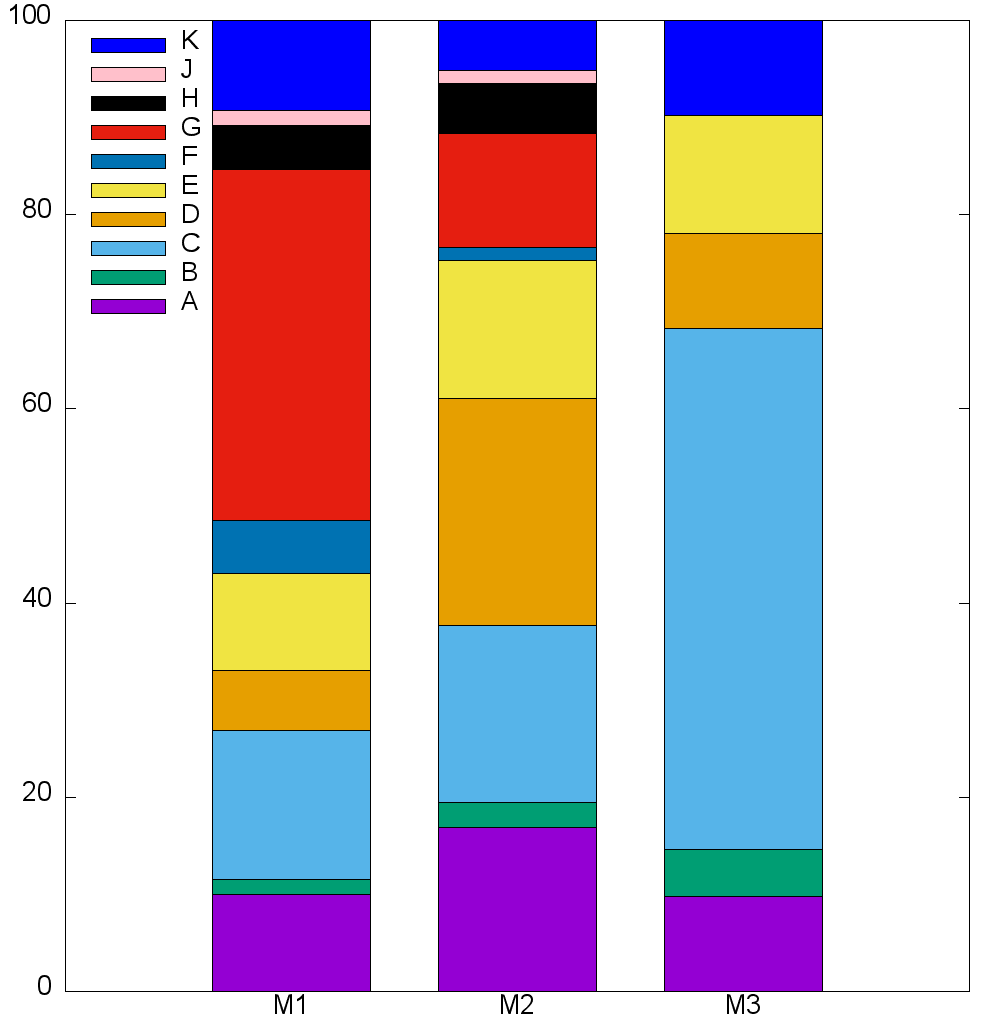}
\caption{ Histograms representing the fraction of (left) neuronal class (SN: sensory neuron, MN: motorneuron and IN: interneuron) and (right) of ganglia (A: anterior ganglion, B: dorsal ganglion, C: lateral ganglion, D: ventral ganglion, E: retrovesicular ganglion, F: posterolateral ganglion, G: ventral cord neuron group, H: pre-anal ganglion, J: dorsorectal ganglion, K: lumbar ganglion) for each module ($M_1$: module 1, $M_2$: module 2 and $M_3$: module 3).}
\label{hist1}
\end{figure}

\begin{figure}[!htpb]
\center
\includegraphics[scale=0.33]{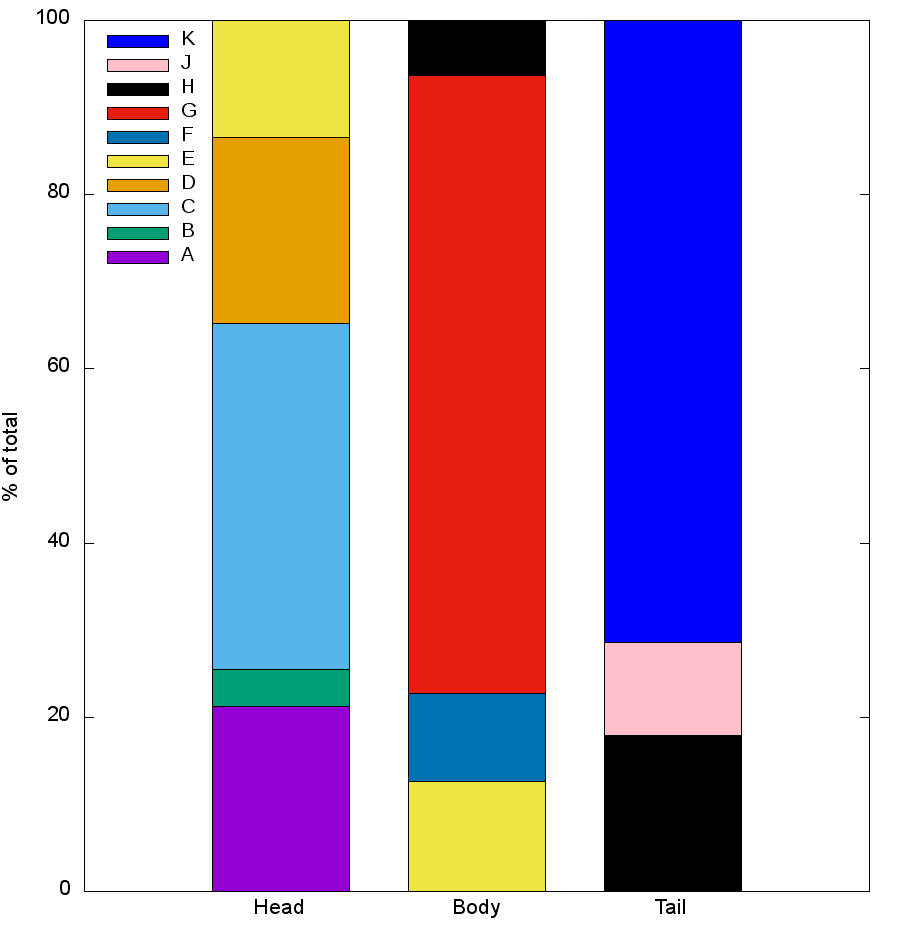}
\includegraphics[scale=0.33]{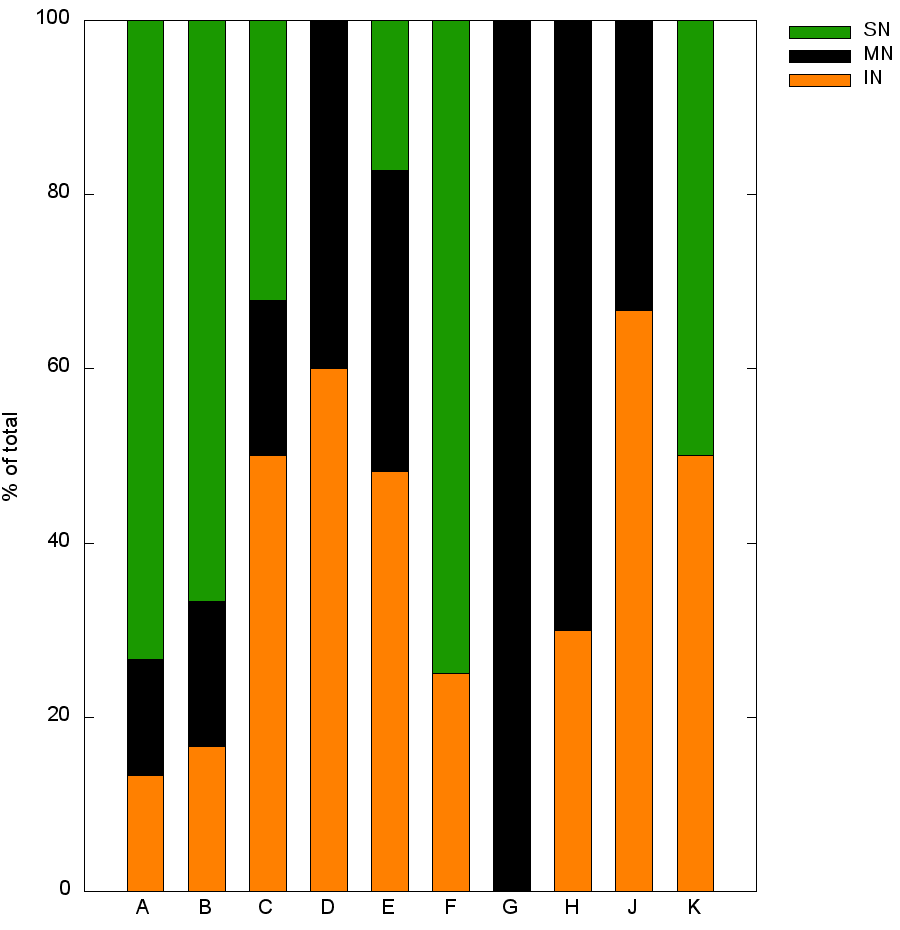}
\caption{ Histograms representing the fraction of (left) different ganglia distributed by physical localization (head, midbody and tail) and (right) the fraction of classes (SN, MN and IN) component in each ganglion.}
\label{hist2}
\end{figure}

Each neuron were further classified as belonging to one of three functional categories (sensory, motor and interneurons) and one of the 10 ganglia (A: anterior ganglion, B: dorsal ganglion, C: lateral ganglion, D: ventral ganglion, E: retrovesicular ganglion, F: posterolateral ganglion, G: ventral cord neuron group, H: pre-anal ganglion, J: dorsorectal ganglion, K: lumbar ganglion \cite{wormatlas}). The compositions of neuronal categories and ganglionic classification in each module are shown in Figure \ref{hist1}. The ganglia are a finer division of the anatomical classification into head (H), body (B) and tail (T). The histograms in Figure \ref{hist2} summarize the information extracted from the WormAtlas showing how ganglia are distributed physically (left panel) and how neuronal functions are represented in each ganglion (right panel). Note that ganglia A, B, C and D belong to the head, G is entirely localized in the body and J and K belong to the tail.

This set of divisions of the neural network into communities can be classified as (i) topological ($M_1$, $M_2$, $M_3$); (ii) anatomical (by ganglion A-K) and; (iii) functional (SN, MN, IN). They are all different and intertwined, showing the complexity of the EJ network.

In the next section we will apply the stimulus to $M_1$ (the largest of the topological modules), to ganglion C (completely located in the head and with mixed types of functional neurons) and to the sensory neurons. Results for modules $M_2$, $M_3$, ganglion G and motoneurons are shown in section \ref{sp}. Previous works  \cite{pan2010} have shown that ganglion C is important in the transmission of information between neurons that receive sensory stimulus and those responsible for motor processing. We also performed simulations on ganglion G (see section \ref{sp}), which is localized completely in the midbody (figure \ref{hist2} left) and is composed only by motoneurons (figure \ref{hist2} right). Finally we note that the sensory neurons are responsible for collecting information from external environment and react to stimuli inside the organism, acting as a input channel. In this sense, \textit{C. elegans} uses these neuronal functions to explore the ambient, navigating over thermal, chemical and oxygen variations, in addition to avoid hostile behavior \cite{wormatlas}.

\subsection{Order parameters and correlations}
\label{op}

The partially forced Kuramoto dynamics will be applied to the {\it C. elegans} as a way to probe its modular structure. Forcing a particular module may or may not induce synchronization with the external frequency on other modules of the system. In order to monitor the behavior of separate modules we define 
\begin{equation}
z_n = \frac{1}{N_n} \sum_{i \in M_n} e^{i \phi_i} \equiv r_n e^{i \psi_n}
\end{equation}
where the subscript $n$ specifies the module $M_n$ of size $N_n$. Therefore, $r_{n}$ is a local order parameter that measures how much the oscillators in the module are synchronized among themselves. The angular velocity $\dot{\psi_n}$ provides information about the motion of the set: $\dot{\psi_n} = 0$ implies sync with the external force, $\dot{\psi_n} = -\sigma$ refers to spontaneous sync whereas  nonconstant values indicate more complex behavior.

Intermodule behavior will also be monitored by the quantities
\begin{equation}
z_{nm} = \frac{1}{N_n+N_m} \sum_{i \in M_n \cup M_m} e^{i \phi_i} \equiv r_{nm} e^{i \psi_{nm}}
\end{equation}
with similar interpretations. Finally we compute the usual order parameter 
\begin{equation}
z = \frac{1}{N} \sum_{i=1}^N e^{i \phi_i} \equiv r_t e^{i \psi_t}
\end{equation}
that provides information on the global network synchrony.

Velocity-velocity correlations between all pairs of oscillators are defined by 
\begin{equation}
\tilde{c}(i,j) = \frac{1}{T} \int_{t_0}^{t_0+T} (\dot{\phi}_i(t) - \langle \dot{\phi_i} \rangle)
(\dot{\phi}_j(t) - \langle \dot{\phi_j} \rangle) \, dt
\label{fluc1}
\end{equation}
where
\begin{equation}
\langle \dot{\phi_i} \rangle = \frac{1}{T} \int_{t_0}^{t_0+T} \dot{\phi}_i(t)  \, dt
\end{equation}
and $t_0$ is a sufficiently long time so that the transient dynamics has passed. 

The normalized velocity-velocity correlation function is then defined as:
\begin{equation}
c(i,j) = \frac{\tilde{c}(i,j)}{\sqrt{\tilde{c}(i,i) \, \tilde{c}(j,j)}},
\label{corr-v-v}
\end{equation}
where $|c(i,j)| \leq 1 $. We note that the correlation is computed in terms of the fluctuations of the average velocity, that was subtracted out in Eq. (\ref{fluc1}). The 248 $\times$ 248 correlation matrix gives direct information about the effect of one neuron over another, irrespective of their synchronization state. If an increase in the velocity of $i$ leads to the average increase in the velocity of $j$ then nodes $i$ and $j$ are positively correlated and $c(i,j) > 0$. If, on the other hand the velocity of $j$ decreases, they are negatively correlated and $c(i,j) < 0$. Finally, if they are uncorrelated $c(i,j) \approx 0$. In the simulations we used $t_0=T/2$ and $T=20$ which was enough for the equilibration of the system. 

The parameters $z_n$ provide information about the synchronization of each module, whereas the average value of the phase velocity $\dot{\psi}_n$ tells whether the module follows the external force or spontaneous collective motion. This information is complemented by the velocity-velocity correlation, which measures the effect of one node over the other even if they synchronize with different frequencies or are not synchronized at all.

\section{Results}
\label{elegansresults}

\begin{figure}[H]
\center
\includegraphics[scale=0.19]{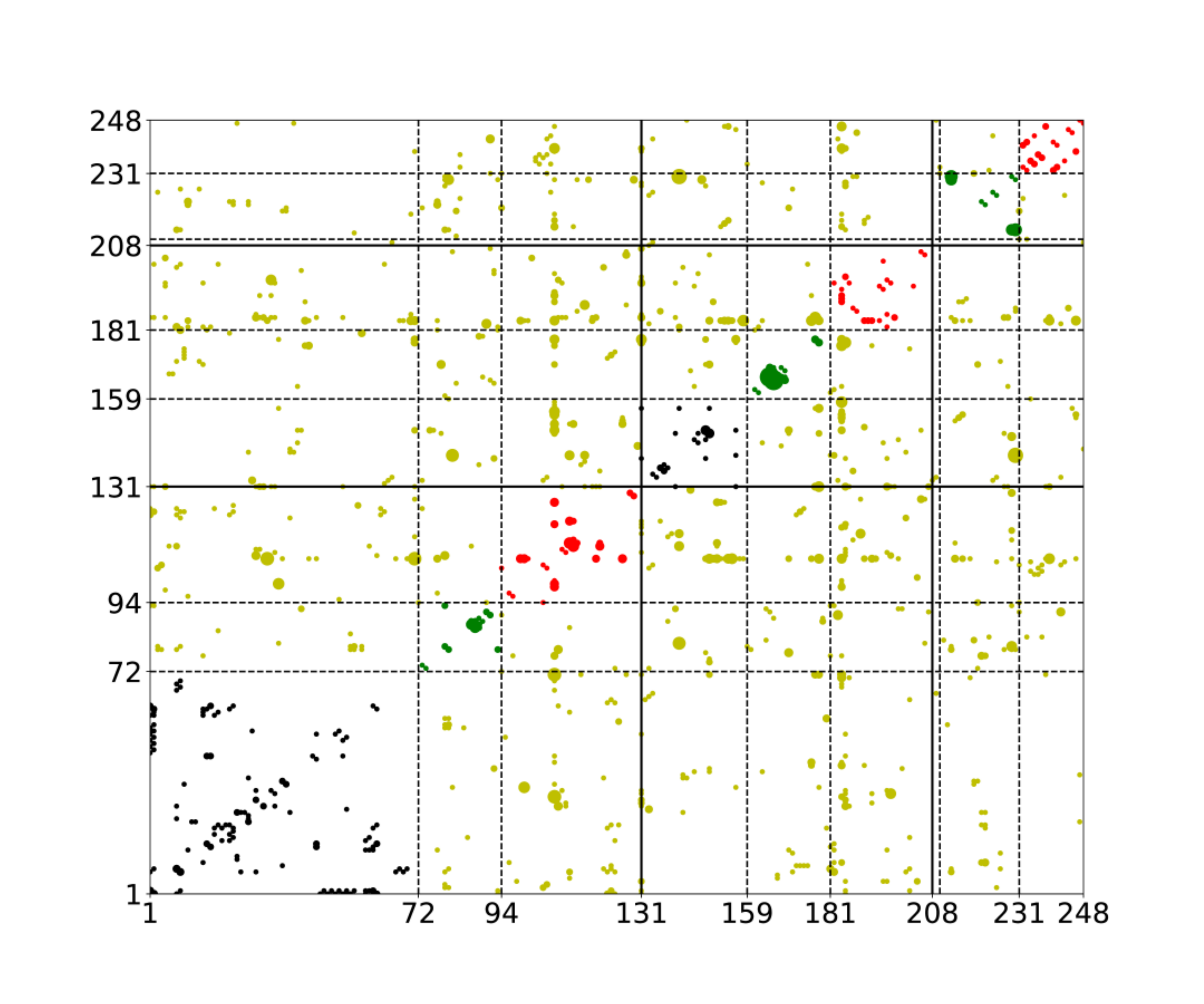}
\includegraphics[scale=0.19]{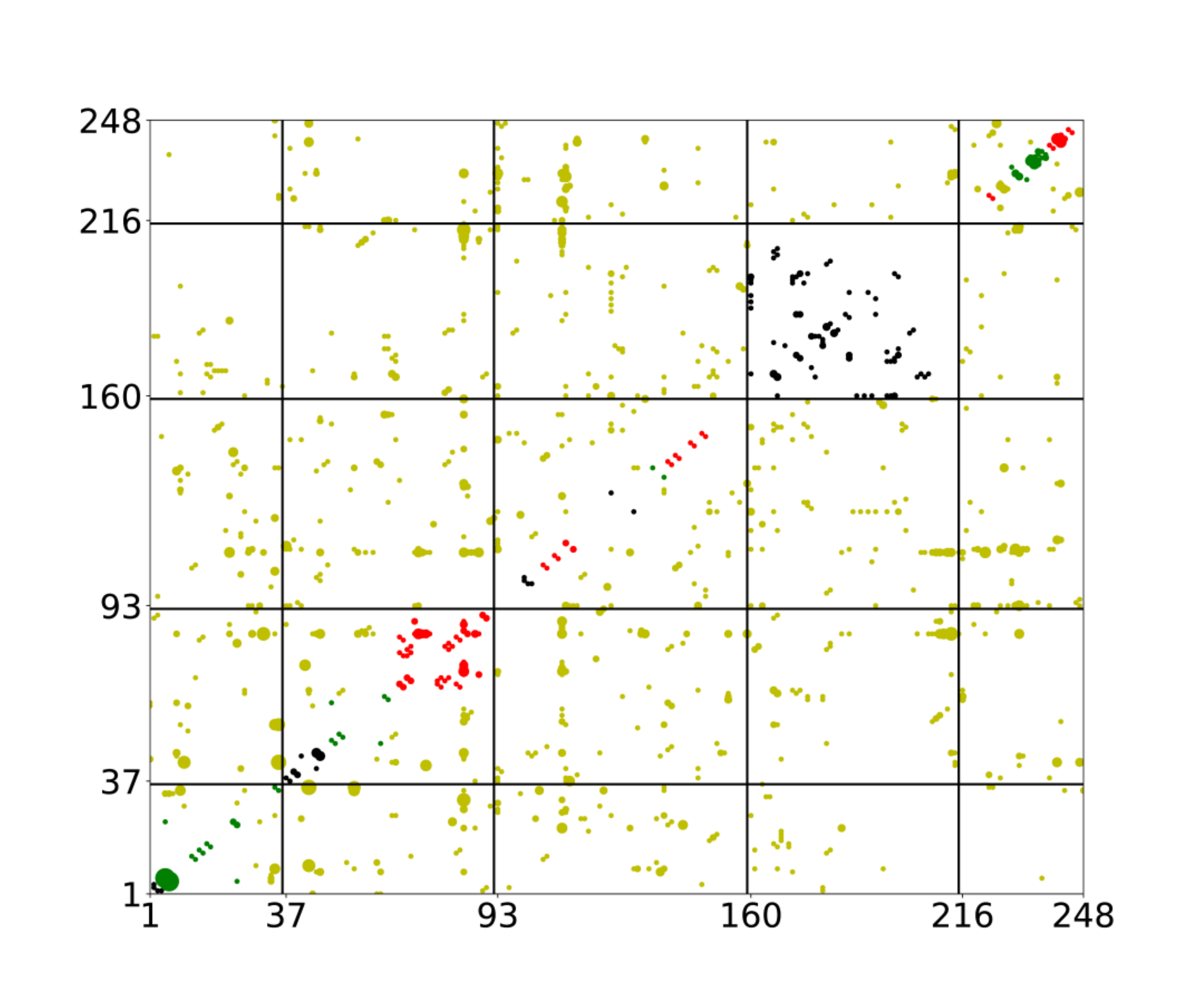}
\caption{ Left panel: weighted adjacency matrix highlighting the 3 topological modules $M_1$, $M_2$ and $M_3$ separated 
by thick black lines and subdivided into motor (black), sensory (green) and interneurons (red) by dashed black lines. Intermodule connections are shown in yellow.
Right panel: weighted adjacency matrix highlighting 5 groups of ganglia separated by thick black lines, \{A, B\}; \{C\}; \{D, E, F\}; \{G\} and \{H, J, K\}. Subdivisions and intermodule connections follow the functional colors of left panel.}
\label{adjs}
\end{figure}

Figure \ref{adjs}  shows the weighted adjacency matrix ordered according to the topological modules $M_1$, $M_2$ and
$M_3$ (left panel). Modules are separated by thick black lines and subdivided into motoneurons (black), sensory neurons
(green) and interneurons (red) by dashed black lines. The size of the dot is proportional to the intensity $A_{ij}$ and
intermodule connections are represented in yellow. The right panel shows the adjacency matrix ordered by ganglion, from
head to tail. The thick black lines highlight 5 groups of ganglia: \{A, B\}; \{C\}; \{D, E, F\}; \{G\} and \{H, J, K\}.
These groupings were defined to facilitate the visualization of the plots and to emphasize the forced ganglia (ganglion
C in figure \ref{corr-C} and ganglion G in figures \ref{GcorrSM} and \ref{GSM}).
Subdivisions and intermodule connections follow the functional colors of left panel. The indexes in both panels delimit
the divisions made.

\begin{figure}[!htpb]
\center
\includegraphics[scale=0.46]{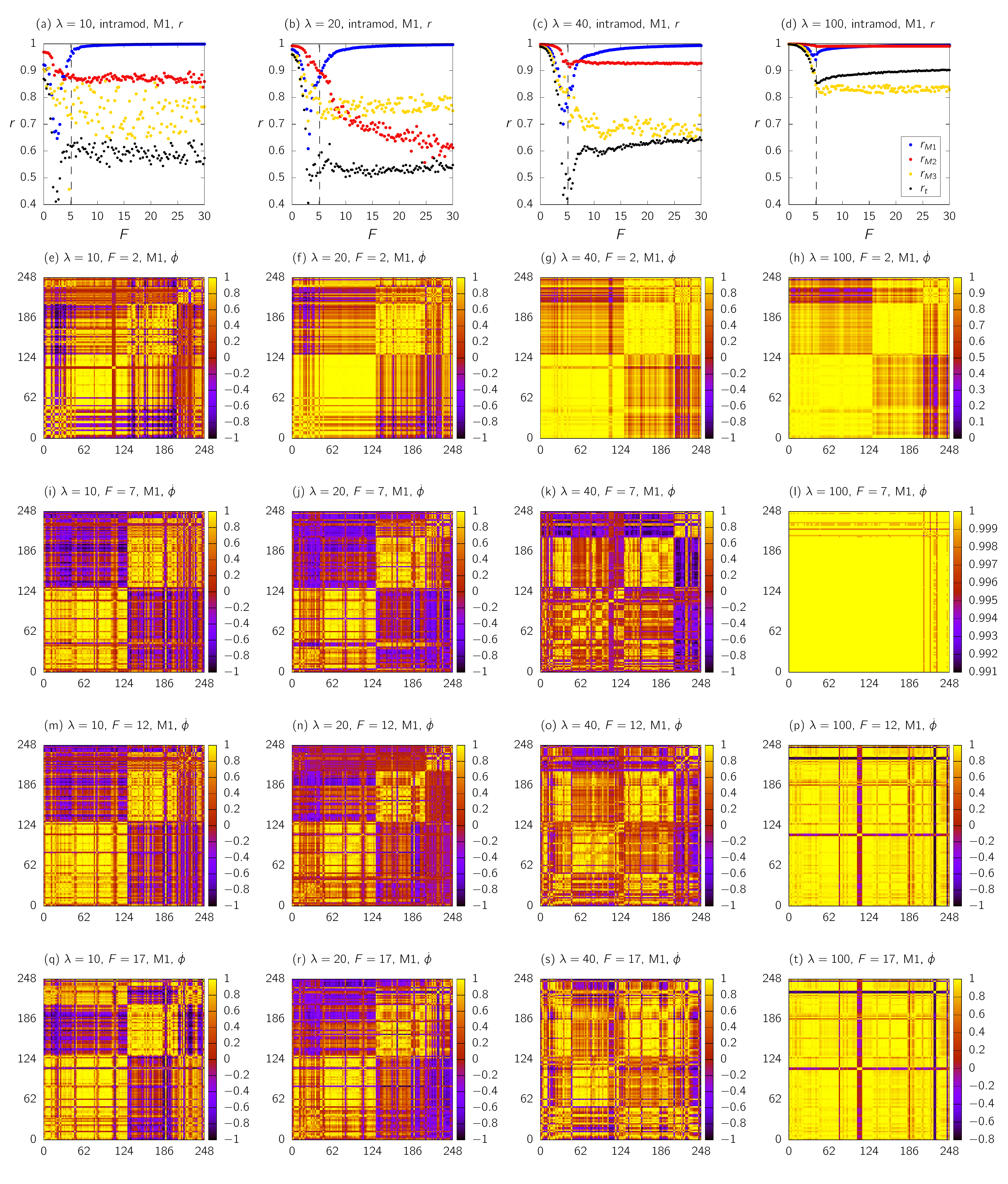}
\caption{ Panels (a)-(d): the global and local order parameters as a function of the external force $F$ acting on
neurons of $M_1$ for $\lambda$ fixed. The dashed lines indicate the critical force, $F_{c,theo}^{M_1} = 5.12$.
Panels (e)-(t): the velocity-velocity correlation matrix 248 $\times$ 248 obtained
using Eq. (\ref{corr-v-v}). In each panel, the fixed parameters $\lambda$ and $F$ are indicated. The $M_1$ neurons are 
indexed by 1 to 130, the $M_2$ neurons by 131 to 207 and the $M_3$ neurons are indexed by 208 to 248.}
\label{corr-M1-r}
\end{figure}

\begin{figure}[!htpb]
\center
\includegraphics[scale=0.45]{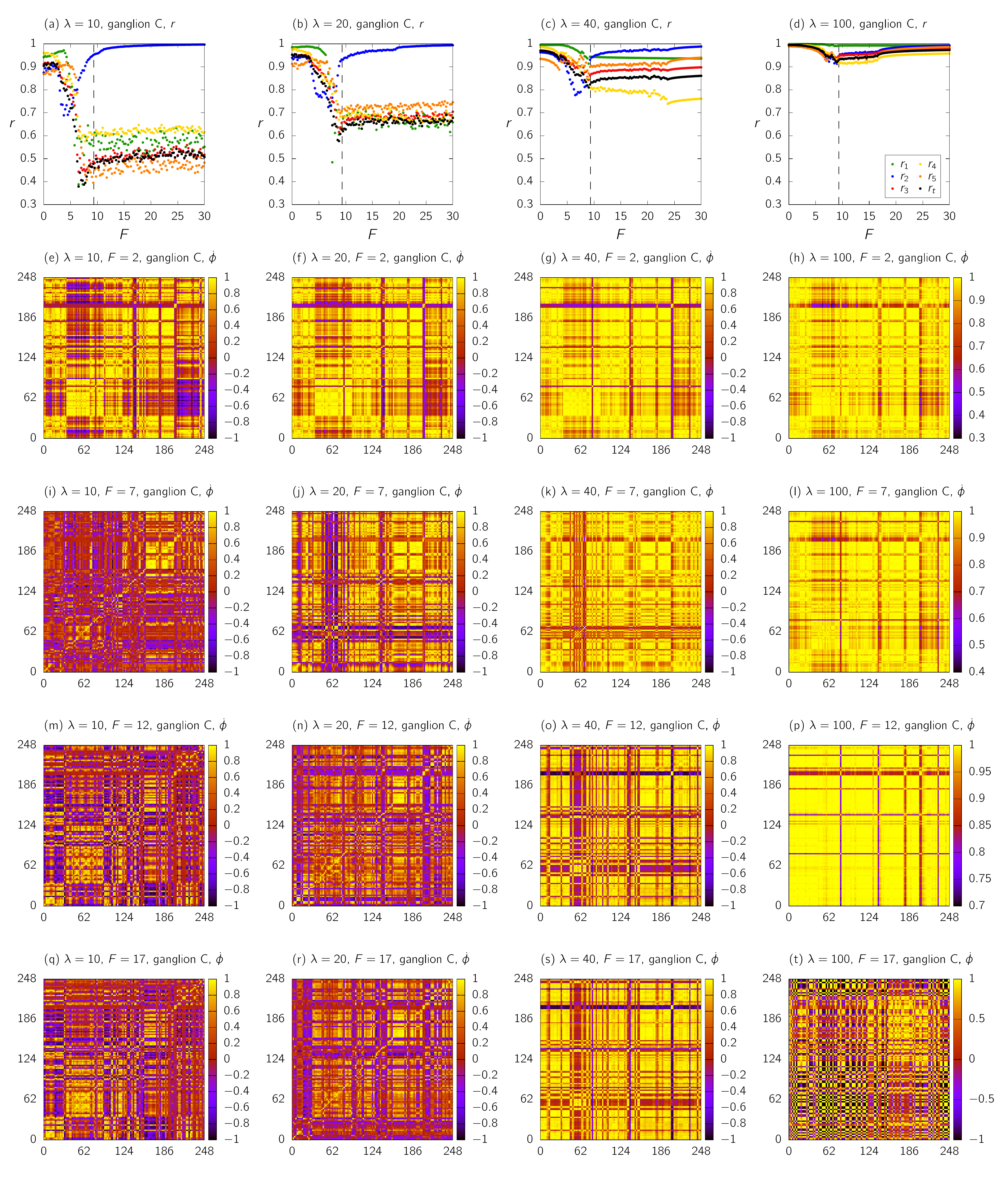}
\caption{ Panels (a)-(d): the global and local order parameters as a function of the external force $F$ acting on
ganglion C for $\lambda$ fixed. The dashed lines indicate the critical force, $F_{c,theo}^{C} = 9.32$. 
Panels (e)-(t): the velocity-velocity correlation matrix 248 $\times$ 248 obtained using Eq. (\ref{corr-v-v}). In each panel,
the fixed parameters $\lambda$ and $F$ are indicated. The group 1 (\{A,B\}) are indexed by 1 to 36, group 2 (\{C\}) 
by 37 to 92, group 3 (\{D, E, F\}) by 93 to 159, group 4 (\{G\}) by 160 to 215 and group 5 (\{H, J, K\}) are indexed by 216 to 248.}
\label{corr-C}
\end{figure}

\begin{figure}[!htpb]
\center
\includegraphics[scale=0.45]{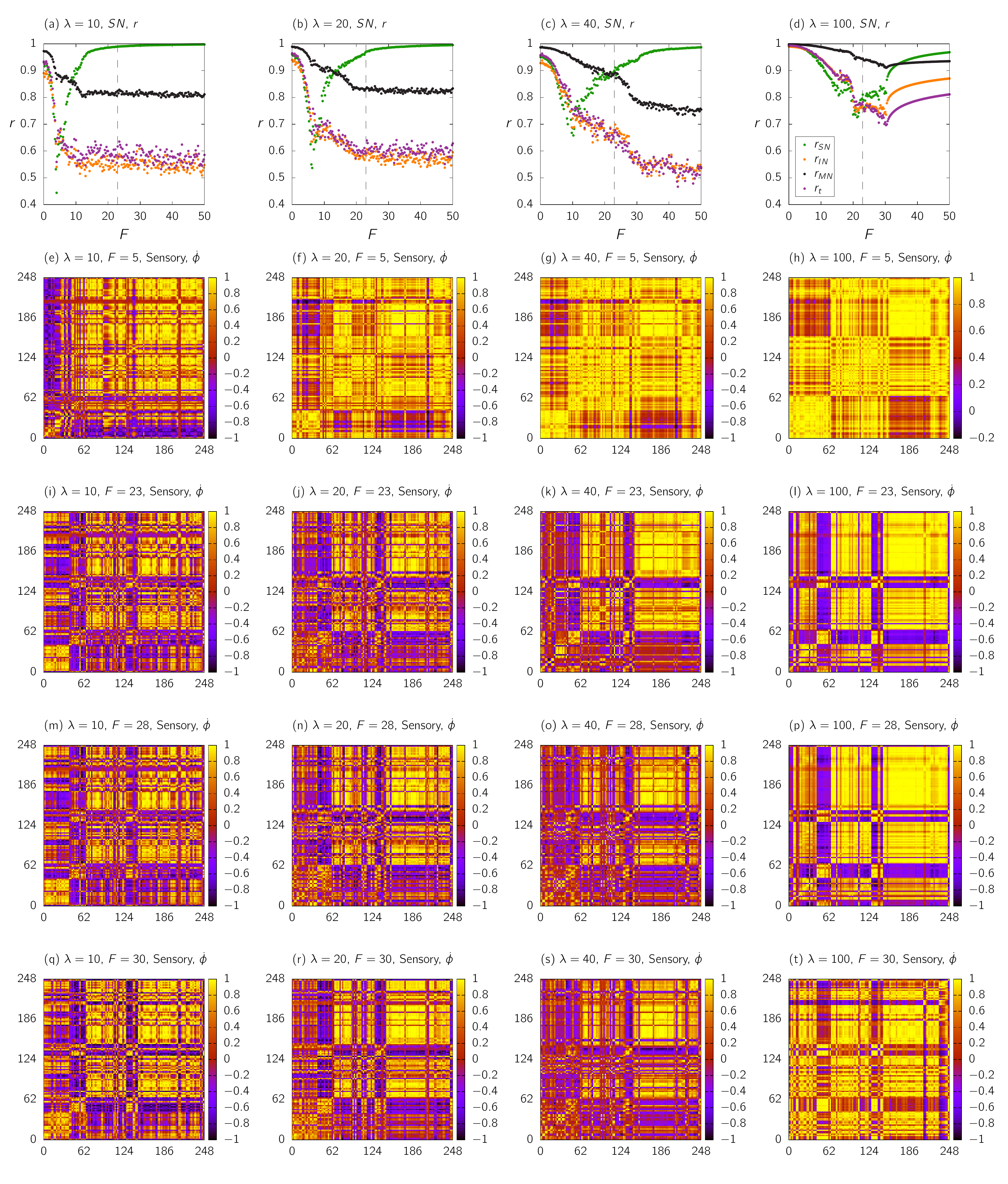}
\caption{ Panels (a)-(d): the global and local order parameters as a function of the external force $F$ acting on sensory
neurons for $\lambda$ fixed. The dashed lines indicate the critical force, $F_{c,theo}^{SN} = 22.96$.
Panels (e)-(t): the velocity-velocity correlation matrix 248 $\times$ 248 obtained using Eq.
(\ref{corr-v-v}). In each panel, the fixed parameters $\lambda$ and $F$ are indicated. The sensory neurons (SN) are indexed 
by 1 to 65, the interneurons (IN) by 66 to 147 and the motoneurons (MN) are indexed by 148 to 248.}
\label{corr-sensory}
\end{figure}

In order to analyze the interdependencies of the modules for  different partitions of the EJ network, we have simulated the application of an external stimulus to one of the modules and observed its effect on the others. The stimulus is
modeled by an external periodic force acting only on the selected module under the Kuramoto dynamics as described in
section \ref{pfk}. In all our simulations we have fixed $\sigma = 3$. The effects on the other modules is measured by
local order parameters, such as $r_{nm}$, and the normalized velocity-velocity correlation function as described in section
\ref{op}. Here we show the numerical results for the cases where the stimulus was applied only to $M_1$, or to ganglion
C or to the sensory neurons, figures 	\ref{corr-M1-r}, \ref{corr-C} and \ref{corr-sensory}, respectively. In all cases
we show  the  global ($r_t$) and local ($r_{nm}$) order parameters as a function of the intensity $F$ of the external force
for four values of the internal coupling $\lambda$, panels (a) to (d), and the velocity-velocity correlation matrices 
in panels (e) to (t).

We used equation (\ref{Fcrit}) to calculate the expected critical force to induce global synchronization in an
equivalent random network. The values of the fraction $f$ of forced nodes, the average degree $\langle k \rangle_C$ of
the forced set and that of the whole network $\langle k \rangle$ are summarized in Table \ref{values-forced}. In chapter \ref{chapter4} \cite{us}, fully synchronized states were defined by the conditions $r_t>0.95$ and $|\dot\psi_t| < 0.01$. Here we also classify
the network as {\it partially synchronized} if $0.8 < r_t \leq 0.95$ and  $|\dot\psi_t| < 0.1$.
Further information can be found in table \ref{critical-forces} of section \ref{sp}. 

\begin{table}[H]
\center
\caption{Basic properties of forced modules: number of neurons, fraction of nodes, average degree and theoretical critical force for full synchronization.}

\begin{tabular}{|c|c|c|c|c|}
\hline
\multicolumn{1}{|l|}{} & Module 1                                           & Ganglion C                                           & Sensory Neurons                                           & Network                                          \\ \hline
Number of neurons      & $N_{M_1}$ = 130                                        & $N_C$ = 56                                          & $N_{SN}$ = 65                                         & $N$ = 248                                              \\ \hline
Fraction of nodes      & $f_{M_1}$ = 52.42 \%                                     & $f_C$ =  22.58 \%                                     & $f_{SN}$ =26.21 \%                                       & $f$ = 100 \%                                           \\ \hline
Average degree         & $\langle s_{M_1} \rangle $ = 7,96 & $\langle s_C \rangle $ = 10,16 & $\langle s_{SN} \rangle $ = 5,27 & $\langle s \rangle $ = 7,13 \\ \hline
Critical force (eq. (\ref{Fcrit}))   & \multicolumn{1}{l|}{$F_{c,theo}^{M_1} $ = 5,12} & \multicolumn{1}{l|}{$F_{c,theo}^{C} $ = 9,32} & \multicolumn{1}{l|}{$F_{c,theo}^{SN} $ = 22,96} & \multicolumn{1}{l|}{$F_{c,theo} $ = 3,00} \\ \hline
\end{tabular}
\label{values-forced}
\end{table}

\textbf{Stimulating the $M_1$: the role of topology} 
\vspace{0.5cm}

\begin{figure}[H]
\center
\includegraphics[scale=0.54]{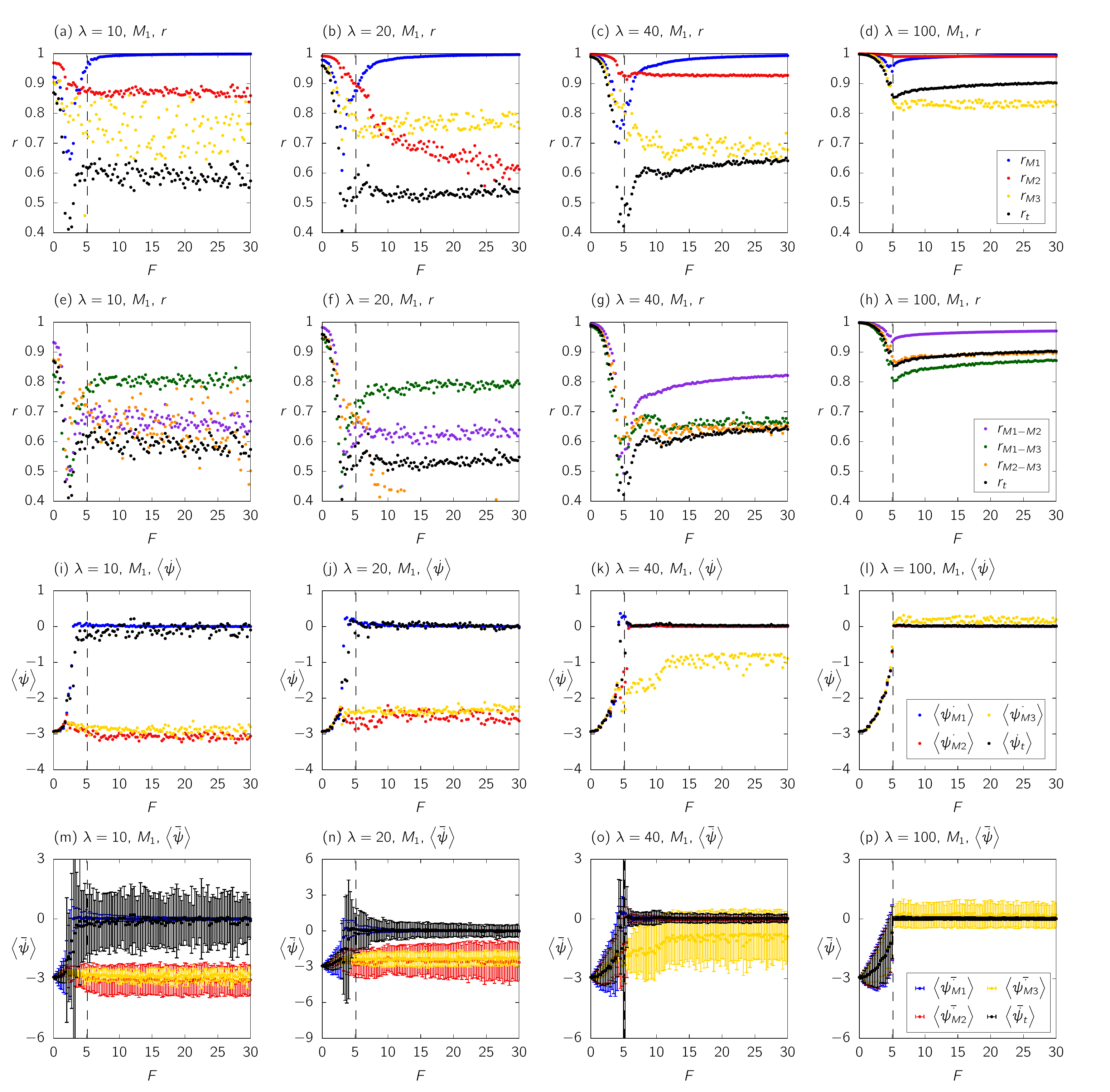}
\caption{Order parameters as a function of $F$ for the \textit{C. elegans} neural network where the external force acts only on $M_1$. Panels (a) to (d) are the same as in the main text and indicate the order parameter $r$ for
intra-modules. Panels (e) to (h) indicate $r$ for inter-module. The results for $\dot\psi$ are exhibited in panels
(i) to (p), with respective error bars, panels (m) to (p), for intra-module case. The coupling $\lambda$ is fixed
and its value is indicated in each panel. We also compute $r_{total}$ and $\dot\psi_{total}$ for whole network. 
The dashed lines indicate the critical force, $F_{c,theo}^{M_1} = 5.12$.}
\label{M1SM}
\end{figure}

Figure \ref{corr-M1-r} shows the results of simulations when only the neurons of $M_1$ are forced (indexes 1 to 130 in
the left panel of figure \ref{adjs}.) As $F$ increases, the neurons go through a region of asynchrony around $F=5$,
which is close to theoretical value for full synchrony $F_{c,theo}^{M_1} $ = 5.12 calculated with eq. (\ref{Fcrit}), and
then they synchronize with the external force ($\dot{\psi}_{M_1}=0$, Fig. \ref{M1SM} on section \ref{sp}) for $F$ larger than about 10, where
$r_{M_1} \rightarrow 1$. For large internal coupling $\lambda$, all modules appear to synchronize with external force
(see panel (d) of Fig. \ref{corr-M1-r} and panel (l) of Fig. \ref{M1SM} on section \ref{sp}), but $M_3$ has large fluctuations in
$\dot\psi_{M_3}$ (panel (p), Fig. \ref{M1SM} on section \ref{sp}). The global order parameter reaches its maximum value at $r_t \approx 0.9$
with $\dot\psi_t \approx 0.0$ for $\lambda = 100$ (tables \ref{table-whole-net} and \ref{table-m1} of section \ref{sp}).

The most striking feature of these simulations is the strong anti-correlation patterns developed between $M_1$ and $M_2$ for $\lambda \leq 20$. From the top panels we notice that, in these cases, $M_1$ is in synchrony with the external force whereas $M_2$ is still synchronized spontaneously. Nevertheless the effects of $M_1$ over $M_2$ are very clearly shown by the purple areas of the correlation plots. This indicates a lower value of the inter-modules order parameter $r_{nm}$, as can be seen between $M_1$-$M_2$ and $M_2$-$M_3$ (panel (k) on Fig. \ref{corr-M1-r} and Fig. \ref{M1SM} on section \ref{sp}). On the other hand, the presence of positive correlations between $M_1$ and $M_3$ (panel (q) on Fig. \ref{corr-M1-r}), is accompanied by an increase of $r_{13}$. 

\vspace{0.5cm}

\textbf{Stimulating ganglion C: the role of anatomy} 
\vspace{0.5cm}

Figures \ref{corr-C} shows the results of simulations when ganglion C is forced. The behavior of the order parameters
$r_t$ and $r_{n}$ as a function of $F$ is similar to that observed
when forcing the neurons of $M_1$, exhibiting a region of asynchrony between $F=5$ and $F=10$, which contains the
theoretical value, $F_{c,theo}^C = 9.32$, followed by stabilization for larger $F$. The
forced neurons are clearly seen as a bright yellow blocks in panels (e) to (h).

For sufficiently large $F$ ganglion C synchronizes with the external force ($r_2 \rightarrow 1$, panels (a) to (d), and
$\dot{\psi}_{2}=0$, Fig. \ref{corr-M3-r} on section \ref{sp}) for all values of $\lambda$ considered. However, the velocity-velocity correlation
matrices show much simpler patterns, displaying either nearly complete correlation (yellow areas in panels (g), (h), (k),
(l) and (p)),  or almost no correlation at all (large red areas in panels
(i), (m), (n), (q) and (r)) with ganglion C itself showing reduced internal correlations. Even for  $F>12$, where $r_2$
indicate that C is nearly fully synchronized for all $\lambda$'s, the correlation matrices show regions of mixed
behavior, especially for small $\lambda$, which means that part of neurons of C are non-correlated with each
other or even anti-correlated (see also Fig. \ref{corr-M3-r}, panels (a), (b), (i) and (j) on section \ref{sp}). Although all ganglia seem to synchronize with the
external force for $\lambda \geq 40$ and $F > 12$, their dynamics are uncorrelated with other ganglia.  The only exception is ganglion G, that shows up as a yellow square in the plots (see also Figs. \ref{GcorrSM} and \ref{GSM} of section \ref{sp}).

We also note that for $\lambda = 40$ the motor part of ganglion C (small yellow squares indexed by 37 to 46) correlates
separately from the rest of C for $F=12$, panel (o), and $F=17$, panel (s), which means that motoneurons respond
differently to external stimuli. For $F=17$, panels (q), (r), (s) and (t), the number of correlated neurons increases
from $\lambda=10$ to $\lambda=40$ but for $\lambda=100$ the entire network goes out of phase, with the exception of
ganglion G: it keeps its internal correlation at all times, maybe because it is entirely a motor ganglion type. Note
that $r_4 \approx 1$ only for $\lambda = 100$, panel (d), which means that full sync requires large
internal coupling. We also note that the global order parameters for $\lambda=100$ are $r_t = 0.98$ and $\dot\psi_t =
0.0$ (tables \ref{table-whole-net} and \ref{table-c} of section \ref{sp}), which means that the network is fully synchronized. 

\begin{table}[H]
\center
\caption{Global order parameters for each forced subset of neurons. The network is considered to be partially synchronized if $0.8 < r_t \leq 0.95$ and fully synchronized if $r_t > 0.95$ and $\dot\psi_t < 10^{-2}$. GS refers to Global synchronization. }
\begin{tabular}{|c||c|c|c||c|c|c||c|c|c|}
\hline
$\lambda$ & $r_{t-M_1}$ & $\dot\psi_{t-M_1}$ & GS ($M_1$) & $r_{t-C}$ & $\dot\psi_{t-C}$ & GS (C) & $r_{t-SN}$ & $\dot\psi_{t-SN}$ & GS (SN) \\ \hline
10        & 0.55      & 0.07             & No	& 0.52    & 0.03         & No  & 0.59     & -2.84           & No \\ \hline
20        & 0.54      & -0.01            & No	& 0.67    & 0.00         & No  & 0.63     & -2.46           & No \\ \hline
40        & 0.65      & 0.05             & No	& 0.87    & 0.00         & Partial  & 0.54     & -2.70           & No \\ \hline
100       & 0.91      & 0.02             & Partial	& 0.98    & 0.00         & Yes	    & 0.81     & 0.00            & Partial \\ \hline
\end{tabular}
\label{table-whole-net}
\end{table}

\textbf{Stimulating the sensory neurons: the role of function} 
\vspace{0.5cm}

Figure \ref{corr-sensory} shows the numerical results when all sensory neurons (SN) receive the external stimuli. Panels (a) to (d) show that the behavior of the network is more complex in this case. The SN synchronize with the external force for: $i)$ $\lambda = 10$ and $F>10$ (panels (a), (i) and Fig. \ref{SNSM}), $ii)$ $\lambda = 20$ and $F>15$ (panels (b), (j) and Fig. \ref{SNSM}), $iii)$ $\lambda = 40$ and $F>20$ (panels (c), (k) and Fig. \ref{SNSM}) and $iv)$ $\lambda = 100$ and $F> 30$ (panels (d), (l) and Fig. \ref{SNSM}). In $iii)$ and $iv)$ the values of $F$ are close to the theoretical value $F_{c,theo}^{SN} = 22.96$. Contrary to all other cases, larger values of $\lambda$ hinders the synchronization of the forced group, since $r_{SN}$ decreases from $\lambda=10$ to $\lambda=100$, although $\dot{\psi_{SN}} = 0$ (Fig. \ref{SNSM}). For $\lambda = 100$, the global order parameters are $r_t \approx 0.81$ and $\dot\psi_t = 0.0$ (table \ref{table-whole-net} and \ref{table-sn}), thus the system synchronizes only partially.

For $\lambda \leq 40$, the motoneurons and almost half of interneurons were in spontaneous sync (panels (a), (b), (i) and (j) on Fig. \ref{SNSM}), while for $\lambda > 40$ and $F>30$ most neurons were synchronized with external stimuli. The velocity-velocity matrices also show regions of anti and non-correlation, as can be seen on purple and red areas of Fig. \ref{corr-sensory}, respectively,  particularly for weak internal coupling, $\lambda < 40$. In these cases, the lack of correlation seems to indicate a lower value of inter-modules order parameter $r_{nm}$, as can be seen in panels (e) to (h)
of Fig. \ref{SNSM}.

\section{Discussion}
\label{elegansconclusion}

Groups of neurons can be defined in many ways, taking into account their anatomical location, their functional role or their topological properties in the network. In this chapter we  investigated the importance of these divisions as targets to stimuli, as well as their roles in spreading the inputs to other parts of the brain. Here we used a much simplified model of synchronization given by the Kuramoto system of phase oscillators subjected to a single stimulus, described by the external force, applied only to a subset of neurons representing a topological module (Fig. \ref{corr-M1-r}), a ganglion composed of different functional neurons (Fig. \ref{corr-C}) or the sensory neurons (Fig. \ref{corr-sensory}). Because the stimulus is permanently turned on in the model, the system behavior converges to an oscillatory state corresponding to an infinite sequence of spikes, which is clearly a simplification. However, the model does provide interesting information about the ability of the group of neurons receiving the input to synchronize among themselves or with other groups, or to develop correlations. 

The modularization procedure applied to EJ network reveals that topological modules do not contain purely anatomical groups or functional classes, but mixes neurons belonging to different ganglia and functional classes. This is illustrated in Fig. \ref{hist1}, where we have analyzed the distribution of neuronal classes and ganglia membership in each module. This corroborates previous studies that employed different modularization techniques \cite{antono2015, pan2010, sohn, chen2011,arenas2008-2} and highlights the complexity of the neuronal wiring regarding their location and function. The response of the electrical neural network to the stimulus was different for each type of neuron grouping, as we summarized in Table \ref{table-whole-net} in terms of synchronization and below in terms of cross correlations with other modules.

Stimulation of the neurons of the largest topological module $M_1$ induced strong anti-correlation in the velocity fluctuations of the neurons in $M_2$ and $M_3$ (purple areas of panels (i), (j) and (m) on Fig. \ref{corr-M1-r} or between $M_2$ and $M_3$ (panel (k) on the same figure), which kept their original state of spontaneous synchronization for moderate values of the internal coupling constant $\lambda$. The smallest topological module $M_3$ remained oblivious to the stimulus even for large values of $\lambda$. Interestingly, for intermediate values of the forcing (panel (j) on Fig. \ref{corr-M1-r}), the neurons of $M_1$ became mostly uncorrelated (red areas on Fig. \ref{corr-M1-r}), indicating a parameter region of poor response to the stimulus. It is possible, however, to identify the modular structure by the presence of three blocks, each of one corresponding to $M_1$, $M_2$ and $M_3$. A very similar behavior is observed when $M_2$ is stimulated (Figs. \ref{corr-M2-r}, \ref{corr-M3-r} and table \ref{table-m2}).  When the stimulus is applied to $M_3$, however, it never spreads to the other modules,  which remain in spontaneous synchrony but develop a pattern of anti-correlation inter-modules for sufficiently large values of $\lambda$ and $F$ (Figs. \ref{corr-M3-r}, \ref{M3SM} and table \ref{table-m3}).

The response of the network to stimulation of ganglion C was quite different from that of $M_1$ displaying essentially two distinct regions with (I) large parameter intervals of almost complete uncorrelated behavior, which occurs for $\lambda \leq 20$ (red areas on panels (i), (m), (n), (q), (r) and (t) on Fig. \ref{corr-C}) and (II) complete correlated behavior, with $\lambda \geq 40$ (yellow areas on Fig. \ref{corr-C}). Effective synchronization of ganglion C with the external force required large values of the coupling constant. Contrary to what occurs when forcing the topological module, the blocks of the correlation matrix corresponding to ganglia groups cannot be clearly distinguished, except for ganglion G (Figs. \ref{corr-M3-r} and \ref{M3SM}), which seems to hold high correlation between its neurons, possibly because it is the only group entirely composed by one class (motoneurons). Stimulation of ganglion G is presented in the section \ref{sp}, Figs. \ref{GcorrSM}, \ref{GSM} and table \ref{table-g}. The patterns of anti-synchronization between modules are again observed in this case, reinforcing the idea that functionality, not spatial location, is the relevant structure in this case.

Finally, stimulation of the sensory neurons leads to synchronization with the external driving for $\lambda \leq 40$, while the other two functional classes remained in spontaneous sync. It was only with strong internal coupling, $\lambda > 40$, and force larger than the theoretical value, $F>30$, that most of neurons were induced to the forced sync. The results also show many regions of anti and non-correlation (purple and red areas on Fig. \ref{corr-sensory}, respectively).  Matrix blocks of similar correlations identify the three classes reasonably well, although displaying visible internal structure, which indicates a more complex relationship between them. When motoneurons are stimulated (Figs. \ref{corr-MN}, \ref{MNSM} and table \ref{table-mn}) partial synchronization is only possible for very large values of $\lambda$ and $F$, but patterns of anti-correlation do appear for small values of $\lambda$, similar to what is observed for ganglion G.

In most cases, the order parameters $r$ and $\dot\psi$ exhibited a jump near critical force (dashed lines in the plots), which is closer to $F_{c,theo}$ as $\lambda$ increases. Global and partial synchronization, however, is only observed in some cases and at much larger values of $F$ and $\lambda$ than predicted by the mean field theory \cite{us}. When the stimulus was applied to ganglion C, in particular, global synchronization happened for $\lambda = 100$ and $F \approx 17$, which is larger than theoretical value found, $F_{c,theo}^C = 9.32$.

Previous studies \cite{us} have shown that the Kuramoto model with external localized stimuli leads to global synchronization on synthetic networks with simple topologies, such as random and scale-free, if $\lambda$ and $F$ are sufficiently large. Here we considered the real neural network of the nematode \textit{C. elegans}  and observed full and partial synchronization in very few cases and for higher values of $\lambda$ and $F$ than predicted. This indicates that the particular modular structure of the network protects the system from `seizures'. We found that the response of the network is highly complex and depends strongly not only on the stimulated group but also on the intensity $F$ and coupling strength $\lambda$. We hypothesize that this complexity  reflects the system flexibility to process and differentiate several types of inputs. The group divisions considered here (topological, functional and anatomical) are natural but not exhaustive and finer subdivisions might be important to understand the system response in more detail. Different types of stimuli, such as non-sinusoidal periodic trains or localized pulses, could also bring up interesting responses that might help distinguish the behavior of the different modules.

\section{Supplementary Material}
\label{sp}

\begin{figure}[H]
\center
\includegraphics[scale=0.46]{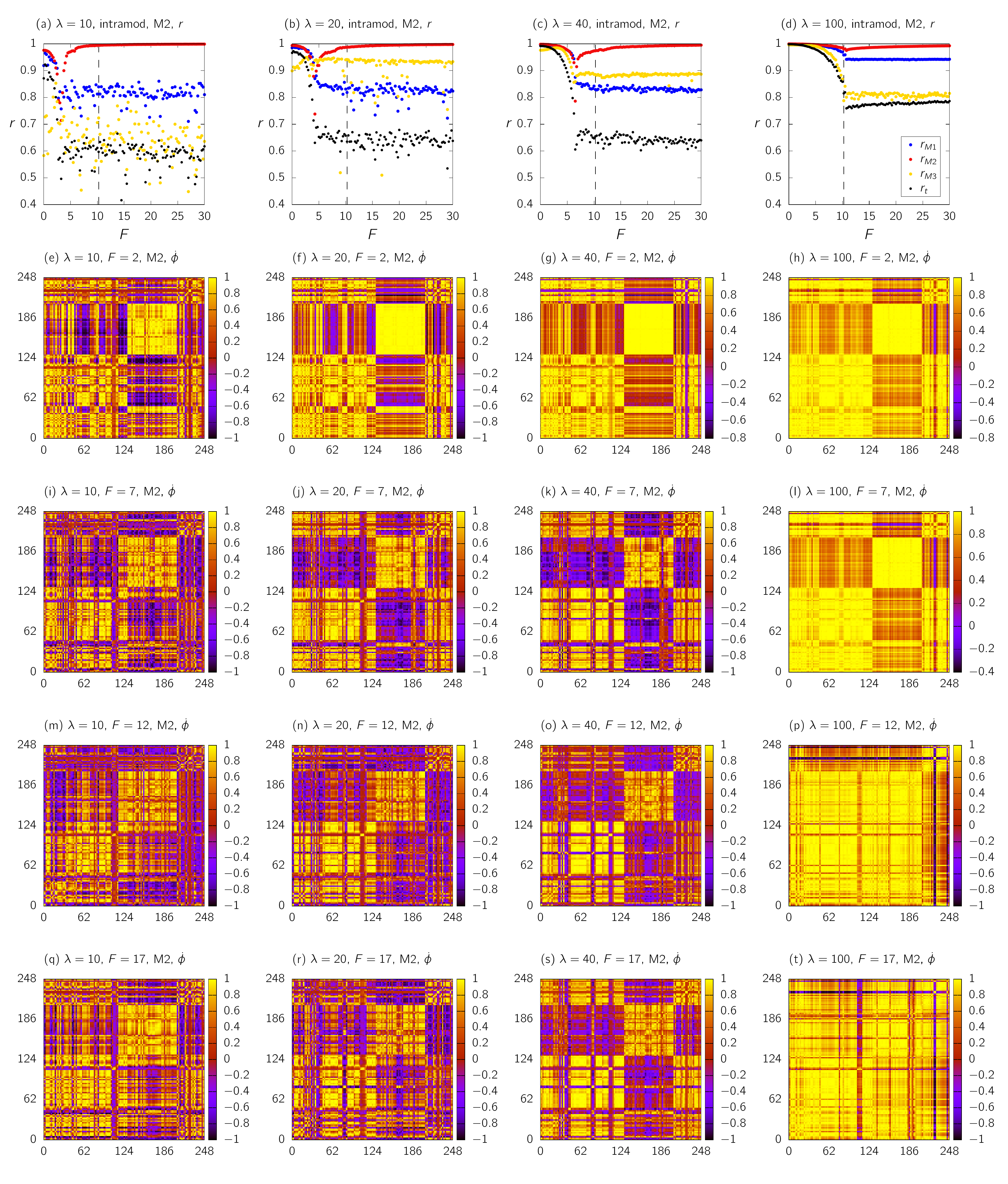}
\caption{ Panels (a)-(d): the global and local order parameters as a function of the external force $F$ acting on
neurons of $M_2$ for $\lambda$ fixed. The dashed lines indicate the critical force, $F_{c,theo}^{M_2} = 10.26$.
Panels (e)-(t): the velocity-velocity correlation matrix 248 $\times$ 248 obtained
using Eq. (\ref{Fcrit}). In each panel, the fixed parameters $\lambda$ and $F$ are indicated. The $M_1$ neurons are 
indexed by 1 to 130, the $M_2$ neurons by 131 to 207 and the $M_3$ neurons are indexed by 208 to 248.}
\label{corr-M2-r}
\end{figure}

\begin{figure}[H]
\center
\includegraphics[scale=0.54]{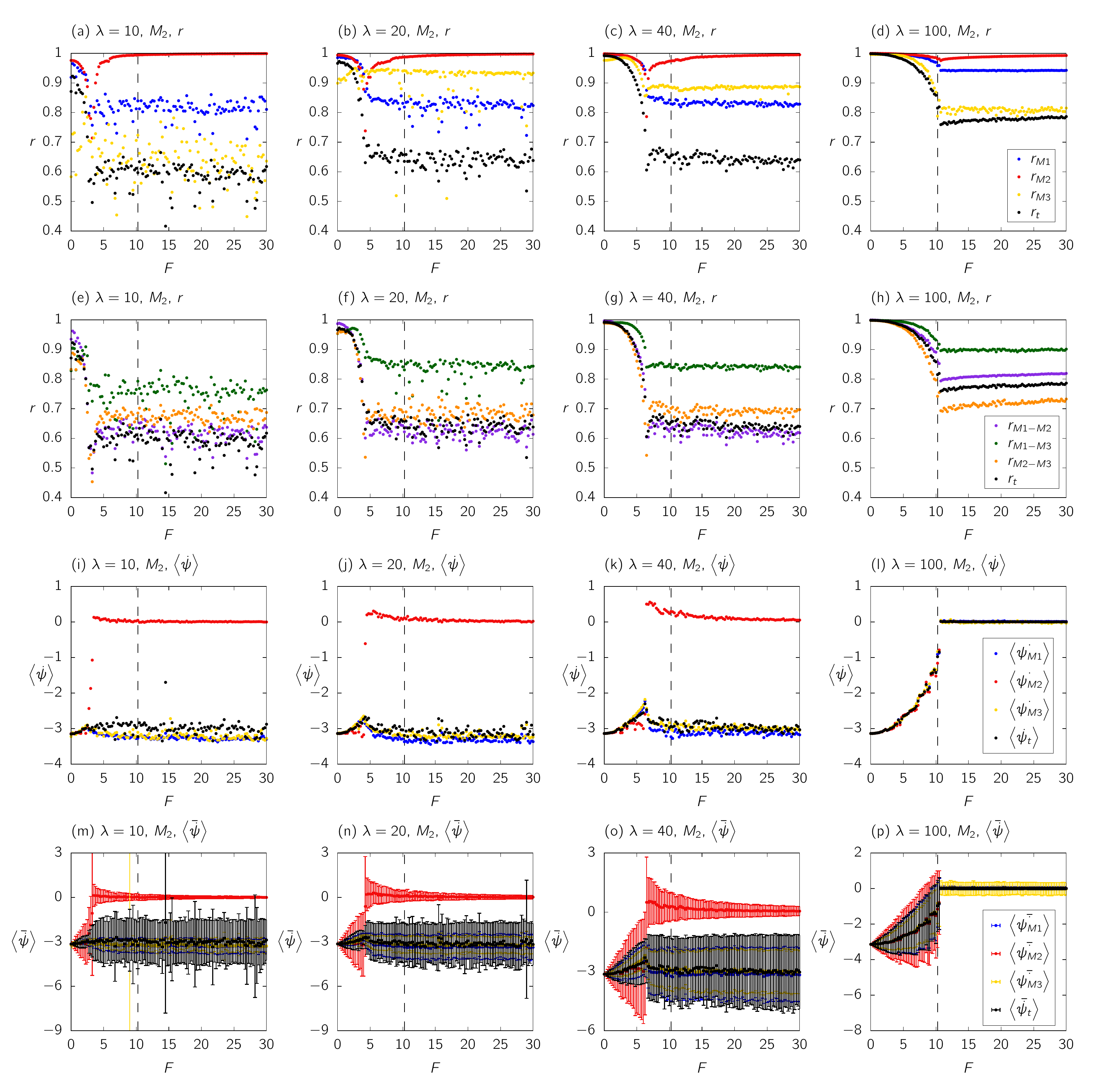}
\caption{Order parameters as a function of $F$ for the \textit{C. elegans} neural network where the external force
acts only on $M_2$. Panels (a) to (d) are the same as in the main text and indicate the order parameter $r$ for
intra-modules. Panels (e) to (h) indicate $r$ for inter-module. The results for $\dot\psi$ are exhibited in panels
(i) to (p), with respective error bars, panels (m) to (p), for intra-module case. The coupling $\lambda$ is fixed 
and its value is indicated in each panel. We also compute $r_{total}$ and $\dot\psi_{total}$ for whole network. 
The dashed lines indicate the critical force, $F_{c,theo}^{M_2} = 10.26$.}
\label{M2SM}
\end{figure}

\begin{figure}[H]
\center
\includegraphics[scale=0.46]{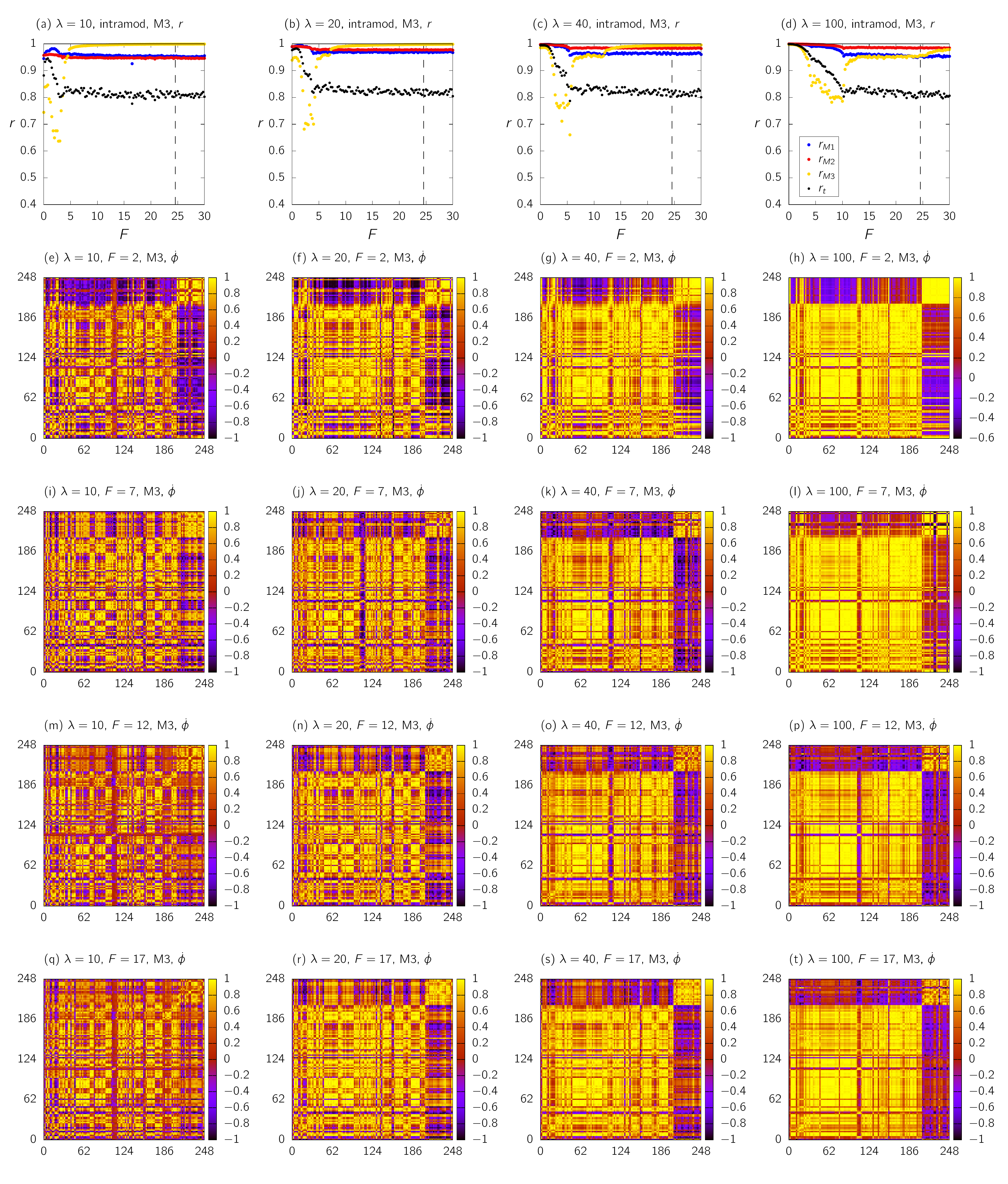}
\caption{ Panels (a)-(d): the global and local order parameters as a function of the external force $F$ acting on
neurons of $M_3$ for $\lambda$ fixed. The dashed lines indicate the critical force, $F_{c,theo}^{M_3} = 24.56$.
Panels (e)-(t): the velocity-velocity correlation matrix 248 $\times$ 248 obtained
using Eq. (\ref{Fcrit}). In each panel, the fixed parameters $\lambda$ and $F$ are indicated. The $M_1$ neurons are 
indexed by 1 to 130, the $M_2$ neurons by 131 to 207 and the $M_3$ neurons are indexed by 208 to 248.}
\label{corr-M3-r}
\end{figure}

\begin{figure}[H]
\center
\includegraphics[scale=0.54]{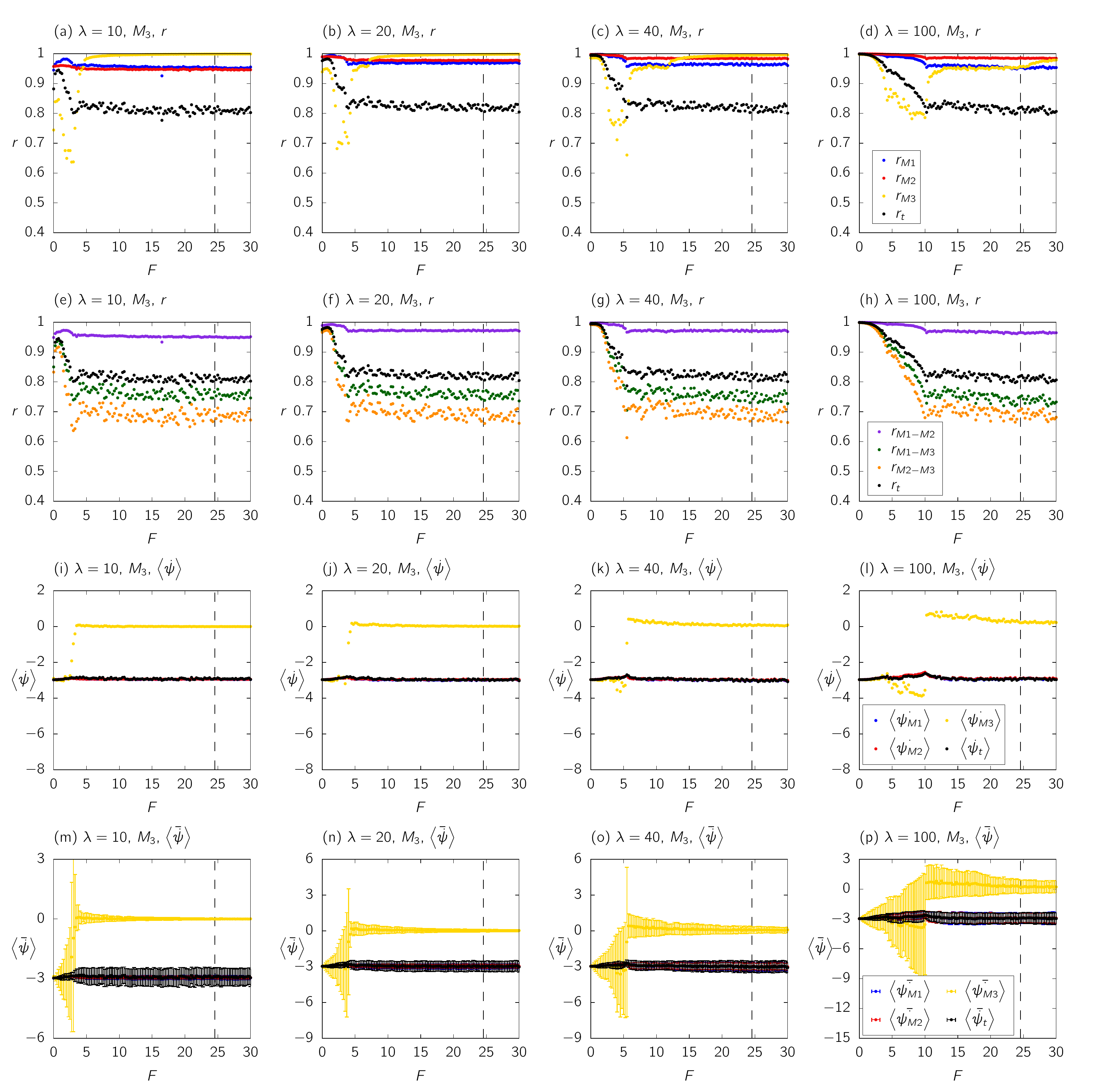}
\caption{Order parameters as a function of $F$ for the \textit{C. elegans} neural network where the external force
acts only on $M_3$. Panels (a) to (d) are the same as in the main text and indicate the order parameter $r$ for 
intra-modules. Panels (e) to (h) indicate $r$ for inter-module. The results for $\dot\psi$ are exhibited in panels 
(i) to (p), with respective error bars, panels (m) to (p), for intra-module case. The coupling $\lambda$ is fixed
and its value is indicated in each panel. We also compute $r_{total}$ and $\dot\psi_{total}$ for whole network. 
The dashed lines indicate the critical force, $F_{c,theo}^{M_3} = 24.56$.}
\label{M3SM}
\end{figure}

\begin{figure}[H]
\center
\includegraphics[scale=0.54]{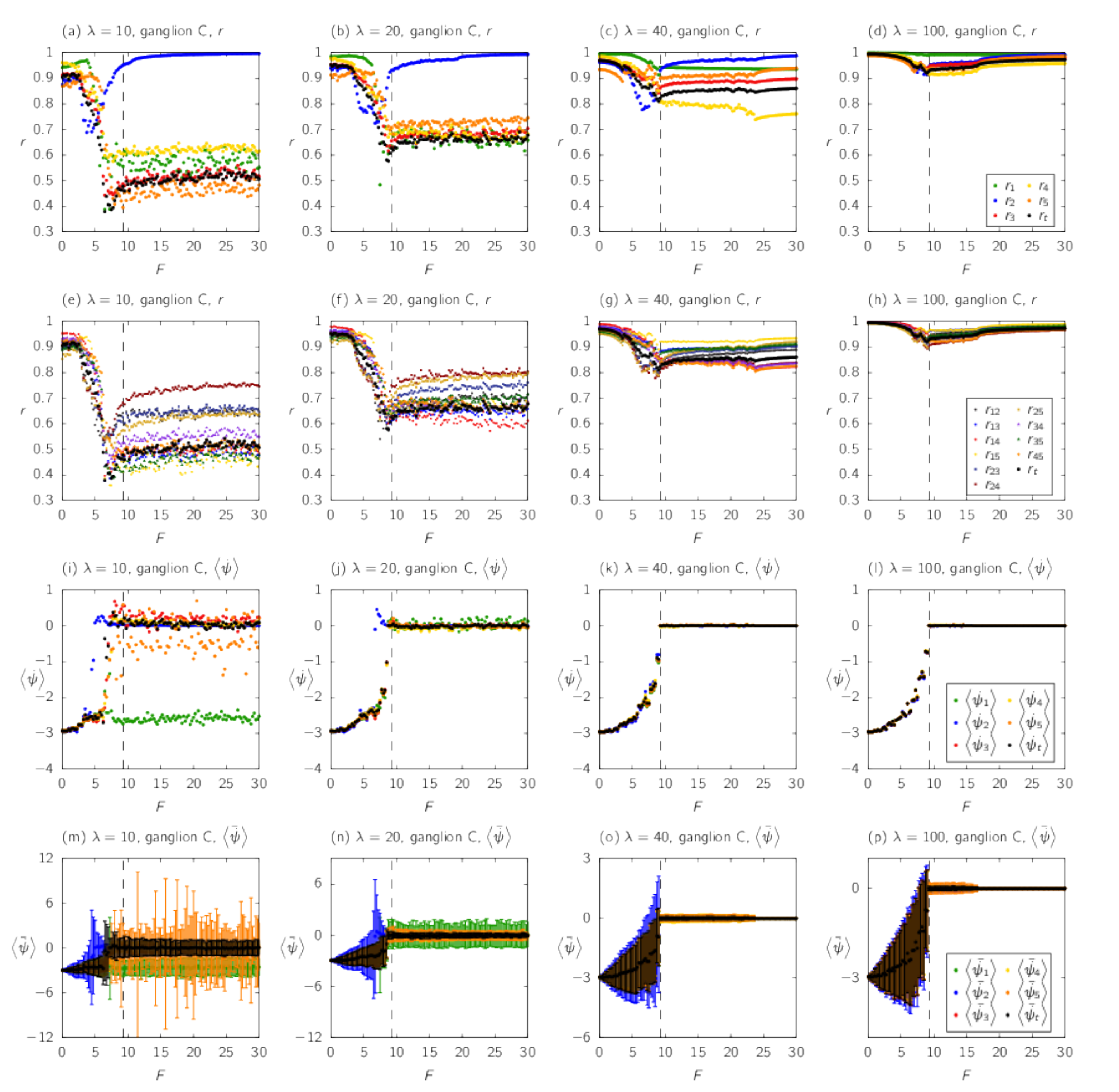}
\caption{Order parameters as a function of $F$ for the \textit{C. elegans} neural network where the external force
acts only on ganglion C. Panels (a) to (d) are the same as in the main text and indicate the order parameter $r$ for
intra-modules. Panels (e) to (h) indicate $r$ for inter-module. The results for $\dot\psi$ are exhibited in panels
(i) to (p), with respective error bars, panels (m) to (p), for intra-module case. The coupling $\lambda$ is fixed
and its value is indicated in each panel. We also compute $r_{total}$ and $\dot\psi_{total}$ for whole network. 
The dashed lines indicate the critical force, $F_{c,theo}^{C} = 9.32$.
The indexes refer to grouping, 1 for (\{A,B\}), 2 for (\{C\}), 3 for (\{D, E, F\}), 4  for (\{G\}) and 5 for (\{H, J, K\}).}
\label{CSM}
\end{figure}

\begin{figure}[H]
\center
\includegraphics[scale=0.46]{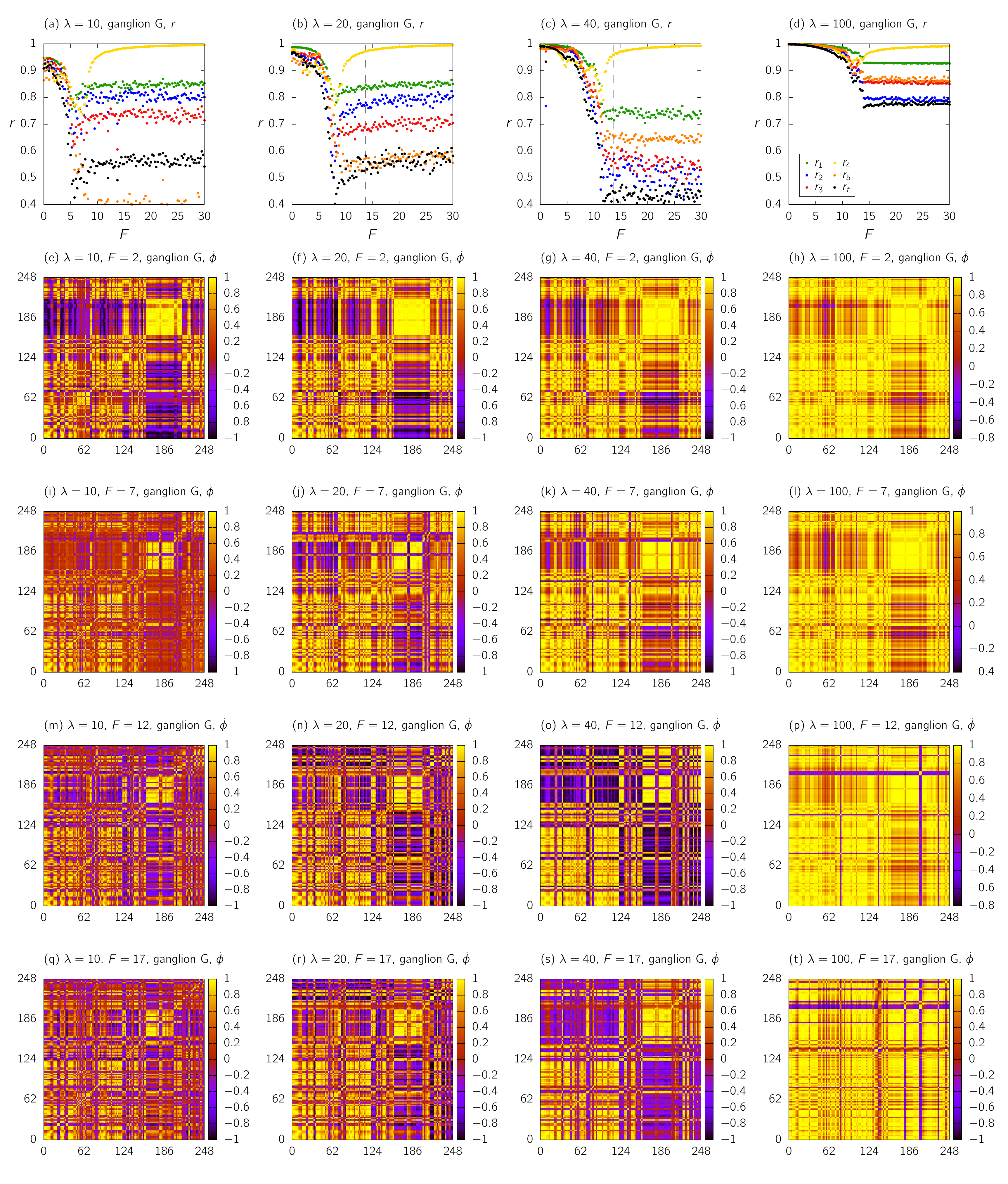}
\caption{ Panels (a)-(d): the local order parameters as a function of the external force
$F$ acting on ganglion G for $\lambda$ fixed. We also compute $r_{total}$ for whole network. 
The dashed lines indicate the critical force, $F_{c,theo}^{G} = 13.67$. Panels (e)-(t): the velocity-velocity correlation matrix 
248 $\times$ 248 obtained using Eq. (\ref{corr-v-v}). In each panel, the fixed parameters $\lambda$ and $F$ are indicated. 
The group 1 (\{A,B\}) are indexed by 1 to 36, group 2 (\{C\}) 
by 37 to 92, group 3 (\{D, E, F\}) by 93 to 159, group 4 (\{G\}, only MN) by 160 to 215 and group 5 (\{H, J, K\}) are indexed by 216 to 248.} 
\label{GcorrSM}
\end{figure}

\begin{figure}[H]
\center
\includegraphics[scale=0.54]{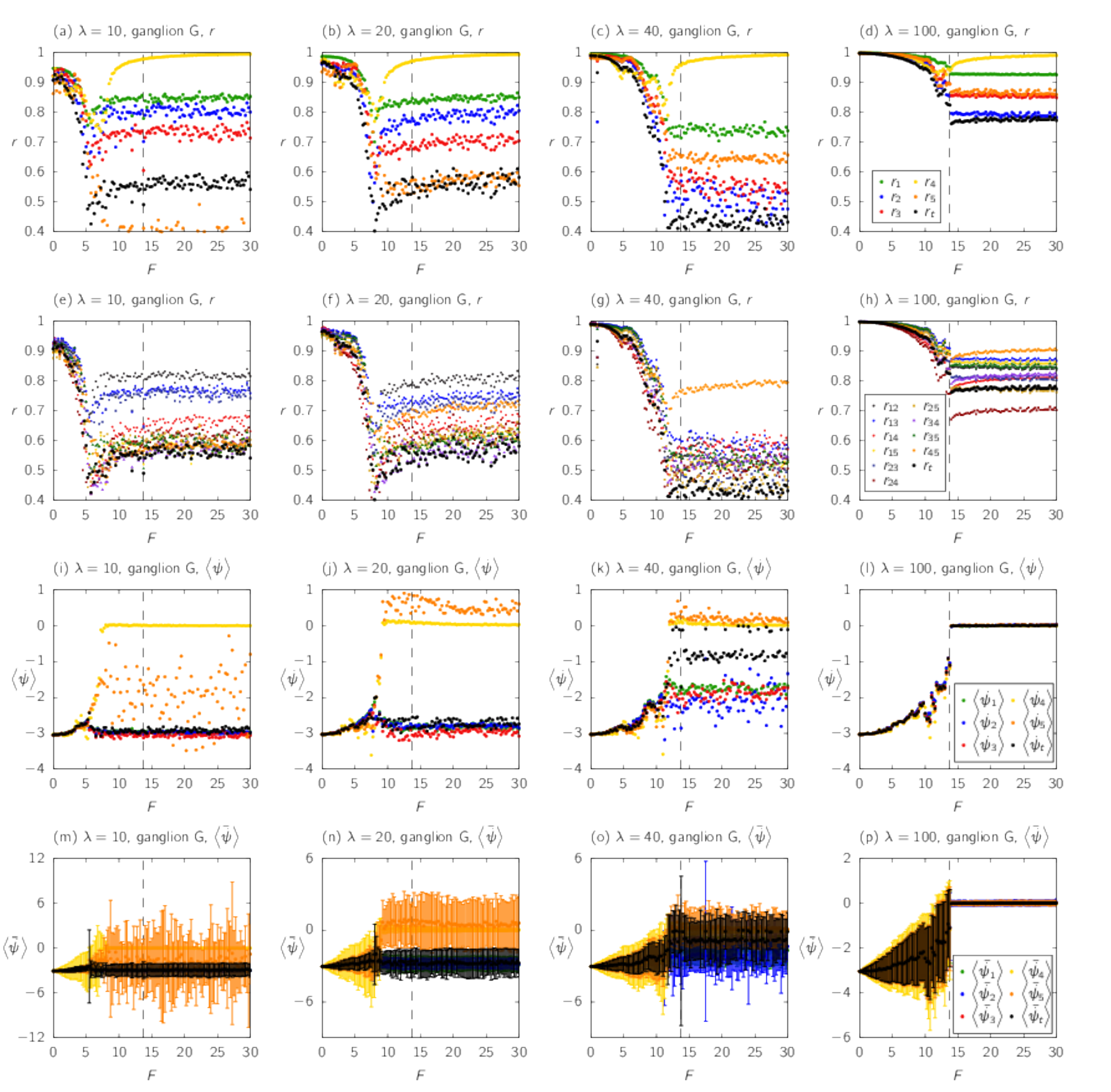}
\caption{Order parameters as a function of $F$ for the \textit{C. elegans} neural network where the external force
acts only on ganglion G. Panels (a) to (d) are the same as in the main text and indicate the order parameter $r$ for 
intra-modules. Panels (e) to (h) indicate $r$ for inter-module. The results for $\dot\psi$ are exhibited in panels 
(i) to (p), with respective error bars, panels (m) to (p), for intra-module case. The coupling $\lambda$ is fixed
and its value is indicated in each panel. We also compute $r_{total}$ and $\dot\psi_{total}$ for whole network. 
The dashed lines indicate the critical force, $F_{c,theo}^{G} = 13.67$.
The indexes refer to grouping, 1 for (\{A,B\}), 2 for (\{C\}), 3 for (\{D, E, F\}), 4  for (\{G\}) and 5 for (\{H, J, K\}).}
\label{GSM}
\end{figure}

\begin{figure}[H]
\center
\includegraphics[scale=0.54]{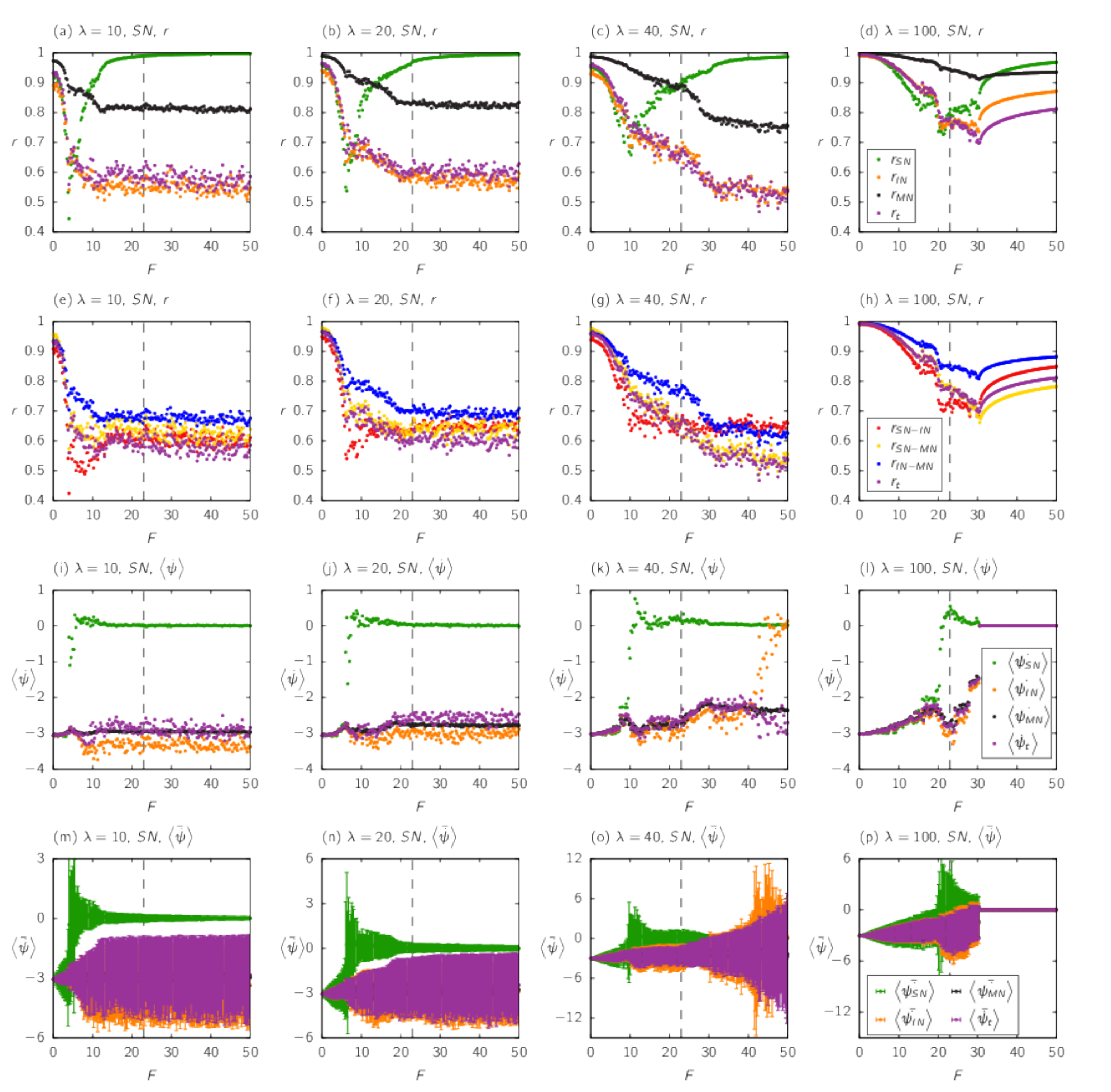}
\caption{Order parameters as a function of $F$ for the \textit{C. elegans} neural network where the external force 
acts only on sensory neurons. Panels (a) to (d) are the same as in the main text and indicate the order parameter $r$ for
intra-modules. Panels (e) to (h) indicate $r$ for inter-module. The results for $\dot\psi$ are exhibited in panels 
(i) to (p), with respective error bars, panels (m) to (p), for intra-module case. The coupling $\lambda$ is fixed 
and its value is indicated in each panel. We also compute $r_{total}$ and $\dot\psi_{total}$ for whole network. 
The dashed lines indicate the critical force, $F_{c,theo}^{SN} = 22.96$.}
\label{SNSM}
\end{figure}

\begin{figure}[H]
\center
\includegraphics[scale=0.46]{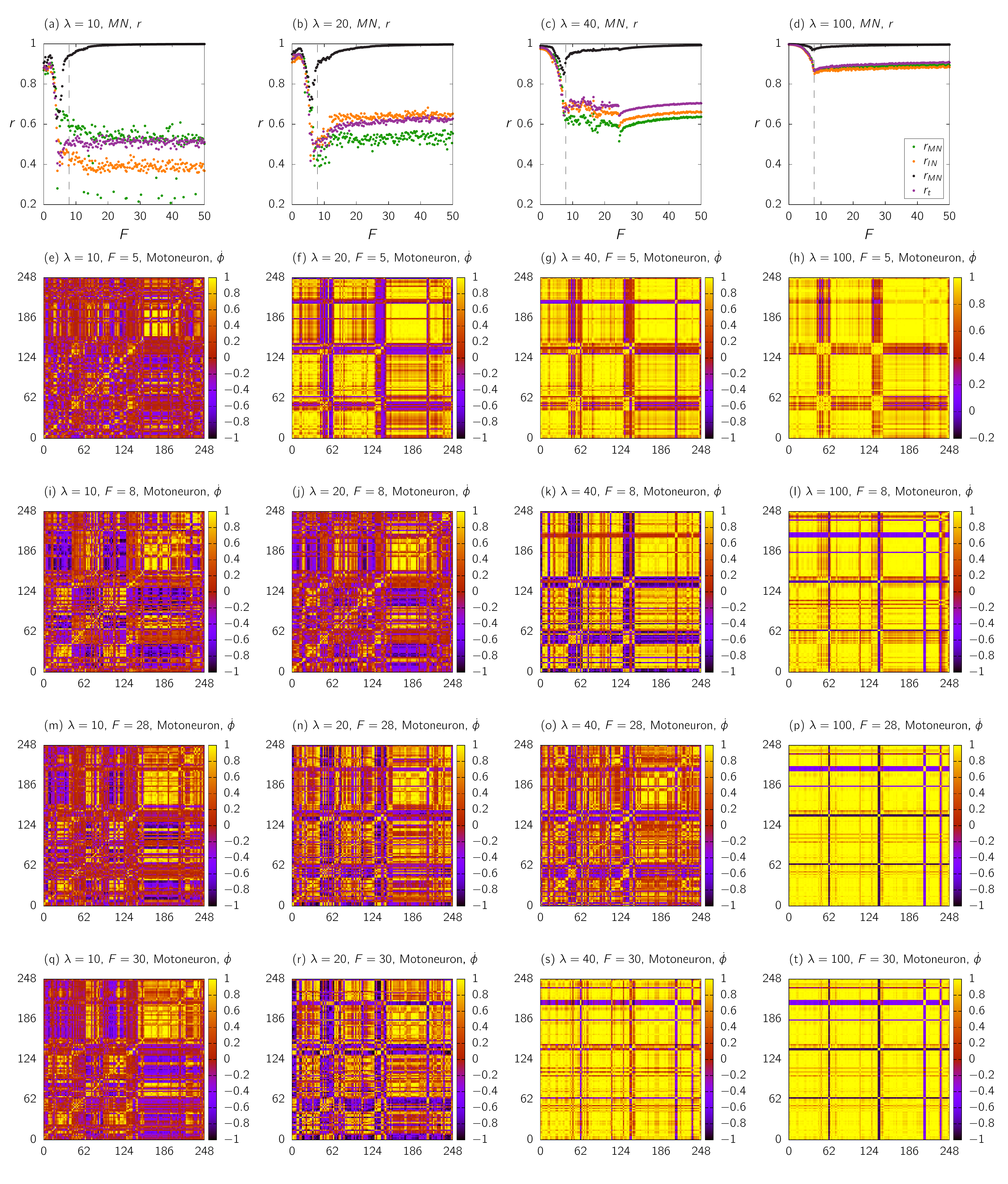}
\caption{ Panels (a)-(d): the local order parameters as a function of the external force
$F$ acting on motoneurons for $\lambda$ fixed. We also compute $r_{total}$ for whole network. 
The dashed lines indicate the critical force, $F_{c,theo}^{MN} = 7.87$. Panels (e)-(t): the velocity-velocity correlation matrix 
248 $\times$ 248 obtained using Eq. (\ref{corr-v-v}). In each panel, the fixed parameters $\lambda$ and $F$ are indicated. 
The sensory neurons (SN) are indexed by 1 to 65, the interneurons (IN) by 66 to 147 and the motoneurons (MN) are indexed by 148 to 248.} 
\label{corr-MN}
\end{figure}

\begin{figure}[H]
\center
\includegraphics[scale=0.54]{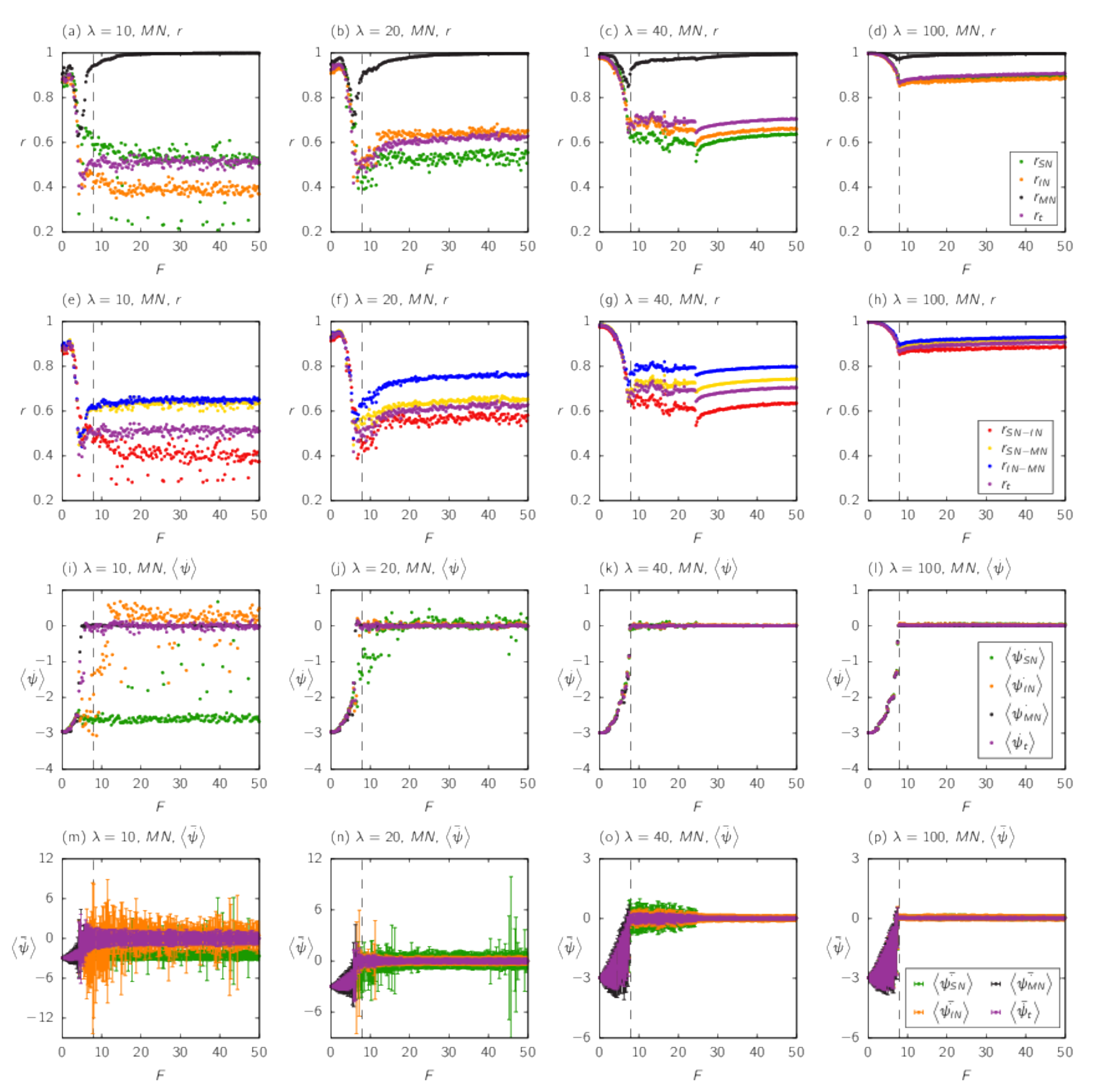}
\caption{Order parameters as a function of $F$ for the \textit{C. elegans} neural network where the external force 
acts only on motoneurons. Panels (a) to (d) are the same as in the main text and indicate the order parameter $r$ for
intra-modules. Panels (e) to (h) indicate $r$ for inter-module. The results for $\dot\psi$ are exhibited in panels 
(i) to (p), with respective error bars, panels (m) to (p), for intra-module case. The coupling $\lambda$ is fixed 
and its value is indicated in each panel. We also compute $r_{total}$ and $\dot\psi_{total}$ for whole network. 
The dashed lines indicate the critical force, $F_{c,theo}^{MN} = 7.87$. }
\label{MNSM}
\end{figure}

This supplementary material contains more detailed data from numerical simulations analyzed on main text and also has additional results of the stimulus application on different subsets of neurons. The material is organized as follows: figures \ref{M1SM}, \ref{M2SM} and \ref{M3SM} provide the curves of order parameters $r$ and $\dot\psi$ for intra and inter-modules and for the whole network when the driving force is applied to modules $M_1$, $M_2$ and $M_3$, respectively; figures \ref{CSM} and \ref{GSM} contains the same parameters when ganglia C and G are forced, respectively, and figures \ref{SNSM} and \ref{MNSM} provide information of forcing the SN and MN, respectively. Figures \ref{corr-M2-r}, \ref{corr-M3-r}, \ref{GcorrSM} and \ref{corr-MN} show the velocity correlation matrix for subsets $M_2$, $M_3$, ganglion G and MN, respectively.

The appendix \ref{app2} contains a list of the 248 neurons we considered in the electric junction network. Neurons are classified according to each class used in the paper: anatomical, functional and topological with 3, 5 and 10 modules.

We also show the values of modularity calculated for each subset (table \ref{table-modularity}) considering the unweighted and weighted networks. We note that the largest value is $Q_w = 0.47$ which was found by the modularization procedure. Table \ref{critical-forces} contains all values of fraction, average degree and critical force calculated by equation (\ref{Fcrit}) for each subset of each division and for the whole network. The values of critical force are highlighted as dashed lines on figures.

Tables \ref{table-m1}, \ref{table-m2} and \ref{table-m3} provide more detailed data of order parameters when forcing the topological modules, $M_1$, $M_2$ and $M_3$, respectively, when $F=50$. Analysing these values together with figures \ref{M1SM}, \ref{M2SM} and \ref{M3SM} we can see that none of them exhibit global synchronization, although each forced module has $r \approx 1.0$ and $\dot\psi \approx 0.0$.

A similar behavior occurs when ganglia C and G receive the driving force. On tables \ref{table-c} and \ref{table-g} (and figures \ref{CSM}  and \ref{GSM}) we can see that intra-module order parameters of forced modules are $r_{2-2} = r_{4-4} \approx 1.0$ and  $\dot\psi_{2-2} = \dot\psi_{4-4} \approx 0.0$, although the global order parameters are not, except for the case $\lambda=100$ when forcing ganglion C, which is the only case where global synchronization occurs, probably because of the relevance of C on receiving external input.

Finally, tables \ref{table-sn} and \ref{table-mn} bring the same parameters when sensory neurons and motoneurons receive the external force, respectively. Together with figures \ref{SNSM} and \ref{MNSM}, we do not observe global synchronization, although partial synchronization occurs only for larger internal coupling ($\lambda = 100$). Forcing all motoneurons have a similar effect when applying the stimulus on ganglion G, which is composed entirely of motoneurons.

We also provide more detailed information about the modularization procedure. In section \ref{sp-2} we described the steps used on ModuLand to generate three different modular networks and then we compare these structures with each other.

\newpage 
\subsection{Results}
\label{sp-1}

\begin{table}[H]
\center
\caption{Number of modules and modularity coefficient for unweighted and weighted adjacency matrices for each division of the network.}
\begin{tabular}{|c|c|c|c|c|c|}
\hline
Division        & \begin{tabular}[c]{@{}c@{}}Topology \\($M_1$, $M_2$, $M_3$)\end{tabular} & \begin{tabular}[c]{@{}c@{}}Anatomy \\ (H, MB, T)\end{tabular} & \begin{tabular}[c]{@{}c@{}}Ganglion (A, B, C, D, \\ E, F, G, H, J, K)\end{tabular} & \begin{tabular}[c]{@{}c@{}}Ganglion \\ (grouped)\end{tabular} & \begin{tabular}[c]{@{}c@{}}Functional \\ (SN, IN, MN)\end{tabular} \\ \hline
\# of modules & 3                                                                        & 3                                                             & 10                                                                                 & 5                                                                                                      & 3                                                                  \\ \hline
$Q$           & 0.44                                                                     & 0.20                                                          & 0.20                                                                               & 0.21                                                                                                   & 0.08                                                               \\ \hline
$Q_w$         & 0.47                                                                     & 0.15                                                          & 0.17                                                                               & 0.18                                                                                                   & 0.04                                                               \\ \hline
\end{tabular}
\label{table-modularity}
\end{table}

\begin{table}[H]
\center
\caption{Fraction, average degree and critical force, calculated by equation (\ref{Fcrit}) for each subset of each division and for the whole network. The values of critical force are highlighted as dashed lines on figures.}
\begin{tabular}{|c|c|c|c|}
\hline
Subset        & Fraction & Average degree & Critical force \\ \hline
M1            & 0.52     & 7.96           & 5.12           \\ \hline
M2            & 0.31     & 6.71           & 10.26          \\ \hline
M3            & 0.17     & 5.27           & 24.56          \\ \hline \hline
\{A, B\}      & 0.14     & 3.14           & 46.94          \\ \hline
\{C\}         & 0.23     & 10.16          & 9.32           \\ \hline
\{D, E, F\}   & 0.27     & 6.82           & 11.61          \\ \hline
\{G\}         & 0.23     & 6.93           & 13.67          \\ \hline
\{H, J, K\}   & 0.13     & 7.30           & 22.01          \\ \hline \hline
SN            & 0.26     & 3.55           & 22.96          \\ \hline
IN            & 0.33     & 10.52          & 6.15           \\ \hline
MN            & 0.41     & 6.67           & 7.87           \\ \hline \hline
Whole network & 1.00     & 7.13           & 3.00           \\ \hline
\end{tabular}
\label{critical-forces}
\end{table}

\begin{table}[H]
\center
\caption{Order parameters $r$ and $\dot\psi$ for intra-module and for the whole network for each $\lambda$ when $M_1$ is stimulated. The values of $r_{total}$ and $\dot\psi_{total}$ indicate if occurs global synchronization. All values were taken for $F=50$.}
\begin{tabular}{|c||c|c||c|c||c|c||c|c|c|}
\hline
$\lambda$ & $r_{M_1-M_1}$ & $\dot\psi_{M_1-M_1}$ & $r_{M_2-M_2}$ & $\dot\psi_{M_2-M_2}$ & $r_{M_3-M_3}$ & $\dot\psi_{M_3-M_3}$ & $r_{total}$ & $\dot\psi_{total}$ & Global sync \\ \hline
10        & 1.00          & 0.01                 & 0.86          & -3.06                & 0.67          & -2.92                & 0.55        & 0.07               & no          \\ \hline
20        & 1.00          & 0.00                 & 0.60          & -2.59                & 0.80          & -2.35                & 0.54        & -0.01               & no          \\ \hline
40        & 1.00          & 0.00                 & 0.93          & 0.00                 & 0.66          & -0.95                & 0.65        & 0.05               & no          \\ \hline
100       & 1.00          & 0.00                 & 0.99          & 0.00                 & 0.83          & 0.21                 & 0.91        & 0.02               & partial          \\ \hline
\end{tabular}
\label{table-m1}
\end{table}

\begin{table}[H]
\center
\caption{Order parameters $r$ and $\dot\psi$ for intra-module and for the whole network for each $\lambda$ when $M_2$ is stimulated. 
The values of $r_{total}$ and $\dot\psi_{total}$ indicate if occurs global synchronization. All values were taken for $F=50$.}
\begin{tabular}{|c||c|c||c|c||c|c||c|c|c|}
\hline
$\lambda$ & $r_{M_1-M_1}$ & $\dot\psi_{M_1-M_1}$ & $r_{M_2-M_2}$ & $\dot\psi_{M_2-M_2}$ & $r_{M_3-M_3}$ & $\dot\psi_{M_3-M_3}$ & $r_{total}$ & $\dot\psi_{total}$ & Global sync \\ \hline
10        & 0.82          & -3.23                & 1.00          & 0.00                 & 0.58          & -3.22                & 0.59        & -2.99              & no          \\ \hline
20        & 0.82          & -3.33                & 1.00          & 0.00                 & 0.92          & -3.25                & 0.65        & -3.04              & no          \\ \hline
40        & 0.83          & -3.11                & 1.00          & 0.10                 & 0.89          & -2.99                & 0.63        & -3.09              & no          \\ \hline
100       & 0.94          & 0.01                 & 1.00          & 0.00                 & 0.81          & -0.03                & 0.79        & 0.00               & no          \\ \hline
\end{tabular}
\label{table-m2}
\end{table}

\begin{table}[H]
\center
\caption{Order parameters $r$ and $\dot\psi$ for intra-module and for the whole network for each $\lambda$ when $M_3$ is stimulated. 
The values of $r_{total}$ and $\dot\psi_{total}$ indicate if occurs global synchronization. All values were taken for $F=50$.}
\begin{tabular}{|c||c|c||c|c||c|c||c|c|c|}
\hline
$\lambda$ & $r_{M_1-M_1}$ & $\dot\psi_{M_1-M_1}$ & $r_{M_2-M_2}$ & $\dot\psi_{M_2-M_2}$ & $r_{M_3-M_3}$ & $\dot\psi_{M_3-M_3}$ & $r_{total}$ & $\dot\psi_{total}$ & Global sync \\ \hline
10        & 0.95          & -2.95                & 0.95          & -2.94                & 1.00          & 0.00                 & 0.80        & -2.96              & no          \\ \hline
20        & 0.97          & -2.96                & 0.98          & -2.95                & 1.00          & 0.01                 & 0.81        & -2.99              & no          \\ \hline
40        & 0.96          & -3.01                & 0.98          & -2.99                & 1.00          & 0.03                 & 0.82        & -3.00              & no          \\ \hline
100       & 0.95          & -2.92                & 0.98          & -2.90                & 0.99          & 0.13                 & 0.79        & -2.99              & no          \\ \hline
\end{tabular}
\label{table-m3}
\end{table}

\begin{table}[H]
\center
\caption{Order parameters $r$ and $\dot\psi$ for intra-module and for the whole network for each $\lambda$ when ganglion C is stimulated. 
The values of $r_{total}$ and $\dot\psi_{total}$ indicate if occurs global synchronization. All values were taken for $F=50$.
The indexes refer to grouping, 1 for (\{A,B\}), 2 for (\{C\}), 3 for (\{D, E, F\}), 4  for (\{G\}) and 5 for (\{H, J, K\}). GS refers to global sync.}
\begin{tabular}{|c||c|c||c|c||c|c||c|c||c|c||c|c|c|}
\hline
$\lambda$ & $r_{1-1}$ & $\dot\psi_{1-1}$ & $r_{2-2}$ & $\dot\psi_{2-2}$ & $r_{3-3}$ & $\dot\psi_{3-3}$ & $r_{4-4}$ & $\dot\psi_{4-4}$ & $r_{5-5}$ & $\dot\psi_{5-5}$ & $r_{total}$ & $\dot\psi_{total}$ & GS \\ \hline
10        & 0.59     & -2.40           & 1.00     & 0.00            & 0.54     & -0.06           & 0.62     & 0.03            & 0.51     & -0.47           & 0.52        & 0.03               & no          \\ \hline
20        & 0.65     & 0.14            & 1.00     & 0.00            & 0.69     & -0.04           & 0.66     & 0.01            & 0.76     & -0.08           & 0.67        & 0.00               & no          \\ \hline
40        & 0.93     & 0.00            & 1.00     & 0.00            & 0.91     & 0.00            & 0.77     & 0.00            & 0.95     & 0.00            & 0.87        & 0.00               & part.          \\ \hline
100       & 0.99     & 0.00            & 1.00     & 0.00            & 0.98     & 0.00            & 0.96     & 0.00            & 0.99     & 0.00            & 0.98        & 0.00               & yes         \\ \hline
\end{tabular}
\label{table-c}
\end{table}

\begin{table}[H]
\center
\caption{Order parameters $r$ and $\dot\psi$ for intra-module and for the whole network for each $\lambda$ when ganglion G is stimulated. 
The values of $r_{total}$ and $\dot\psi_{total}$ indicate if occurs global synchronization. All values were taken for $F=50$.
The indexes refer to grouping, 1 for (\{A,B\}), 2 for (\{C\}), 3 for (\{D, E, F\}), 4  for (\{G\}) and 5 for (\{H, J, K\}). GS refers to global sync.}
\begin{tabular}{|c||c|c||c|c||c|c||c|c||c|c||c|c|c|}
\hline
$\lambda$ & $r_{1-1}$ & $\dot\psi_{1-1}$ & $r_{2-2}$ & $\dot\psi_{2-2}$ & $r_{3-3}$ & $\dot\psi_{3-3}$ & $r_{4-4}$ & $\dot\psi_{4-4}$ & $r_{5-5}$ & $\dot\psi_{5-5}$ & $r_{total}$ & $\dot\psi_{total}$ & GS \\ \hline
10        & 0.86     & -3.00           & 0.81     & -2.97           & 0.74     & -3.03           & 1.00     & 0.00            & 0.39     & -2.63           & 0.58        & -2.85              & no          \\ \hline
20        & 0.86     & -2.81           & 0.80     & -2.76           & 0.71     & -2.90           & 1.00     & 0.01            & 0.60     & 0.21            & 0.58        & -2.71              & no          \\ \hline
40        & 0.74     & -1.67           & 0.52     & -1.48           & 0.57     & -1.75           & 1.00     & 0.00            & 0.67     & 0.08            & 0.47        & -0.84              & no          \\ \hline
100       & 0.93     & 0.01            & 0.79     & 0.05            & 0.85     & 0.02            & 1.00     & 0.00            & 0.86     & 0.03            & 0.78        & 0.02               & no         \\ \hline
\end{tabular}
\label{table-g}
\end{table}

\begin{table}[H]
\center
\caption{Order parameters $r$ and $\dot\psi$ for intra-module and for the whole network for each $\lambda$ when sensory neurons are stimulated. 
The values of $r_{total}$ and $\dot\psi_{total}$ indicate if occurs global synchronization. All values were taken for $F=50$.}
\begin{tabular}{|c||c|c||c|c||c|c||c|c|c|}
\hline
$\lambda$ & $r_{SN-SN}$ & $\dot\psi_{SN-SN}$ & $r_{IN-IN}$ & $\dot\psi_{IN-IN}$ & $r_{MN-MN}$ & $\dot\psi_{MN-MN}$ & $r_{total}$ & $\dot\psi_{total}$ & Global sync \\ \hline
10        & 1.00          & 0.00                 & 0.55          & -3.37                & 0.81          & -2.97                & 0.59        & -2.84              & no          \\ \hline
20        & 1.00          & -0.01                & 0.60          & -2.88                & 0.83          & -2.78                & 0.63        & -2.46              & no          \\ \hline
40        & 0.99          & 0.02                 & 0.54          & 0.14                 & 0.75          & -2.36                & 0.54        & -2.70              & no          \\ \hline
100       & 0.97          & 0.00                 & 0.87          & 0.00                 & 0.94          & 0.00                 & 0.81        &  0.00               & part.          \\ \hline
\end{tabular}
\label{table-sn}
\end{table}

\begin{table}[H]
\center
\caption{Order parameters $r$ and $\dot\psi$ for intra-module and for the whole network for each $\lambda$ when motoneurons are stimulated. 
The values of $r_{total}$ and $\dot\psi_{total}$ indicate if occurs global synchronization. All values were taken for $F=50$. GS refers to global sync.}

\begin{tabular}{|c||c|c||c|c||c|c||c|c|c|}
\hline
$\lambda$ & $r_{SN-SN}$ & $\dot\psi_{SN-SN}$ & $r_{IN-IN}$ & $\dot\psi_{IN-IN}$ & $r_{MN-MN}$ & $\dot\psi_{MN-MN}$ & $r_{total}$ & $\dot\psi_{total}$ & GS \\ \hline
10        & 0.50          & -2.67                 & 0.37          & 0.49                & 1.00          & 0.00                & 0.51        & -0.04              & no          \\ \hline
20        & 0.55          & 0.00                  & 0.65          & -0.01                & 1.00          & 0.00                & 0.63        & -0.01              & no          \\ \hline
40        & 0.64          & 0.01                 & 0.66          & 0.03                 & 0.99          & 0.00                & 0.70        & 0.01              & no          \\ \hline
100       & 0.89          & 0.05                 & 0.88          & 0.04                 & 1.00          & 0.00                 & 0.91        &  0.02               & part.          \\ \hline
\end{tabular}
\label{table-mn}
\end{table}

\newpage
\subsection{Modularization}
\label{sp-2}

The ModuLand plug-in is a tool that uses a hierarchical algorithm
to detect multiple layers of communities in networks, where nodes of the higher 
hierarchical layer are communities of the lower one. It consists of a  
family of algorithms based on the following four common steps:

1. calculation of influence functions, attributing a value for each node or link \footnote{For undirected networks, the NodeLand and LinkLand algorithms calculate the influence functions for a
given node or link, respectively.} based 
on its influence over the network. For example, a node that belongs to a module has larger
influence on the links within that module than on the links of the entire network;

2. construction of a community landscape by summing, for each link,  the influence function 
values over all nodes and links and plotting contour lines;

3. determination of the modules of the network, identifying them as the hills of the contour line plot;
 
4. determination of a hierarchy of higher level networks, where each module is considered as a single node on the next step.
 
The method also allows the
merging of some nodes before starting a new run to generate the next  hierarchical level. The choice of initial
parameters and the decision to join or not nodes before the next run changes the sequence of modules created.

We constructed three different modular networks using ModuLand and we found networks with 3, 5 and 10 modules 
with large values of $Q$. Table \ref{5-10-mod} summarizes the values of modularity for unweighted and weighted
adjacency matrices of each topological modularization.

To generate the network with 3 modules, we run the algorithm creating 36 modules on the first level. Next, we merge them
using a threshold of 0.9, as recommended by the authors \cite{moduland2}, reducing the
modules to 32. Then, we run the algorithm again, resulting in 3 modules. As
recommended \cite{moduland2}, in both runs we used the LinkLand algorithm, since the network is undirected. To
construct the network with 5 modules we slightly changed the initial conditions. We first run the LinkLand algorithm on
the first level obtaining 36 modules. Next, we choose the NodeLand algorithm and we obtained the 5 modules. The case of
10 modules was similar, but we started by running the algorithm using NodeLand, creating 36 modules, and then we apply
the same algorithm again, without joining nodes, obtaining 10 modules.

\begin{table}[H]
\center
\caption{Modularity coefficient for unweighted and weighted adjacency matrices for each topological modularization of the network.}
\begin{tabular}{c|c|c|c|}
\cline{2-4}
\multicolumn{1}{l|}{}       & 3 modules & 5 modules & 10 modules \\ \hline
\multicolumn{1}{|c|}{$Q$}   & 0.44      & 0.44      & 0.55       \\ \hline
\multicolumn{1}{|c|}{$Q_w$} & 0.47      & 0.48      & 0.60       \\ \hline
\end{tabular}
\label{5-10-mod}
\end{table}

\begin{figure}[H]
\center
\includegraphics[scale=0.3]{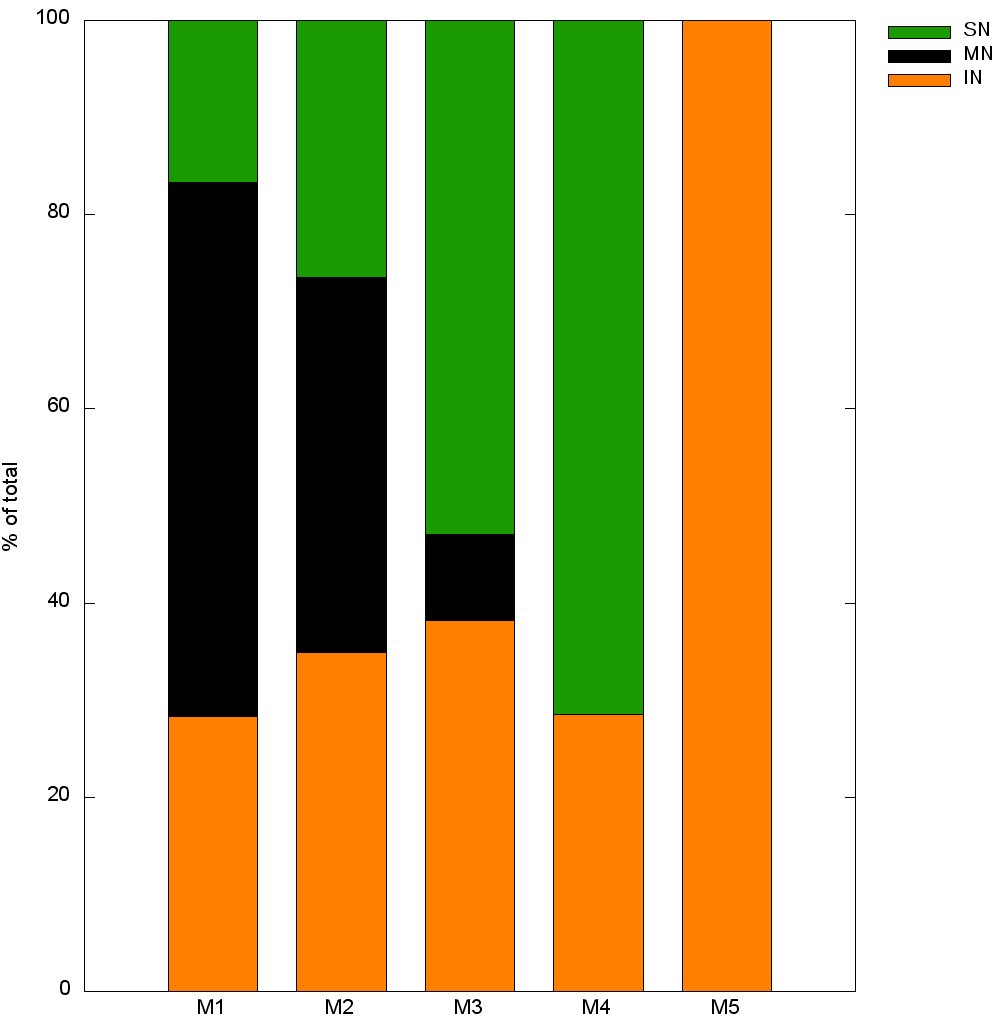}
\includegraphics[scale=0.3]{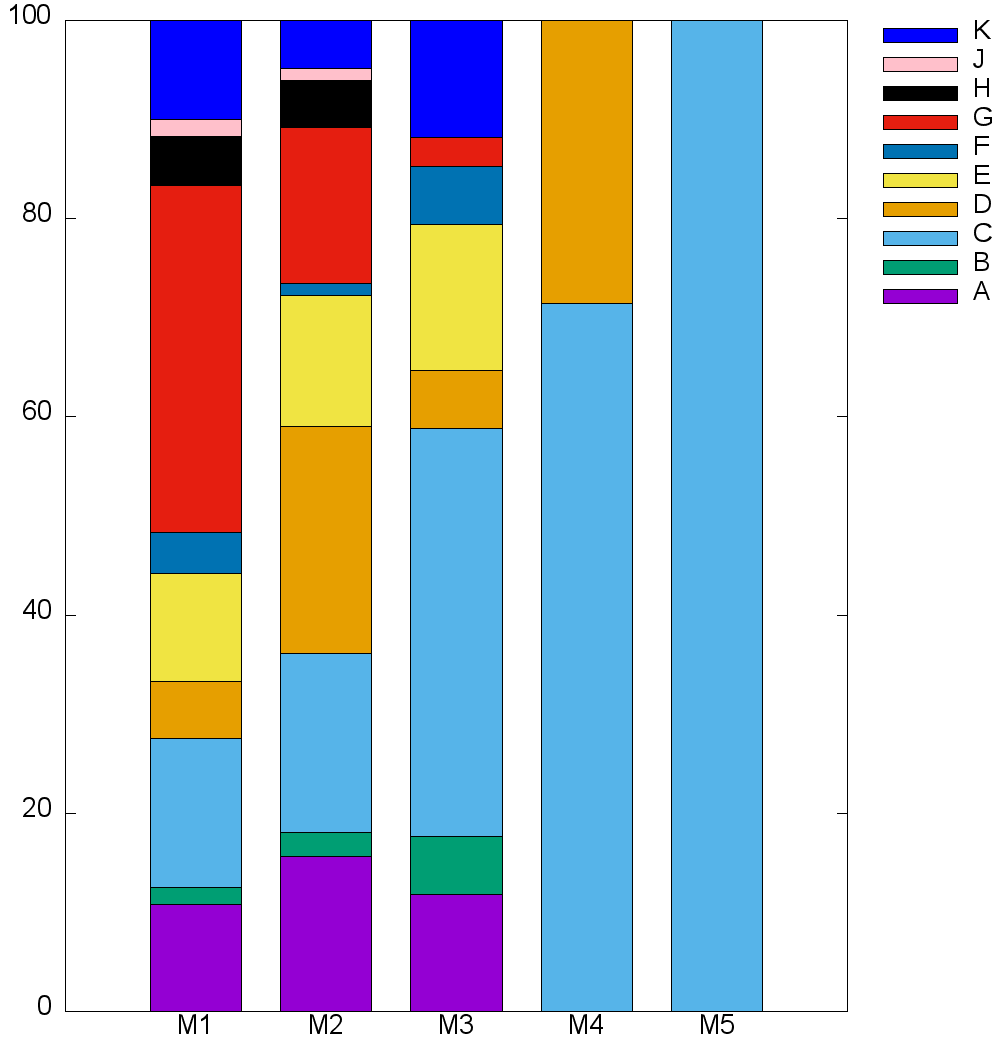}
\caption{ Histograms representing the fraction of (left) neuronal class (SN: sensory neuron, MN: motorneuron 
and IN: interneuron) and (right) of ganglia (A: anterior ganglion, B: dorsal ganglion, C: lateral ganglion, D: ventral ganglion, 
E: retrovesicular ganglion, F: posterolateral ganglion, G: ventral cord neuron group, H: pre-anal ganglion, J: dorsorectal 
ganglion, K: lumbar ganglion) for each module ($M_i$: module $i$, where $i = 1, 2, 3, 4, 5$).}
\label{hist5mod}
\end{figure}

\begin{figure}[H]
\center
\includegraphics[scale=0.3]{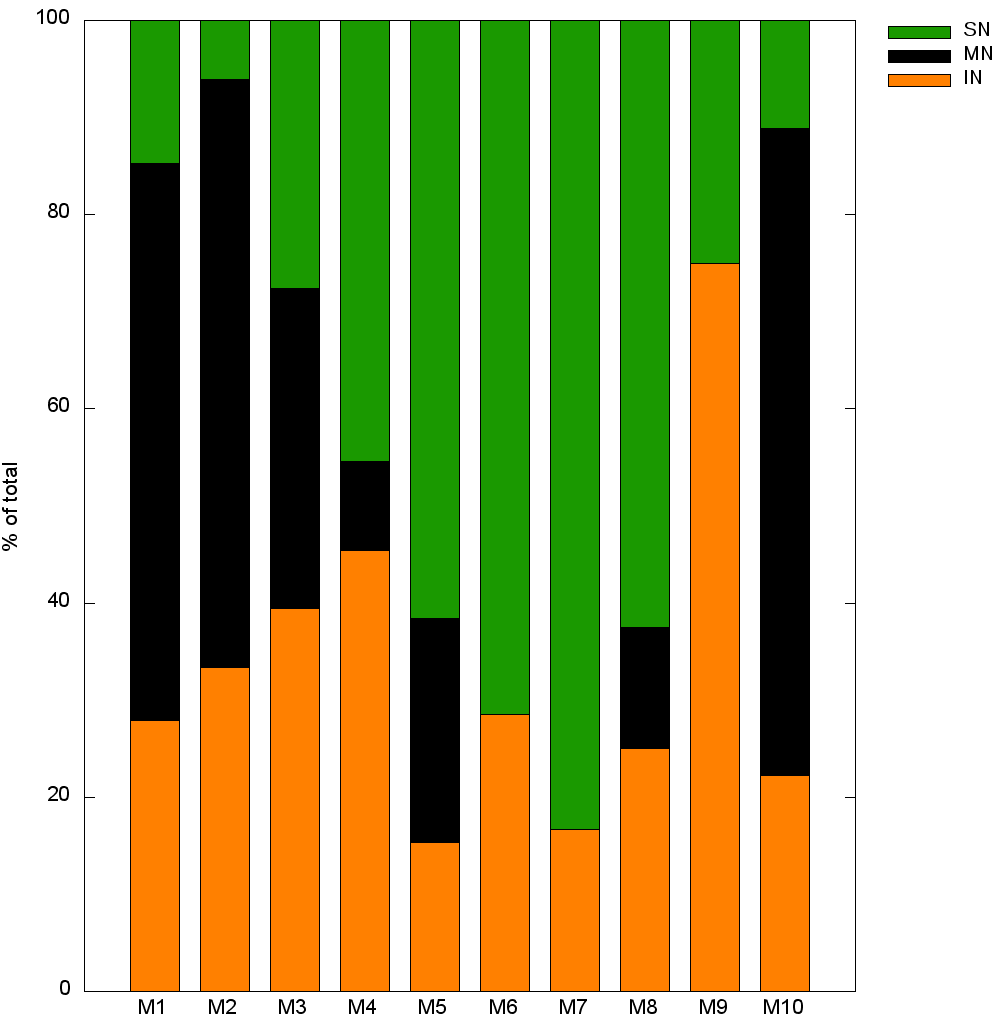}
\includegraphics[scale=0.3]{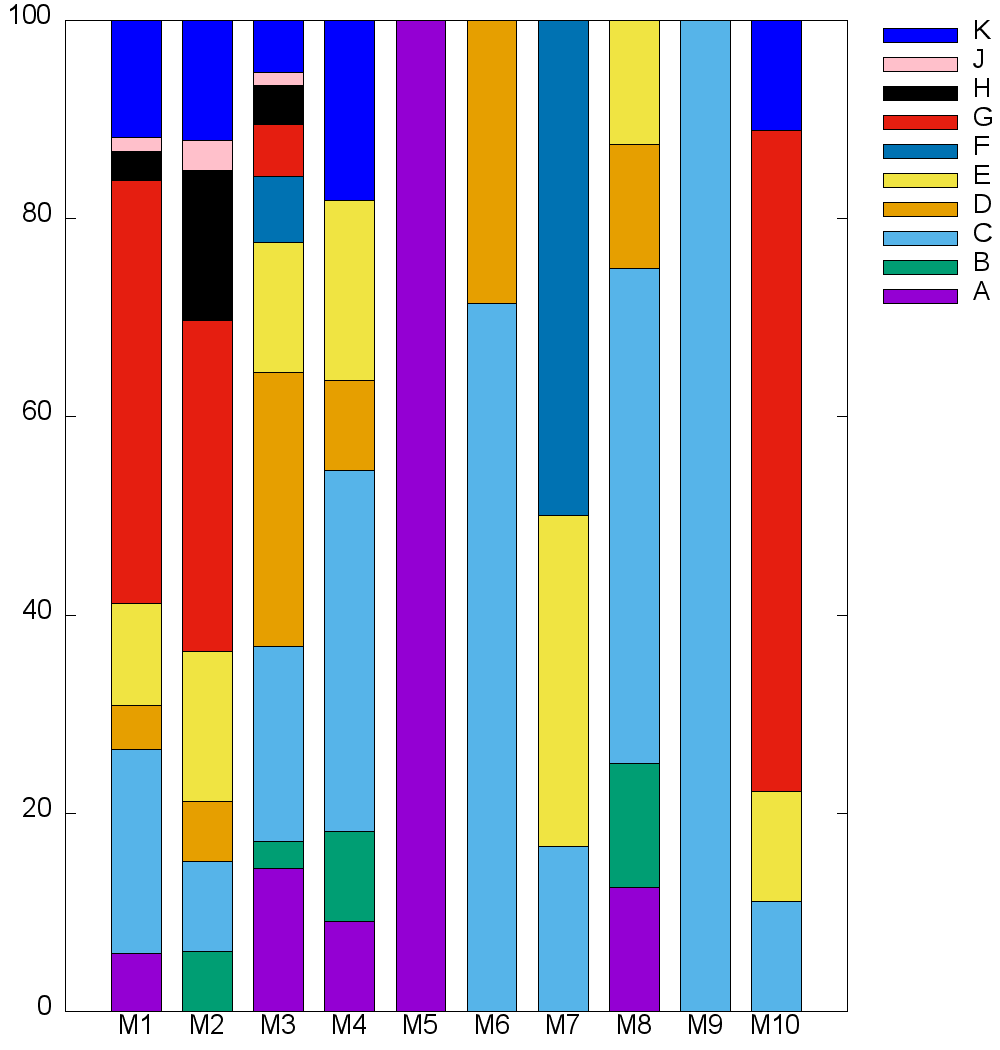}
\caption{ Histograms representing the fraction of (left) neuronal class (SN: sensory neuron, MN: motorneuron 
and IN: interneuron) and (right) of ganglia (A: anterior ganglion, B: dorsal ganglion, C: lateral ganglion, D: ventral ganglion, 
E: retrovesicular ganglion, F: posterolateral ganglion, G: ventral cord neuron group, H: pre-anal ganglion, J: dorsorectal 
ganglion, K: lumbar ganglion) for each module ($M_i$: module $i$, where $i = 1$ to 10).}
\label{hist10mod}
\end{figure}

The choice of modular partition is rather arbitrary, as long as the
modularity coefficient is not too low. We performed our analysis on the network with three 
modules to conform with the number of functional (SN, MN, IN) and anatomical 
(head, body, tail) neurons. We found that the topological modules do not separate these classifications 
(histograms of main text). We performed the same analysis for the other two topological 
divisions (figures \ref{hist5mod} and \ref{hist10mod}), and we found that these modular structures also mix the neurons.

\begin{figure}[H]
\center
\includegraphics[scale=0.9]{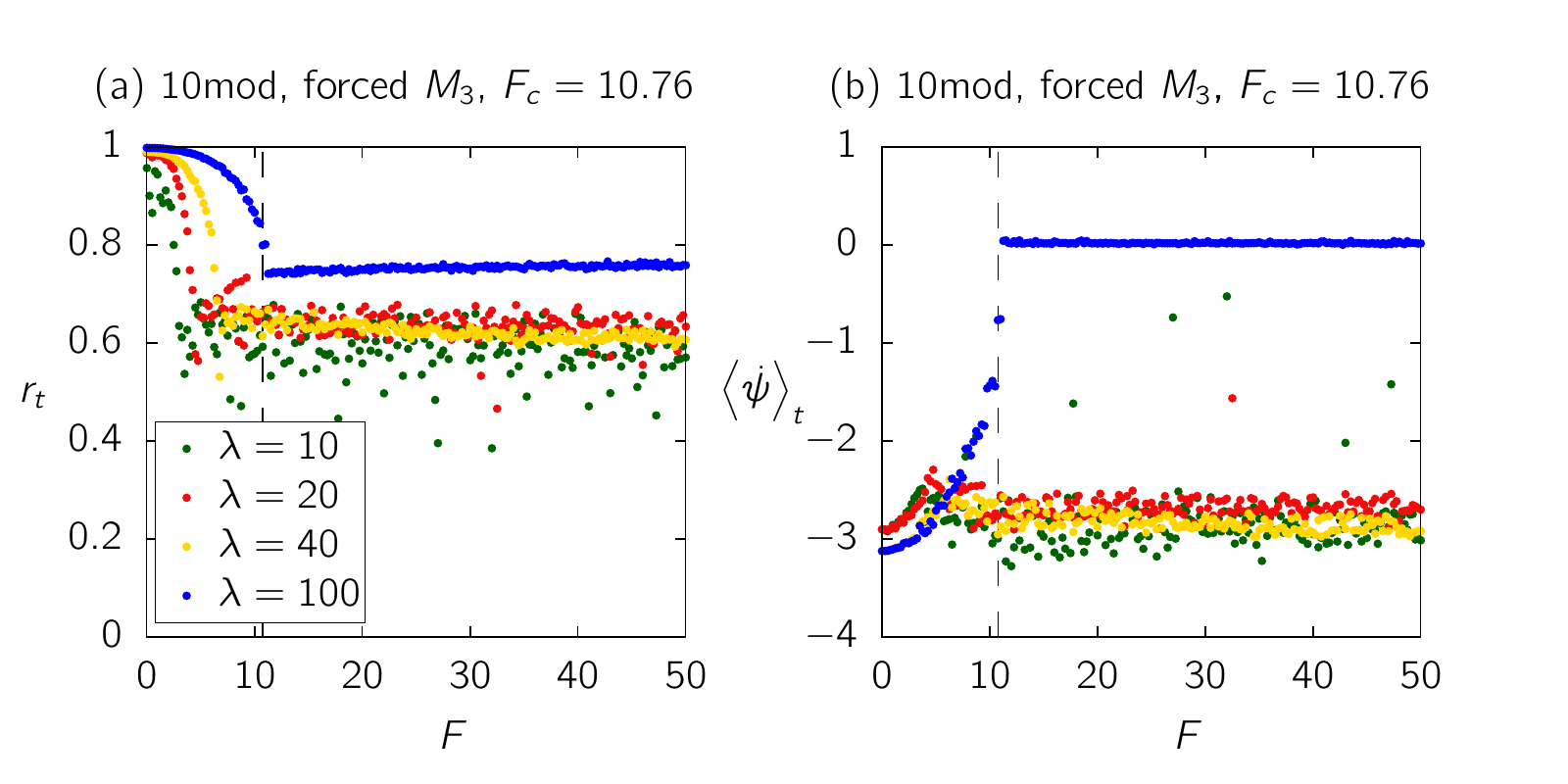}
\caption{ Order parameters as a function of $F$ for the \textit{C. elegans} neural network where the external 
force acts only on the largest module of the network divided in 10 communities. Panel (a) computes $r_{total}$ and 
(b) $\dot\psi_{total}$ for fixed $\lambda$. The dashed lines indicate the critical force, $F_c = 10.76$. }
\label{10modsimu}
\end{figure}

Lastly, we have also performed dynamical simulations on the network with 10 modules. We applied the stimulus on the
largest module, which contains 76 neurons. The results in figure \ref{10modsimu} show the total order parameters $r_t$
and $\dot\psi_t$ as a function of $F$ for different values of $\lambda$. The expected critical force to reach global
synchronization is $F_c = 10.76$ (dashed lines), but numerical simulations show that the maximum value of $r_t$ is
$\approx 0.77$. This corroborates our results obtained with the 3-modules partition and indicates that global
synchronization is hindered by the modular structure of the network, independent on the specific way the modules are
defined. The results are qualitatively similar even using the present modular partition, which has the highest value of
$Q$ we could find for the {\it C. elegans} electric junction network.

\chapter{Conclusions and perspectives}

The Kuramoto model is perhaps the simplest dynamical system that allows the study of synchronization and has become a paradigm, being extensively explored in the last years in connection with biological systems, neural networks and the social sciences. In chapter \ref{chapter2} we reviewed the analytical calculations made by Kuramoto, where he considered a system composed by identical units (oscillators) interacting with each other via a coupling parameter. We saw that for small values of the coupling strength the units move as if they were independent, but as the coupling increased beyond a critical value, a finite fraction of oscillators started to move together. This fraction increased smoothly until the coupling reached a large enough value, where the whole system oscillates on the same frequency, leading to global synchronization. The crossover between these two regimes characterizes a second order phase transition. Kuramoto showed an exact analytical expression for the minimum value of the coupling strength in a system composed of infinite oscillators. The main idea of the mathematical approach is to define a probability density function in order to pass from discrete to continuous limit and analyse the cases where the system is desynchronized and partially synchronized. 

The Kuramoto system can be easily extended to complex networks if we allow the oscillators to interact via an adjacency matrix. We showed in chapter \ref{chapter2} that the theoretical curve behavior of the original model is satisfied in fully connected, random and scale-free networks and that the larger the number of elements, the better is the result. The scenario of the second order phase transition is, however, abruptly changed under specific conditions. We reviewed some cases where the system went into a first order phase transition, a behavior termed explosive synchronization. This phenomenon appears in many applications ranging from neuroscience, where it is observed on epileptic seizures and waking from anesthesia, to electronic devices, as the Rössler units.

The original Kuramoto model exhibits spontaneous synchronization. However, in many biological systems, we can see that the synchronization phenomena are frequently dependent of external stimuli. Information processing in the brain requires the synchronous firing of specific groups of neurons to respond to external stimuli \cite{brette,usrey,salinas}. In the retina, neighboring cells synchronize at a very fine timescale to keep up with the constant motion of the eyes and the head \cite{brivanlou,meister} and information about visual stimuli is contained in the relative spike timing \cite{gollish}. In the auditory system, sound localization is determined by phase locking in the auditory nerve fibers \cite{joris94}, producing correlations in spike timing that encodes the physical location of the sound source \cite{joris98}. Although synchronization is ubiquitous in neural systems \cite{brette} the specific group of neurons that synchronize depends on the type of stimulus and the  time scale of the synchronization might vary from milliseconds \cite{gollish} to rates up to 170 spikes per second \cite{joris94}. We can use the Kuramoto system to model all these examples, but it is necessary to include the influence of an external force on the mathematical equations. 

In chapter \ref{chapter3} we studied the forced Kuramoto model. We added a periodic external drive on the original equations reviewing the work of Childs and Strogatz \cite{Childs2008} using similar techniques of chapter \ref{chapter2}. In this sense, we defined a density function and took the continuum limit. It was possible to expand the density function in Fourier series and then we used the Ott and Antonsen ansatz to restrict the analysis to a special family of densities - which obey analytical conditions to perform the calculations. This technique allowed us to reduce the infinite dimensional system into a bi-dimensional problem. We saw that the analytical results exhibit rich dynamics: the stability diagram shows a set of curves composed of saddle-node, SNIPER, Hopf and homoclinic bifurcations. We also reproduced the phase portraits showing the transitions between the five regions that appear in the stability diagram. As we discussed in the end of chapter \ref{chapter3}, the stability diagram in a zoom out scale is essentially divided in two regions, concerning the competition between the regimes where the system is synchronized with the external force and spontaneously.

The idea introduced in chapters \ref{chapter2} and \ref{chapter3} allowed us to study the forced Kuramoto model on networks. We considered the analytical results of Childs and Strogatz and applied the external force only on a fraction of the oscillators interacting via an adjacency matrix. The problem was inspired by artificial heart pacemakers \cite{Reece2012} and information processing in the brain induced by an external stimulus \cite{Pikovsky2003}. In both cases the stimulus is perceived by a subset of the system (a heart chamber or photo-receptor cells in the eye) and propagates to other parts of the network structure. In chapter \ref{chapter4} we have explored the conditions for global synchronization as a function of the fraction of nodes being forced and how these conditions depend on network topology, strength of internal coupling and intensity of external forcing. The numerical calculations showed that the force required to synchronize the network with the external drive increases as the inverse of the fraction of forced nodes. However, for a given coupling strength, synchronization did not occur  below a critical fraction, no matter how large was the force. Network topology and properties of the forced nodes also affected the critical force for synchronization. In scale-free networks, for example, when the external force is applied to nodes with highest degree, the critical force for synchronization is smaller than when applied to the same number of randomly nodes or to the nodes with the lowest degrees. We also developed analytical calculations for the critical force for synchronization as a function of the fraction of forced oscillators and for the critical fraction as a function of coupling strength.

The numerical and analytical results of chapter \ref{chapter4} led us to apply the forced Kuramoto model on a real complex network. We chose the \textit{C. elegans} electrical junction network because of its small size, containing only 302 neurons, and because its whole nervous system is mapped and available online \cite{wormatlas, openworm}. In this sense, we studied in chapter \ref{chapter5} the response of the nematode's neural network to external stimuli using the partially forced Kuramoto model. We applied the force to specific groups of neurons, classified in topological modules, physical distribution and functional classes.  We found that topological modules do not contain purely anatomical groups (ganglia) or functional classes, corroborating previous results, and that stimulating different classes of neurons led to very different responses, measured in terms of synchronization and phase velocity correlations. In all cases the modular structure hindered full synchronization, protecting the system from seizures. The responses to stimuli applied to topological and functional modules showed pronounced patterns of correlation or anti-correlation with other modules that were not observed when the stimulus was applied to a ganglion with mixed functional neurons.

So far we have seen many examples in real systems that exhibit spontaneous or induced synchronization. In this work we studied the Kuramoto model on synthetic networks and in the \textit{C. elegans} electrical junction network, as a simple application on real systems. We want to go further and test our model on more complex systems such as the cat cerebral cortex and even the human brain. In the latter case, the comprehension of how the brain stores and processes information is one of the major scientific challenges of this century. The difficulties of this task are the large number of neurons in the network, in the order of billions, the complex form with which they are connected and the dynamics of chemical and electrical impulses that occur continuously. This highly complex system gives rise to sensory perceptions, coordinates decision making and, at least in humans, stores consciousness. Many efforts have been made in order to understand these processes and one of the paths that has been used is the construction of simplified models to study specific mechanisms.

In order to study how the synchronization processes take place on neural networks one can use a more realistic model instead of the Kuramoto system. There are several models on the literature that describe the dynamics of biological neurons involving a system of nonlinear coupled differential equations. One of the most known examples is the Hodgkin-Huxley model \cite{HH1952} which relates the different ionic currents that flow through the membrane of the neuron. However, this model is computationally costly, since it involves four complex differential equations, which makes the numerical integration very slow, specially for large systems, containing many neurons \cite{Viana2015}. In computational terms the model proposed by Rulkov \cite{Rulkov2001,Rulkov2002} is an excellent alternative to the neuronal dynamics because it essentially describes the same mechanisms as the Hodgkin-Huxley, but is simples than its typical dimensionally reduced versions as, for instance, the FitzHugh-Nagumo system \cite{FHNagumo}. The model is defined by a discrete two-dimensional map describing the phenomenological aspects of the so called bursting neurons \footnote{``Bursting neurons'' are neurons that repeatedly fires discrete groups or burst of spikes. Each burst is characterized by a followed period of quiescence before the next occurs. These neurons are important for motor pattern generation and synchronization.}. Thus, one of the possible extensions of this work is to use the Rulkov model on real networks and include the influence of an external force, such as the periodic term we have introduced in equation (\ref{forced2}). In this sense we could proceed as we have done in chapters \ref{chapter4} and \ref{chapter5} applying the stimulus on a specific group of neurons and then test the response of the network on them. If we work with a modular network, we could also study if its structure protects the systems from ``failures''. This could allow us to compare our results of the \textit{C. elegans} neural network obtained with the Kuramoto forced model with the Rulkov neurons applied on the same system.

Another interesting phenomenon in network synchronization appears in the so called Janus oscillators. This system was studied recently by Nicolaou et al. \cite{janus} and exhibits simultaneously chimera states, where incoherence and synchronization coexists in identically coupled oscillators, explosive synchronization, where the transition to synchronization is abrupt, and asymmetry-induced synchronization (AIS) \cite{AIS1, AIS2, AIS3, AIS4},  characterized by the existence of specific asymmetries when the oscillators or their couplings are set to be nonidentical. The occurrence of these phenomena has been observed in other systems independently, which turns the Janus oscillators the first system that concurrently show all these behaviors. The units are set as a pair of phase oscillators distributed on a ring network topology. We can label each Janus oscillator $i$ as a pair of phases $(\theta_i^1, \theta_i^2)$, where 1 and 2 denotes each unit of one oscillator. The dynamic is defined by a pair of differential equations $(\dot \theta_i^1, \dot \theta_i^2)$ where each component has a natural frequency, $(\omega_i^1, \omega_i^2)$, an internal coupling $(\sin(\theta_i^2-\theta_i^1), (\sin(\theta_i^1-\theta_i^2))$ with intensity $\beta$ and an external coupling $(\sin(\theta_{i-1}^2-\theta_i^1), \sin(\theta_{i+1}^1-\theta_i^2))$ with intensity $\sigma$. The numerical results show that, in the space defined by the Kuramoto order parameter versus the coupling strength $\sigma$ for fixed $\omega$ and $\beta$, there are several stable solution branches where the dynamic is very sensitive to initial conditions. The transitions between the chimera states, the explosive synchronization and to the AIS are analysed in the context of the bifurcation theory.

The Janus oscillators, as pointed by the authors \cite{janus}, can be used to model the oscillatory dynamics emerging in driven systems that exhibit  anti-ferromagnetic order when subject to an external magnetic field, which can lead to applications in spintronics. Another interesting system is the rotating flagella in the cells of specific groups of algae, which have internal cellular and environmental interactions that can be modeled as the Janus oscillators. The system can be also applied on complex networks. In this case, it is possible to insert the adjacency matrix in the external coupling term. The element $a_{ij}$ of the matrix can be defined in many ways: one can assume, for example, that $a_{ij}=1$ if the first component of node $i$ is connected with the second oscillator of node $j$, and zero otherwise; another possible scenario is to consider that $a_{ij}=1$ if the first components of nodes $i$ and $j$ are connected, and zero otherwise. A feasible extension is to introduce an external drive forcing, as we did in this work and analyse if the stable solution branches are modified with the intensity and frequency of the external signal.

Throughout this work we saw several examples of how the original Kuramoto system can be used to model theoretical and real complex systems. All these examples are described in the context of two dimensions, where the attribute of each element is a single scalar variable $\theta_i$. However, there are many other applications for which the higher-dimensional space is important. In this sense, the work of Chandra et al. \cite{3Dkuramoto} developed an extension of the Kuramoto model in $D$ dimensions. The motivation is to study three-dimensional systems like the swarm of flying drones, flock of birds and school of fish, as examples. Mathematically, the model can be generalized to higher dimensions if we rewrite equation (\ref{original}) for $\theta_i$ as a function of the evolution of unit vectors. The main result of the paper is the striking difference between the usual two-dimensional Kuramoto model and its generalization to three-dimensions: while the first case exhibits a continuous phase transition to the synchronized state, the second shows a discontinuous phase transition. For negative values of the coupling strength, which can be seen as repulsive interactions, the elements are completely desynchronized, but as we increase the coupling strength through zero, a discontinuous jump to the synchronized state is measured by the order parameter. This behavior occurs also for large odd-dimensions, all of them exhibiting the critical coupling at zero.

The work presented here made contributions to the understanding of synchronization processes in neural networks subjected to external perturbations. We hope this thesis will stimulate the readers and convince them that there is sill much interesting work to be done in this field.

\appendix
\chapter{Phase portraits}
\label{app1}

\begin{figure}[H]
\center
\includegraphics[scale=0.5]{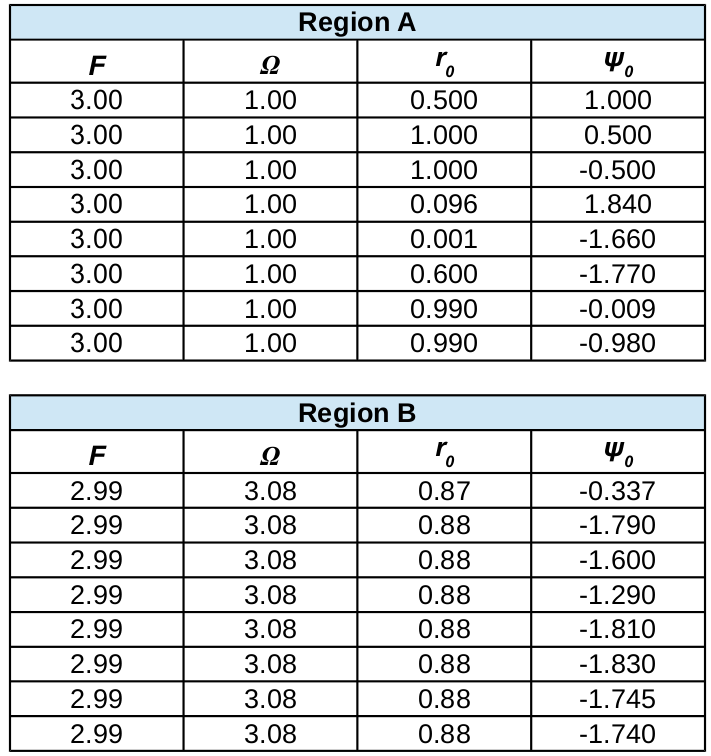}
\end{figure}

\begin{figure}[H]
\center
\includegraphics[scale=0.5]{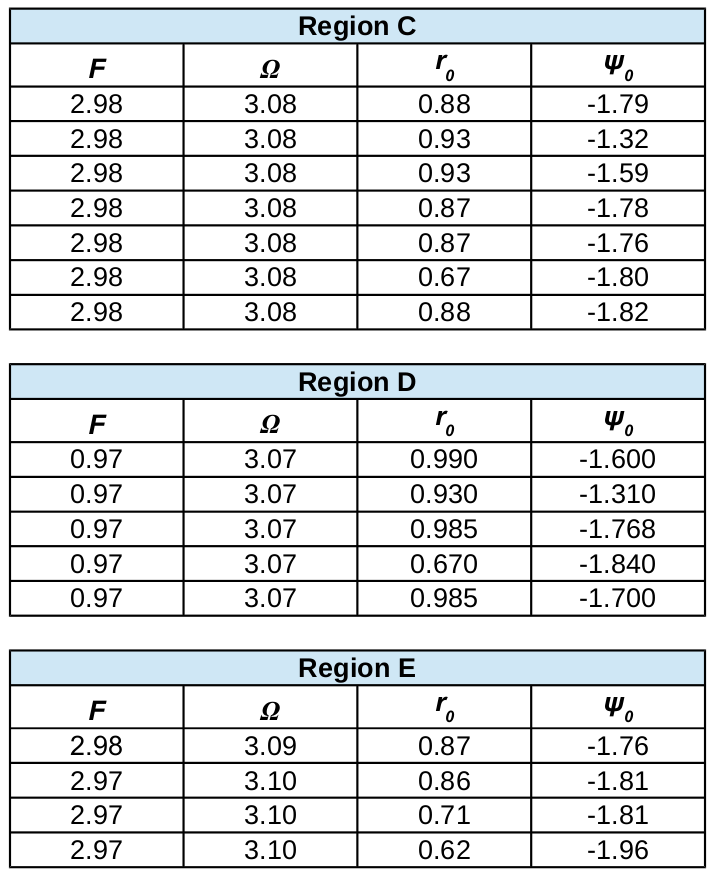}
\end{figure}

\chapter{Data EJ248}
\label{app2}

This appendix contains a list of the 248 neurons we considered in the electric junction network. Neurons are classified according to each class used in the paper \cite{Arruda2019}: anatomical, (classified as head, mid body and tail and also organized by ganglia), functional (classified as sensory neurons, interneurons and motoneurons), and topological with 3, 5 and 10 modules. It is worth noting that the classification into anatomical and functional classes were obtained in WormAtlas \cite{dataexcel}. We downloaded the file Connectivity Data-download (excel file) and we filtered by the Electric junction (EJ) connection. We performed the classification into topological modules using the app ModuLand \cite{moduland1, moduland2} available in the software Cytoscape \cite{cytoscape}. All codes and data used in this thesis are available in \cite{github}.

\begin{figure}[H]
\center
\includegraphics[scale=0.6]{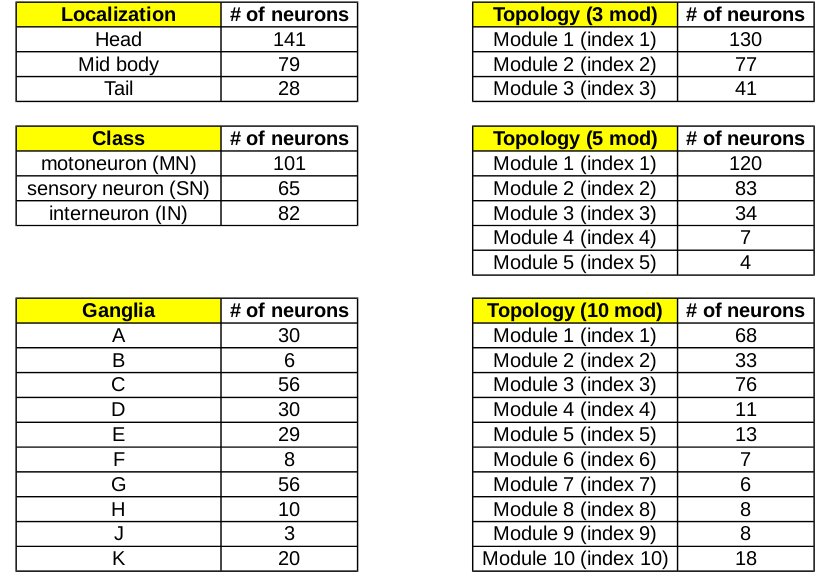}
\end{figure}

\includepdf[pages={1-5},scale=1.0]{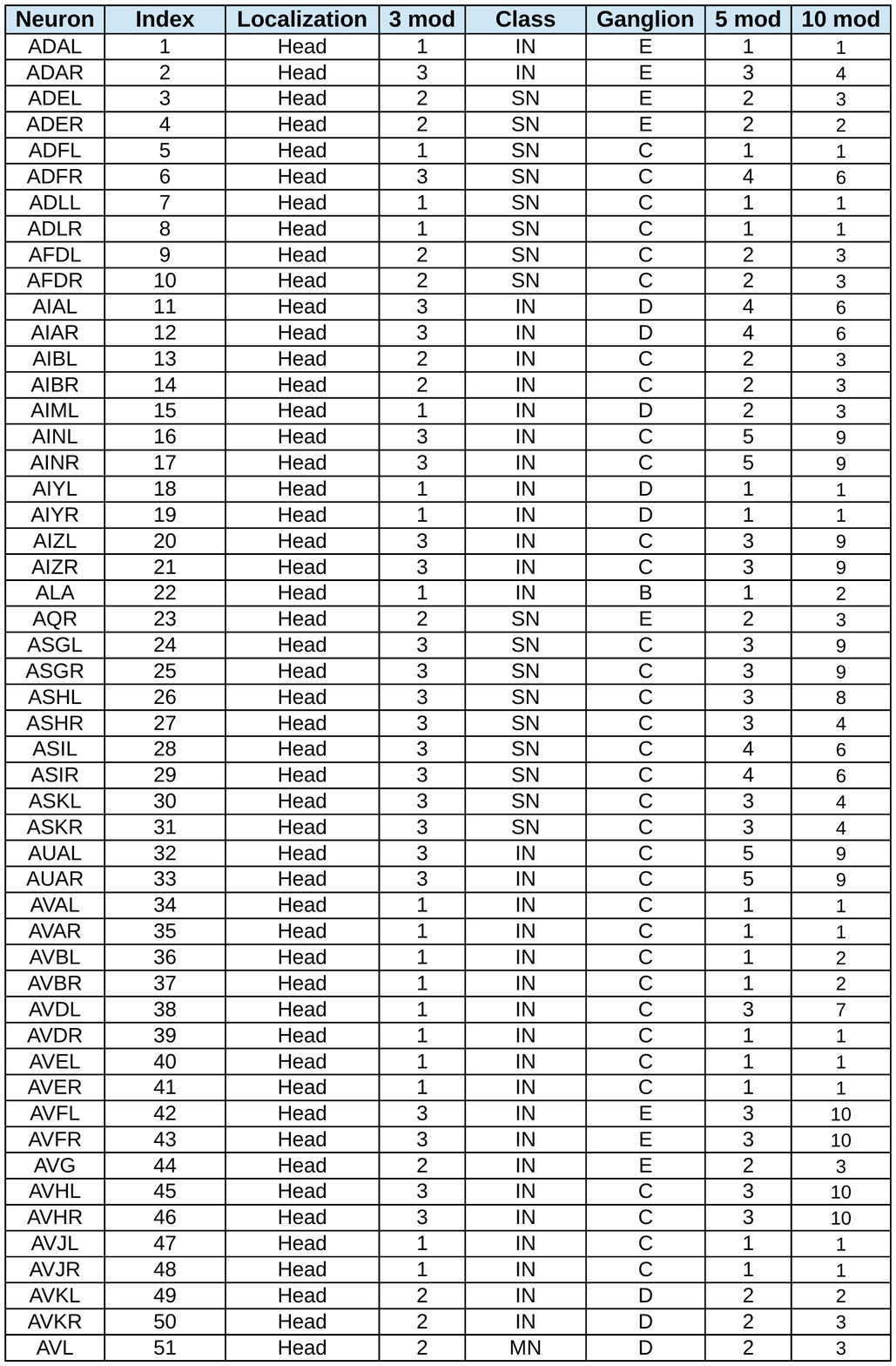}


\begin{thebibliography}{100}
\addcontentsline{toc}{chapter}{Bibliography}

\bibitem{github}
C.elegans-sync github repository.[Online; updated March 2020]. URL: https://github.com/carolina-arruda/C.elegans-sync.

\bibitem{Michaels704}
D. C. Michaels et al. “Mechanisms of sinoatrial pacemaker synchronization: a new hypothesis.” In: Circulation Research 61 (1987), pp. 704–714.

\bibitem{Moiseff181}
A. Moiseff and J. Copeland. “Firefly Synchrony: A Behavioral Strategy to Minimize Visual Clutter”. In: Science 329 (2010), pp. 181–181.

\bibitem{Buck1976}
J. Buck and E. Buck. “Synchronous Fireflies”. In: Scientific American 234 (1976), pp. 74–85.

\bibitem{steriade1997}
M. Steriade. “Synchronized activities of coupled oscillators in the cerebral cortex and thalamus at different levels of vigilance.” In: Cerebral Cortex 7 (1997), pp. 583–604.

\bibitem{kelso2014}
E. Tognoli et al. “The metastable brain”. In: Neuron 81 (2014), pp. 35–48.

\bibitem{Pikovsky2003}
A. Pikovsky et al. Synchronization: A Universal Concept in Nonlinear Sciences Series. 1st ed. 12. Cambridge Nonlinear Science, 2003.

\bibitem{gray1994}
C.M. Gray. “Synchronous Oscillations in Neuronal Systems: Mechanisms and Functions”. In: Journal of Computational Neuroscience 1 (1994), pp. 11–38.

\bibitem{Mackay1997}
W. A. MacKay. “Synchronized neuronal oscillations and their role in motor processes”. In: Trends in Cognitive Sciences 1 (1997), pp. 176–183.

\bibitem{Deco2011}
G. Deco et al. “The role of rhythmic neural synchronization in rest and task conditions”. In: Frontiers in human neuroscience 5 (2011), pp. 4–4.

\bibitem{Kiss2008}
I. Kiss et al. “Resonance clustering in globally coupled electrochemical oscillators with external forcing”. In: Phys. Rev. E 77 (2008), p. 046204.

\bibitem{Pantaleone2002}
J.Pantaleone. “Synchronization of metronomes”. In: American Journal of Physics 70 (2002), pp. 992–1000.

\bibitem{Winfree}
A. T. Winfree. “Biological rhythms and the behavior of populations of coupled oscillators”. In: Journal of Theoretical Biology 16 (1967), pp. 15–42.

\bibitem{Kuramoto1975}
Y. Kuramoto. “Self-entrainment of a population of coupled non-linear oscillators”. In: International Symposium on Mathematical Problems in Theoretical Physics, Berlin/Heidelberg: Springer-Verlag (1975), pp. 420–422.95.

\bibitem{Rodrigues2016}
F. A. Rodrigues et al. “The Kuramoto model in complex networks”. In: Physics Reports 610 (2016), pp. 1–98.


\bibitem{Acebron2005}
J. A. Acebrón et al. “The Kuramoto model: A simple paradigm for synchronization phenomena”. In: Rev. Mod. Phys. 77 (1 2005), pp. 137–185.

\bibitem{Kim2017}
M. Kim et al. “Relationship of Topology, Multiscale Phase Synchronization, and State Transitions in Human Brain Networks”. In: Frontiers in computational neuroscience 11 (2017), pp. 55–55.

\bibitem{Kim2016}
M. Kim et al. “Functional and Topological Conditions for Explosive Synchronization Develop in Human Brain Networks with the Onset of Anesthetic-Induced Unconsciousness”. In: Frontiers in Computational Neuroscience 10 (2016), p. 1.

\bibitem{Wang2016}
C-Q. Wang et al. “Explosive synchronization enhances selectivity: Example of the cochlea”. In: Frontiers of Physics 12 (2016), p. 128901.

\bibitem{Wang2017}
Z. Wang et al. “A small change in neuronal network topology can induce explosive synchronization transition and activity propagation in the entire network”. In: Scientific Reports 7 (2017), p. 561.

\bibitem{Leyva2012}
I. Leyva et al. “Explosive First-Order Transition to Synchrony in Networked Chaotic Oscillators”. In: Phys. Rev. Lett. 108 (16 2012), p. 168702.

\bibitem{Deco2013}
G. Deco et al. “Resting-State Functional Connectivity Emerges from Structurally and Dynamically Shaped Slow Linear Fluctuations”. In: 33 (2013), pp. 11239–11252.

\bibitem{Uhlhaas2006}
P.J. Uhlhaas and W. Singer. “Neural Synchrony in Brain Disorders: Relevance for Cognitive Dysfunctions and Pathophysiology”. In: Neuron 52 (2006), pp. 155–168.

\bibitem{Schmidt2015}
R. Schmidt et al. “Kuramoto model simulation of neural hubs and dynamic synchrony in the human cerebral connectome”. In: BMC Neuroscience 16 (2015), p. 54.

\bibitem{Reece2012}
J. B. Reece et al. Campbell biology : concepts and connections. 7th ed. Pearson; Custom, 2012.

\bibitem{Liu1997}
C. Liu et al. “Cellular Construction of a Circadian Clock: Period Determination in the Suprachiasmatic Nuclei”. In: Cell 91 (1997), pp. 855–860.

\bibitem{Liu2011}
Y-Y. Liu et al. “Controllability of complex networks”. In: Nature 473 (2011), p. 167.

\bibitem{Sakaguchi1988}
H. Sakaguchi. “Cooperative Phenomena in Coupled Oscillator Systems under External Fields”. In: Progress of Theoretical Physics 79 (1988), pp. 39–46.

\bibitem{Ott2008}
E. Ott and T. M. Antonsen. “Low dimensional behavior of large systems of globally coupled oscillators”. In: Chaos: An Interdisciplinary Journal of Nonlinear Science 18 (2008), p. 037113.

\bibitem{Antonsen2008}
T. M. et al. “External periodic driving of large systems of globally coupled phase oscillators”. In: Chaos: An Interdisciplinary Journal of Nonlinear Science 18 (2008), p. 037112.

\bibitem{Childs2008}
L. M. Childs and S. H. Strogatz. “Stability diagram for the forced Kuramoto model”. In: Chaos: An Interdisciplinary Journal of Nonlinear Science 18 (2008), p. 043128.96

\bibitem{Hindes2015}
J. Hindes and C. R. Myers. “Driven synchronization in random networks of oscillators”. In: Chaos: An Interdisciplinary Journal of Nonlinear Science 25 (2015), p. 073119.

\bibitem{Sporns2013}
M. P. van den Heuvel and O. Sporns. “Network hubs in the human brain”. In: Trends in Cognitive Sciences 17 (2013), pp. 683–696.

\bibitem{Rubinov2010}
M. Rubinov and O. Sporns. “Complex network measures of brain connectivity: Uses and interpretations”. In: NeuroImage 52 (2010), pp. 1059–1069.

\bibitem{sporns-betzel2016}
O. Sporns and R. F. Betzel. “Modular Brain Networks”. In: Annu. Rev. Psychol. 67 (2016), pp. 613–640.

\bibitem{bacik2016}
K. A. Bacik et al. “Flow-Based Network Analysis of the Caenorhabditis elegans Connectome”. In: PLOS Computational Biology 12 (2016), pp. 1–27.

\bibitem{antono2015}
C. G. Antonopoulos. “Dynamic range in the C. elegans brain network”. In: Chaos 26 (2016), pp. 1–9.

\bibitem{kim2014}
J.S. Kim and M. Kaiser. “From Caenorhabditis elegans to the human connectome: a specific modular organization increases metabolic, functional and developmental efficiency”. In: Philos. Trans. R. Soc. Lond B Biol. Sci. 369 (2014), pp. 1–9.

\bibitem{global2012}
P. Jiruska et al. “Synchronization and desynchronization in epilepsy: controversies and hypotheses”. In: The Journal of Physiology 591 (2013), pp. 787–797.

\bibitem{global2016}
C. Babiloni et al. “Brain neural synchronization and functional coupling in Alzheimer’s disease as revealed by resting state EEG rhythms”. In: International Journal of Psychophysiology 103 (2016), pp. 88–102.

\bibitem{helfrich2018}
R. F. Helfrich et al. “Old Brains Come Uncoupled in Sleep: Slow Wave-Spindle Synchrony, Brain Atrophy, and Forgetting”. In: Neuron 97 (2018), 221–230.e4.

\bibitem{sowinskia2013}
J. Sowińskia and S. D. Bella. “Poor synchronization to the beat may result from deficient auditory-motor mapping”. In: Neuropsychologia 51 (2013), pp. 1952–1963.

\bibitem{autism2013}
J. Salmi et al. “The brains of high functioning autistic individuals do not synchronize with those of others”. In: NeuroImage: Clinical 3 (2013), pp. 489–497. issn: 2213-1582.

\bibitem{autism2011}
I. Dinstein et al. “Disrupted Neural Synchronization in Toddlers with Autism”. In: Neuron 70 (2011), pp. 1218–1225.

\bibitem{Saa2019}
J. S. Climaco and A. Saa. “Optimal global synchronization of partially forced Kuramoto oscillators”. In: Chaos: An Interdisciplinary Journal of Nonlinear Science 29.7 (2019), p. 073115.

\bibitem{Oh2005}
E. Oh et al. “Modular synchronization in complex networks”. In: Phys. Rev. E 72 (4 2005), p. 047101.

\bibitem{FARODRIGUES2012}
T. K. D. .K.D. Peron and F. A. Rodrigues. “Determining the critical coupling of explosive synchronization transitions in scale-free networks by mean-field approximations”. In: Phys. Rev. E 86 (5 2012), p. 056108.

\bibitem{SAARAFAEL2015}
R. S. Pinto and A. Saa. “Explosive synchronization with partial degree-frequency correlation”. In: Phys. Rev. E 91 (2015), p. 022818.97

\bibitem{GomezGardenes}
J. Gómez-Gardeñes et al. “Explosive Synchronization Transitions in Scale-Free Networks”. In: Phys. Rev. Lett. 106 (12 2011), p. 128701.

\bibitem{Raissa2019}
R. M. D’Souza et al. “Explosive Phenomena in Complex Networks”. In: arXiv e-prints, arXiv:1907.09957 (July 2019), arXiv:1907.09957.

\bibitem{us}
C. A. Moreira and M. A. M. de Aguiar. “Global synchronization of partially forced Kuramoto oscillators on networks”. In: Physica A: Statistical Mechanics and its Applications 514 (2019), pp. 487–496.

\bibitem{Arenas2008}
A. Arenas et al. “Synchronization in complex networks”. In: Physics Reports 469 (2008), pp. 93–153.

\bibitem{Mcgraw2005}
P. N. McGraw and M. Menzinger. “Clustering and the synchronization of oscillator networks”. In: Phys. Rev. E 72 (1 2005), p. 015101.

\bibitem{Mcgraw2007}
P. N. McGraw and M. Menzinger. “Analysis of nonlinear synchronization dynamics of oscillator networks by Laplacian spectral methods”. In: Phys. Rev. E 75 (2 2007), p. 027104.

\bibitem{Mcgraw2008}
P. N. McGraw and M. Menzinger. “Laplacian spectra as a diagnostic tool for network structure and dynamics”. In: Phys. Rev. E 77 (3 2008), p. 031102.

\bibitem{Restrepo2014}
J. G. Restrepo and E. Ott. “Mean-field theory of assortative networks of phase oscillators”. In: EPL (Europhysics Letters) 107 (2014), p. 60006.

\bibitem{Arenas2006}
A. Arenas et al. “Synchronization processes in complex networks”. In: Physica D: Nonlinear Phenomena 224 (2006), pp. 27–34.

\bibitem{Arenas2007}
A. Arenas and A. Diaz-Guilera. “Synchronization and modularity in complex networks”.In: The European Physical Journal Special Topics 143 (2007), pp. 19–25.

\bibitem{wang2015}
C. Wang et al. “One node driving synchronisation”. In: Scientific Reports 5 (2015), p. 18091.

\bibitem{Adler1973}
R. Adler. “A Study of Locking Phenomena in Oscillators”. In: Proceedings of the IRE 34 (1946), pp. 351–357.

\bibitem{Ermentrout1984}
G. B. Ermentrout and J. Rinzel. “Beyond a pacemaker’s entrainment limit: phase walkthrough”. In: American Journal of Physiology-Regulatory, Integrative and Comparative Physiology 246 (1984), R102–R106.

\bibitem{Jensen2002}
R. V. Jensen. “Synchronization of driven nonlinear oscillators”. In: American Journal of Physics 70 (2002), pp. 607–619.

\bibitem{Latora2003}
V. Latora and M. Marchiori. “Economic small-world behavior in weighted networks”. In: The European Physical Journal B - Condensed Matter and Complex Systems 32 (2003), pp. 249–263.

\bibitem{epi1}
S.N. Williams et al. “Epileptic-like convulsions associated with LIS-1 in the cytoskeletal control of neurotransmitter signaling in Caenorhabditis elegans”. In: Hum. Mol. Genet. 13 (2004), pp. 2043–2059.

\bibitem{epi2}
M. G. Risley et al. “Modulating Behavior in C. elegans Using Electroshock and Antiepileptic Drugs”. In: PLOS ONE 11 (2016), pp. 1–13.98.

\bibitem{parkinson1}
B.A. Martinez et a. “C. elegans as a model system to accelerate discovery for Parkinson disease”. In: Curr. Opin. Genet. Dev. 44 (2017), pp. 102–109.

\bibitem{parkinson2}
J. F. Cooper and J.M. Van Raamsdonka. “Modeling Parkinson’s Disease in C. elegans”. In: J. Parkinsons Dis. 8 (2018), pp. 17–32.

\bibitem{wormatlas}
Z.F. Altun et al. “WormAtlas”. In: (2002-2018). Online; accessed December 2018. url: http://www.wormatlas.org/.

\bibitem{openworm}
The Open Worm Project. [Online; accessed December 2018]. url: http://openworm.org/.

\bibitem{moduland1}
M. Szalay-Bekő et al. “ModuLand plug-in for Cytoscape: determination of hierarchical layers of overlapping network modules and community centrality”. In: Bioinformatics 28 (2012), pp. 2202–2204.

\bibitem{moduland2}
I. Kovács. “Community Landscapes: An Integrative Approach to Determine Overlapping Network Module Hierarchy, Identify Key Nodes and Predict Network Dynamics”. In: PLOS ONE 5 (2010), pp. 1–14.

\bibitem{Arruda2019} 
C. A. Moreira and M. A. M. de Aguiar. “Modular structure in C. elegans neural network and its response to external localized stimuli”. In: Physica A: Statistical Mechanics and its Applications 533 (2019), p. 122051.

\bibitem{Baptista2010}
M.S. Baptista et al. “Combined effect of chemical and electrical synapses in Hindmarsh-Rose neural networkson synchronization and the rate of information”. In: Phys. Rev. E 82
(2010), pp. 1–12.

\bibitem{Antonopoulos2015}
C. G. Antonopoulos et al. “Do Brain Networks Evolve by Maximizing Their Information Flow Capacity?” In: PLOS Computational Biology 11 (2015), pp. 1–29.

\bibitem{Borges2017}
R.R. Borges et al. “Spike timing-dependent plasticity induces non-trivial topology in the brain”. In: Neural Networks 88 (2017), pp. 58–64.

\bibitem{Fortunato2010}
A. Fortunato. “Community detection in graphs”. In: Physics Reports 486 (2010), pp. 75–174.

\bibitem{newman2004}
M. E. J. Newman and M. Girvan. “Finding and evaluating community structure in networks”. In: Phys. Rev. E 69 (2 2004), p. 026113.

\bibitem{newman2006}
M. E. J. Newman. “Finding community structure in networks using the eigenvectors of matrices”. In: Phys. Rev. E 74 (3 2006), p. 036104.

\bibitem{cytoscape}
P. Shannon et al. “Cytoscape: A Software Environment for Integrated Models of Biomolecular Interaction Networks”. In: Genome Research 13 (2003), pp. 2498–2504.

\bibitem{sohn}
Y. Sohn et al. “Topological Cluster Analysis Reveals the Systemic Organization of the Caenorhabditis elegans Connectome”. In: PLOS Computational Biology 7 (2011), pp. 1–10.

\bibitem{pan2010}
R. K. Pan et al. “Mesoscopic Organization Reveals the Constraints Governing Caenorhabditis elegans Nervous System”. In: PLOS ONE 5 (Feb. 2010), pp. 1–15.

\bibitem{chen2011}
L. R. Varshney et al. “Structural Properties of the Caenorhabditis elegans Neuronal Network”. In: PLOS Computational Biology 7 (2011), pp. 1–21.

\bibitem{nature}
G. Yan et al. “Network control principles predict neuron function in the Caenorhabditis elegans connectome”. In: Nature 550 (2017), p. 519.

\bibitem{arenas2008-2}
A. Arenas et al. “A Complex Network Approach to the Determination of Functional Groups in the Neural System of C. elegans”. In: Bio-Inspired Computing and Communication. Berlin, Heidelberg, 2008, pp. 9–18.

\bibitem{brette}
R. Brette. “Computing with Neural Synchrony”. In: PLOS Computational Biology 8 (2012), pp. 1–18.

\bibitem{usrey}
W. M. Usrey and R. C. Reid. “Synchronous activity in the visual system”. In: Annual Review of Physiology 61 (1999), pp. 435–456.

\bibitem{salinas}
E. Salinas and J. Terrence. “Correlated neuronal activity and the flow of neural information”. In: Nature Reviews Neuroscience 2 (2001), pp. 539–550.

\bibitem{brivanlou}
I. H. Brivanlou, D. K. Warland, and M. Meister. “Mechanisms of Concerted Firing among Retinal Ganglion Cells”. In: Neuron 20 (1998), pp. 527–539.

\bibitem{meister}
M. Meister and J. M. Berry. “The Neural Code of the Retina”. In: Neuron 22 (1999), pp. 435–450.

\bibitem{gollish}
T. Gollisch and M. Meister. “Rapid Neural Coding in the Retina with Relative Spike Latencies”. In: Science 319 (2008), pp. 1108–1111.

\bibitem{joris94}
P. X. Joris et al. “Enhancement of neural synchronization in the anteroventral cochlear nucleus. I. Responses to tones at the characteristic frequency”. In: Journal of Neurophysiology
71 (1994), pp. 1022–1036.

\bibitem{joris98}
P. X. Joris et al. “Coincidence Detection in the Auditory System: 50 Years after Jeffress”. In: Neuron 21 (1998), pp. 1235–1238.

\bibitem{HH1952}
A. L. Hodgkin and A. F. Huxley. “A quantitative description of membrane current and its application to conduction and excitation in nerve”. In: The Journal of physiology 117 (1952).

\bibitem{Viana2015}
F.A.S. Ferrari et al. “Phase synchronization of coupled bursting neurons and the generalized Kuramoto model”. In: Neural Networks 66 (2015), pp. 107–118.

\bibitem{Rulkov2001}
Nikolai F. Rulkov. “Regularization of Synchronized Chaotic Bursts”. In: Phys. Rev. Lett. 86 (2001), pp. 183–186.

\bibitem{Rulkov2002}
Nikolai F. Rulkov. “Modeling of spiking-bursting neural behavior using two-dimensional map”. In: Phys. Rev. E 65 (2002), p. 041922.

\bibitem{FHNagumo}
P.T.C. Barbosa and A. Saa. “Chaotic oscillations in singularly perturbed FitzHugh-Nagumo systems”. In: Chaos, Solitons \& Fractals 59 (2014), p. 28.

\bibitem{janus}
Zachary G. Nicolaou, Deniz Eroglu, and Adilson E. Motter. “Multifaceted Dynamics of Janus Oscillator Networks”. In: Phys. Rev. X 9 (2019), p. 011017.

\bibitem{AIS1}
T. Nishikawa and A. E. Motter. “Symmetric States Requiring System Asymmetry”. In: Phys. Rev. Lett. 117 (2016), p. 114101.100

\bibitem{AIS2}
T. Nishikawa Y. Zhang and A. E. Motter. “Asymmetry-Induced Synchronization in Oscillator Networks”. In: Phys. Rev. E 95 (2017), p. 062215.

\bibitem{AIS3}
A. Saa. “Symmetries and synchronization in multilayer random networks”. In: Phys. Rev. E 97 (2018), p. 042304.

\bibitem{AIS4}
Y. Zhang and A. E. Motter. “Identical synchronization of nonidentical oscillators: when only birds of different feathers flock together”. In: Nonlinearity 31 (2018), R31.

\bibitem{3Dkuramoto}
Sarthak Chandra, Michelle Girvan, and Edward Ott. “Continuous versus Discontinuous Transitions in the $D$-Dimensional Generalized Kuramoto Model: Odd $D$ is Different”. In: Phys. Rev. X 9 (2019), p. 011002.

\bibitem{dataexcel}
Z.F. Altun et al. “WormAtlas”. In: (2002-2020). Online; accessed March 2020. url: https://www.wormatlas.org/neuronalwiring.html\#Connectivitydata.

\end{thebibliography}
\end{document}